\documentclass[twocolumn]{aastex61}

\usepackage{natbib}
\usepackage{graphicx}

\newcommand{\vjrot}{$S_\mathrm{SED}$}
\newcommand{\uvrot}{$C_\mathrm{SED}$}
\newcommand{\ssed}{$S_\mathrm{SED}$}
\newcommand{\csed}{$C_\mathrm{SED}$}
\newcommand{\ssfruv}{SSFR$_\mathrm{UV,corr}$}
\newcommand{\sfruv}{SFR$_\mathrm{UV,corr}$}

\newcommand{\ssfruvir}{SSFR$_\mathrm{UV+IR}$}
\newcommand{\sfruvir}{SFR$_\mathrm{UV+IR}$}
\newcommand{\delssfr}{$\Delta\log$\,SSFR$_\mathrm{UV,corr}$}
\newcommand{\delsma}{$\Delta\log$~SMA}
\newcommand{\lir}{$L_\mathrm{IR}$}

\shorttitle{THE $UVJ$ DIAGRAM SINCE $z\sim2.5$ IN CANDELS}
\shortauthors{FANG ET AL.}

\begin{document}


\title{Demographics of Star-forming Galaxies since $\lowercase{z}\sim2.5$. I. The $UVJ$ Diagram in CANDELS}

\author{Jerome J.~Fang}
\affiliation{UCO/Lick Observatory, Department of Astronomy and Astrophysics, University of California, Santa Cruz, CA 95064, USA}
\affiliation{Astronomy Department, Orange Coast College, Costa Mesa, CA, 92626, USA}

\author{S.~M.~Faber}
\affiliation{UCO/Lick Observatory, Department of Astronomy and Astrophysics, University of California, Santa Cruz, CA 95064, USA}

\author{David C.~Koo}
\affiliation{UCO/Lick Observatory, Department of Astronomy and Astrophysics, University of California, Santa Cruz, CA 95064, USA}

\author{Aldo Rodr\'iguez-Puebla}
\affiliation{Instituto de Astronom\'{i}a, Universidad Nacional Aut\'{o}noma de M\'{e}xico, A. P. 70-264, 04510, M\'{e}xico, D.F., M\'{e}xico}

\author{Yicheng Guo}
\affiliation{Department of Physics and Astronomy, University of Missouri, Columbia, MO, 65211, USA}

\author{Guillermo Barro}
\affiliation{Department of Astronomy, University of California, Berkeley, CA 94720, USA}

\author{Peter Behroozi}
\affiliation{Steward Observatory, Department of Astronomy, University of Arizona, Tucson, AZ 85719, USA}

\author{Gabriel Brammer}
\affiliation{Space Telescope Science Institute, Baltimore, MD 21218, USA}

\author{Zhu Chen}
\affiliation{Shanghai Key Lab for Astrophysics, Shanghai Normal University, 100 Guilin Road, Shanghai 200234, China}

\author{Avishai Dekel}
\affiliation{Racah Institute of Physics, The Hebrew University, Jerusalem 91904, Israel}

\author{Henry C.~Ferguson}
\affiliation{Space Telescope Science Institute, Baltimore, MD 21218, USA}

\author{Eric Gawiser}
\affiliation{Department of Physics and Astronomy, Rutgers University, Piscataway, NJ 08854, USA}

\author{Mauro Giavalisco}
\affiliation{Department of Astronomy, University of Massachusetts, Amherst, MA 01003, USA}

\author{Jeyhan Kartaltepe}
\affiliation{School of Physics and Astronomy, Rochester Institute of Technology, Rochester, NY 14623, USA}

\author{Dale D.~Kocevski}
\affiliation{Department of Physics and Astronomy, Colby College, Waterville, ME 04901, USA}

\author{Anton M.~Koekemoer}
\affiliation{Space Telescope Science Institute, Baltimore, MD 21218, USA}

\author{Elizabeth J.~McGrath}
\affiliation{Department of Physics and Astronomy, Colby College, Waterville, ME 04901, USA}

\author{Daniel McIntosh}
\affiliation{Department of Physics and Astronomy, University of Missouri, Kansas City, MO 64110, USA}

\author{Jeffrey A.~Newman}
\affiliation{Department of Physics and Astronomy, University of Pittsburgh, Pittsburgh, PA 15260, USA}

\author{Camilla Pacifici}
\affiliation{Space Telescope Science Institute, Baltimore, MD 21218, USA}

\author{Viraj Pandya}
\affiliation{UCO/Lick Observatory, Department of Astronomy and Astrophysics, University of California, Santa Cruz, CA 95064, USA}

\author{Pablo G.~P\'erez-Gonz\'alez}
\affiliation{Universidad Complutense de Madrid, 28040 Madrid, Spain}

\author{Joel R.~Primack}
\affiliation{Department of Physics, University of California, Santa Cruz, CA 95064, USA}

\author{Brett Salmon}
\affiliation{Space Telescope Science Institute, Baltimore, MD 21218, USA}

\author{Jonathan R.~Trump}
\affiliation{Department of Physics, University of Connecticut, Storrs, CT 06269, USA}

\author{Benjamin Weiner}
\affiliation{Steward Observatory, Department of Astronomy, University of Arizona, Tucson, AZ 85719, USA}

\author{S.~P.~Willner}
\affiliation{Harvard-Smithsonian Center for Astrophysics, Cambridge, MA 02138, USA}

\author{Viviana Acquaviva}
\affiliation{Department of Physics, New York City College of Technology, Brooklyn, NY 11201, USA}

\author{Tomas Dahlen}
\affiliation{Space Telescope Science Institute, Baltimore, MD 21218, USA}

\author{Steven L.~Finkelstein}
\affiliation{Department of Astronomy, The University of Texas at Austin, Austin, TX 78712, USA}

\author{Kristian Finlator}
\affiliation{Department of Astronomy, New Mexico State University, Las Cruces, NM 88003, USA}

\author{Adriano Fontana}
\affiliation{INAF - Osservatorio di Roma, I-00040 Monte Porzio Catone, Rome, Italy}

\author{Audrey Galametz}
\affiliation{Max-Planck-Institut f\"ur Extraterrestrische Physik, D-85741 Garching, Germany}

\author{Norman A.~Grogin}
\affiliation{Space Telescope Science Institute, Baltimore, MD 21218, USA}

\author{Ruth Gruetzbauch}
\affiliation{Center for Astronomy and Astrophysics, University of Lisbon, 1600-276 Lisboa, Portugal}

\author{Seth Johnson}
\affiliation{Department of Astronomy, University of Massachusetts, Amherst, MA 01003, USA}

\author{Bahram Mobasher}
\affiliation{Department of Physics and Astronomy, University of California, Riverside, CA 92521, USA}

\author{Casey J.~Papovich}
\affiliation{Department of Physics and Astronomy, Texas A\&M University, College Station, TX 77843, USA}

\author{Janine Pforr}
\affiliation{ESA/ESTEC, Noordwijk, The Netherlands}

\author{Mara Salvato}
\affiliation{Max-Planck-Institut f\"ur Extraterrestrische Physik, D-85741 Garching, Germany}

\author{P.~Santini}
\affiliation{INAF - Osservatorio di Roma, I-00040 Monte Porzio Catone, Rome, Italy}

\author{Arjen van der Wel}
\affiliation{Max-Planck Institut f\"ur Astronomie, D-69117 Heidelberg, Germany}

\author{Tommy Wiklind}
\affiliation{Department of Physics, The Catholic University of America, Washington, DC 20064, USA}

\author{Stijn Wuyts}
\affiliation{Max-Planck-Institut f\"ur Extraterrestrische Physik, D-85741 Garching, Germany}

\correspondingauthor{Jerome Fang}
\email{jjfang@ucolick.org}


\begin{abstract}

This is the first in a series of papers examining the demographics of star-forming galaxies at $0.2<z<2.5$ in CANDELS. We study 9,100 galaxies from GOODS-S and UDS having published values of redshifts, masses, star-formation rates (SFRs), and dust attenuation ($A_V$) derived from UV--optical SED fitting. {In agreement with previous works,} we find that the $UVJ$ colors of a galaxy are closely correlated with its specific star-formation rate (SSFR) and $A_V$. We define rotated $UVJ$ coordinate axes, termed \ssed\ and \csed, that are parallel and perpendicular to the star-forming sequence and derive a quantitative calibration that predicts SSFR from \csed\ with an accuracy of $\sim0.2$~dex. {SFRs from UV--optical fitting and from UV+IR values based on \emph{Spitzer}/MIPS $24\,\mu$m agree well overall, but systematic differences of order 0.2~dex exist at high and low redshifts. A novel plotting scheme conveys the evolution of multiple galaxy properties simultaneously,} and dust growth, as well as star-formation decline and quenching, exhibit ``mass-accelerated evolution'' (``downsizing''). A population of transition galaxies below the star-forming main sequence is identified. These objects are located between star-forming and quiescent galaxies in $UVJ$ space and have lower $A_V$ and smaller radii than galaxies on the main sequence. Their properties are consistent with their being in transit between the two regions. {The relative numbers of quenched, transition, and star-forming galaxies are given as a function of mass and redshift.  } 

\end{abstract}


\section{Introduction}\label{intro}

Understanding galaxy evolution is challenging in part because galaxy properties are so rich.  Galaxies have baryonic mass, dark matter mass, and radial mass profiles of both quantities.  Their spectral energy distributions (SEDs) reflect different star-formation histories, and they have different color and luminosity profiles.  Structure is a key parameter, including flattening, irregularity, and bulge-to-disk ratio.  Added to these are black hole masses, AGN activity, and environmental properties.  Formulating a coherent vision for the evolution of all of these properties and their interrelationships is a formidable task.

Two-color diagrams have emerged as a simple visual tool for understanding galaxies, especially the $UVJ$ diagram (rest-frame $U-V$ vs.~rest-frame $V-J$). An early version of $UVJ$ was introduced by \citet{labbe05}, who used observed $I-K_s$ vs.~$K_s-[4.5]$, which translate to rest-frame $UVJ$ at $z\sim3$. Large scatter was detected in $K_s-[4.5]$ at fixed $I-K_s$, which suggested (from models) that two kinds of galaxies were present: galaxies reddened by old age and galaxies reddened by dust. The first use of $UVJ$ was therefore to discriminate age from dust. Interest in this use gained impetus when star-formation quenching was identified as a key phase in the life of massive galaxies \citep[e.g.,][]{bell04,faber07}, and questions arose as to where, when, and why quenching happens. Because massive galaxies are also dusty \citep[e.g.,][]{reddy06}, having a tool to identify the location of an individual galaxy along the evolutionary track from dusty/star-forming (SF) to quenched became important.

The first rest-frame $UVJ$ diagram of distant ($z\sim2.5$) galaxies was presented by \citet{wuyts07}.  A large spread in $V-J$ at fixed $U-V$ was again seen, and the identification of the high-$(V-J)$ objects as dusty and SF was confirmed by $24\,\micron$ detections.  The elongated, slanting locus of SF galaxies in the $UVJ$ diagram was attributed to different amounts of dust reddening.

\citet{williams09} presented the first richly populated high-redshift $UVJ$ diagram, based on deep IRAC data in the UKIDSS/UDS field. Higher object numbers revealed two separate clumps of red galaxies for the first time: quiescent and dusty/SF.  The border between the two regions was determined, a robust prescription that remains effective today at separating these two galaxy types. \citet{williams09} also studied how the $UVJ$ diagram evolves with redshift. An increase in quiescent objects with time was clearly seen, with the first quenched red galaxies appearing at $z\sim2.0-2.5$. It is now commonplace to use the $UVJ$ diagram to identify quiescent galaxies in high-redshift samples.

Evidence of additional richness in $UVJ$ emerged in subsequent studies. \citet{williams10} combined SED-derived specific star-formation rates (SSFRs) with galaxy $UVJ$ colors to map out the distribution of SSFR in $UVJ$ space.  ``Stripes'' of constant SSFR were seen within the SF population. These stripes ran roughly parallel to the long axis of the SF locus, with higher SSFR in stripes toward the bottom of the distribution, i.e., bluer $U-V$. Similar stripes in SSFR were seen by \citet{patel11} for galaxies at $z\sim0.8$, and stripes in stellar age were seen by \citet{whitaker12} using data from the NEWFIRM survey.  {Ages and SSFRs in these studies were determined by fitting stellar population models to UV--optical SEDs only.   \citet{arnouts13} demonstrated a close link between the dust attenuation $A_V$ determined from UV--optical colors and the infrared excess based on \emph{Spitzer}/MIPS $24\,\mu$m, and \citet{straatman16} demonstrated SSFR stripes using MIPS-based star formation rates (SFRs) out to $z = 2.5$.}       

This paper is the first in a series that combines new CANDELS estimates of dust content and star formation from \citet{santini15} with comprehensive structural data by \citet{vdw12} based on CANDELS imaging. For this, we employ the official CANDELS multiwavelength photometry catalogs in GOODS-S \citep{guo13} and UDS \citep{galametz13}. {The depth of these catalogs permits extending the useful mass limit down to $\sim10^{9.5} M_\odot$ at $z\sim2.5$, which is the estimated mass of the Milky Way at this redshift \citep{vdk13,papovich15}.  Because our aim is to establish accurate trends and correlations, our sample is magnitude-limited at the bright level $H=24.5$ to ensure excellent-quality data.  (The completeness of the sample is discussed in Section \ref{samp_selection}.)}

{This first paper concentrates on the $UVJ$ systematics of SF galaxies.  These are presented using a grid of diagrams laid out by mass and redshift on which evolutionary paths are superimposed. ``Downsizing'' in SSFR, dust, and quenching are clearly visible.  The aforementioned stripes in SSFR are clearly visible, and their stability with mass and redshift is examined.  A quantitative calibration is presented that estimates UV--optical SSFRs from $UVJ$ to an accuracy of 0.2 dex for most galaxies. Such a calibration is useful for quick estimates and for instances where a full SED is not available \citep[e.g., gradient measurements,][]{wang17}. The SEDs of galaxies with similar $UVJ$ colors are shown to be similar from $FUV$ to $K$, and a check is made on the consistency of SED modeling assumptions ($\tau$-models plus Calzetti foreground-screen dust are assumed) by comparing the de-reddened colors of galaxies to the original $\tau$-model tracks. Transition galaxies below the main sequence are identified and are located near the quenched/SF boundary in $UVJ$, as expected. Their low $A_V$ and small radii further signify fading star formation.  Their numbers are given relative to quenched and SF galaxies for testing future theoretical models.  Finally, a comparison is made of UV--optical SSFRs to $24\,\mu$m values that considers \emph{residuals} about the SF main-sequence (SFMS), not just absolute values, as have been used previously.  Overall agreement from these various checks is good, but discrepancies in SSFR of order 0.2~dex in zero point are found at high and low redshift that merit future follow-up.} 

This paper is organized as follows. Section \ref{uvj_data} describes the sources of data, sample selection, and the method used to calculate residual quantities used in the paper. How SSFR, dust attenuation, and SED shape vary across the $UVJ$ diagram and the empirical calibration to estimate SSFR from $UVJ$ colors are shown in Section \ref{uvj_trends}. Section \ref{uvj_dc} examines the \emph{dust-corrected} $UVJ$ diagram and the clues that it offers {to the accuracy of the SED modeling assumptions} and how star formation proceeds in galaxies on the SFMS. Transition galaxies are discussed in Section \ref{fading_gal}, and the relative numbers of SF, transition, and quiescent galaxies as a function of mass and redshift are presented in Section \ref{fractions}. Our summary and conclusions are given in Section \ref{conclusions}. {UV--optical SFRs are compared to $24\,\mu$m SFRs in the Appendix.}

In this paper, all magnitudes are on the AB system \citep{oke74}, and the following cosmology has been adopted: $H_0 = 70\ \mathrm{km\ s^{-1}\ Mpc^{-1}}$, $\Omega_\mathrm{m}=0.3$, and $\Omega_\Lambda=0.7$.

\section{Data and Sample Selection}\label{uvj_data}

This study makes use of the rich multi-wavelength and ancillary datasets produced by the Cosmic Assembly Near-Infrared Deep Extragalactic Legacy Survey \citep[CANDELS;][]{grogin11,koekemoer11}. Out of the five fields targeted in the survey, we use data from the first two available fields of the survey, the southern field of the Great Observatories Origins Deep Survey \citep[GOODS-S,][]{giavalisco04} and the UKIDSS Ultra-Deep Survey \citep[UDS,][]{lawrence07}. The data can be retrieved from the \emph{Rainbow} database \citep{barro11a}, a central repository of CANDELS-related data that can be accessed via a web-based interface.\footnote{The \emph{Rainbow} database can be accessed at \texttt{http://rainbowx.fis.ucm.es}.} Below, we summarize the catalogs as well as our sample selection criteria.

\subsection{Multi-wavelength Photometric Catalogs}

Multi-wavelength photometric catalogs exist for both GOODS-S \citep{guo13} and UDS \citep{galametz13}, and the reader is referred to the cited papers for more details on source identification and measurement. Briefly, for both fields, the catalogs were constructed from a combination of ground- and space-based observations, spanning the $U$-band through to $8\,\mu$m. Objects were selected from the \emph{HST}/WFC3 F160W ($H$-band; $1.6\,\mu$m) images and cross-matched to the other datasets. Consistent multi-wavelength photometry was measured using TFIT \citep{laidler07}.

\subsection{Redshifts and Rest-frame Photometry}
The redshifts used in this study include a combination of broadband photometric, moderate-resolution spectroscopic, and grism redshifts. Our first choice, when possible, is to use reliable-quality spectroscopic redshifts from the literature, which are available for both GOODS-S and UDS, or redshifts based on \emph{HST}/WFC3 grism spectroscopy \citep[for GOODS-S only;][]{morris15}. Photometric redshifts were taken from the catalog of \citet{dahlen13}, which provides median values of $z$ based on SED fitting outputs from 11 different methods. In all, spectroscopic redshifts were used for 22.3\% of our final sample, grism redshifts for 5.1\%, and photometric redshifts for 72.6\%. The consistency among all three redshift sources has been previously demonstrated \citep{dahlen13,morris15}. Rest-frame magnitudes in various standard filters, from FUV to $K$, were computed from the redshifts and multi-wavelength observations (D.~Kocevski et al., in preparation) using the EAZY code \citep{brammer08}, which fits a set of galaxy SED templates to the observed photometry. Uncertainties for a given rest-frame magnitude were estimated by combining in quadrature the flux error in the nearest observed-frame bandpass with the template mismatch error determined by \citet[][their Figure 3]{brammer08}. 

\subsection{Stellar Masses and Dust Attenuation}\label{sect2dot3}

The stellar masses, $M_*$, and the visual attenuation, $A_V$, used here were derived from SED fitting procedures applied to the NUV--NIR photometry. Recently, the CANDELS collaboration released a catalog of ``official'' stellar masses for the GOODS-S and UDS fields that combine the results from ten separate SED fitting methods \citep{santini15}. These median masses are more robust than any individual mass determination, as they average over variations in the assumptions used in each method (e.g., star-formation histories, dust prescription, and metallicity). A \citet{chabrier03} initial mass function (IMF) is assumed. The typical formal uncertainty in the median stellar masses is $\sim0.1$~dex, based on the scatter of the methods. A detailed assessment of the methods used to derive stellar masses is presented by \citet{mobasher15}.

To ensure more robust values of $A_V$, we combined results from five methods \citep[labeled 2a$_\tau$, 2d$_\tau$, 12a, 13a$_\tau$, and 14a$_\tau$ by][]{santini15} and computed the median $A_V$. The methods were chosen based on their similar simplifying assumptions ($\tau$-models and the Calzetti dust law applied as a foreground screen). The typical formal uncertainty in the median $A_V$ is $\sim0.1$~mag based on the scatter of the methods.

\subsection{Structural Parameters}

Galaxy structural parameters, as measured by GALFIT \citep{peng02}, are available for all CANDELS galaxies. Details on the measurement procedure and catalog construction were presented by \citet{vdw12}. Briefly, GALFIT was applied to the \emph{HST}/WFC3 F160W ($H$-band) images. Each galaxy was fit with a single-S\'ersic model, and the best-fitting S\'ersic index, semi-major axis (SMA), ellipticity, axis ratio, and position angle were computed along with uncertainty estimates. {The typical uncertainty in these quantities is $\lesssim10\%$ for galaxies in our sample \citep{vdw12}.} In this work, we use the effective radius \emph{along the major axis} (i.e., SMA).  SMA is used as the indicator of galaxy size, rather than circularized effective radius, $R_\mathrm{eff}$, because the latter depends on the axis ratio $b/a$ ($R_\mathrm{eff}\equiv\sqrt{b/a}\times\mathrm{SMA}$), while SMA is a more faithful indicator of \emph{intrinsic} size for inclined disks. 

{Because we used GALFIT measurements based only on the $H$-band images, which correspond to different rest-frame wavelengths as a function of redshift, our structural parameters may be affected by color evolution. However, this and future papers are primarily concerned with \emph{relative} values of SMA for galaxies in narrow bins of mass and redshift. Therefore we need not correct for color evolution. Corrections are likely to be small. \citet{vdw14a} offered corrections as a function of $M_*$ and $z$ to standardize observed galaxy sizes to $V$-band. For a SF galaxy with redshift between $1<z<2$, the correction is $\lesssim10\%$.}

\subsection{Mid- and Far-infrared Data}\label{irdata}

Infrared (IR) observations from \emph{Spitzer}/MIPS 24\,$\mu$m are available for both fields  as part of the FIDEL survey \citep{pgp08}. In addition, \emph{Herschel} observations of GOODS-S were taken as part of the GOODS-\emph{Herschel} \citep{elbaz11}, HerMES \citep{oliver12}, and PEP \citep{magnelli13} surveys, while \emph{Herschel} data for UDS were obtained as part of the CANDELS-\emph{Herschel} campaign (Inami et al., in preparation).  {The MIPS and \emph{Herschel} data were re-reduced by \citet{rawle16}, merging all available data in the archive. Reductions were compared to GOODS-\emph{Herschel} and PEP public catalogs, and images and fluxes are similar. For UDS there is no public release to compare to. Catalogs for MIPS were created with direct detections in several passes, using a PSF-fitting algorithm \citep{pgp05}.  The \emph{Herschel} bands used a prior-based algorithm (using positions from MIPS and IRAC) including deletion of non-resolved neighbors (a difference with the PEP and GOODS-\emph{Herschel} catalogs). Fluxes were measured with a PSF-fitting method, as explained by \citet{pgp10}. This gave individual catalogs in the five \emph{Herschel} bands plus $70\,\mu$m in GOODS-S. Merged PACS and SPIRE catalogs were then produced as explained by \citet{rawle16}, assigning to each PACS or SPIRE source the coordinates of its most probable IRAC/MIPS counterpart. Altogether, we have complete wavelength coverage at \emph{Spitzer}/MIPS 24\,$\mu$m and \emph{Herschel} PACS 100 and 160\,$\mu$m and SPIRE 250, 350, and 500\,$\mu$m.}

{MIPS 24\,$\mu$m sources will prove to be our major source of IR fluxes because they are available for the largest number of galaxies.  In GOODS-S, the typical rms uncertainty is $4\,\mu$Jy, and the faintest sources are $20\,\mu$Jy, making them 5-$\sigma$ detections.  In UDS, the typical rms uncertainty is $14\,\mu$Jy and the faintest sources are $50\,\mu$Jy, making them 3.6-$\sigma$ detections.  Source detection is limited by confusion at the faintest levels. }

{The far-IR merged catalogs were  cross-correlated with the CANDELS optical--IRAC catalog (G.~Barro, private communication) using a $2''$ search radius for MIPS and PACS. The large radius meant that several CANDELS sources were often seen within the search region, and the one with smallest projected distance was selected as the counterpart, not the brightest IRAC 8$\,\mu$m source. }

\subsection{Star-formation Rates}\label{sfr_defn}

The UV--optical SFRs used in this work do \emph{not} come directly from the SED fitting results  \citep{santini15} but were rather derived from the rest-frame near-ultraviolet (NUV; $\lambda\approx2800\,\mathrm{\AA}$) luminosities after correcting for dust using $A_V$ from the \citet{santini15} SED fits. We originally preferred this approach because of its simplicity, more direct relation to the observed SED, and less dependence (we thought) on the assumed star-formation history. However, comparison to the Santini results shows virtually no differences, and so our SSFR values are effectively the same as standard SED-fitting values using $\tau$-models and a \citet{calzetti00} foreground screen.  We will therefore often refer to the two methods interchangeably as ``UV--optical SED fitting''.  The UV absorption assumed at $\lambda\approx2800$ \AA\ is $A_\mathrm{NUV}=1.8A_V$. After correcting the NUV luminosity by this, we converted the NUV luminosity to SFR using the \citet{kennicutt12} calibration:
\begin{equation}\label{sfr_uv}
\mathrm{SFR_{UV,corr}}\ [M_\odot\,\mathrm{yr^{-1}}]=2.59\times10^{-10}L_\mathrm{NUV,corr}\ [L_\odot],
\end{equation}
where the calibration constant assumes a \citet{kroupa01} IMF (essentially identical to the Chabrier IMF used by \citet{santini15}) and $L_\mathrm{NUV,corr}\equiv\nu L_\nu(2800\,\mathrm{\AA})\times10^{0.4A_\mathrm{NUV}}$.

{The above UV--optical rates form the backbone of SSFRs used in this paper, but SSFRs based on adding raw UV and IR luminosities are also considered.  \emph{Herschel} data are far too sparse to make a comparison at high redshift. {Indeed, as shown in the Appendix, only $27\%$ of the sample has photometry from \emph{Spitzer} and only 35\% of these objects are detected by \emph{Herschel}. } The only statistically meaningful test we can make is with MIPS $24\,\mu$m. An extensive comparison is presented in the Appendix.} The main result is that $\mathrm{SFR_{UV,corr}}$ is broadly consistent with these UV+IR rates, so we adopt $\mathrm{SFR_{UV,corr}}$ as our fiducial measure of star-formation activity because it is available for all galaxies. 
 
Given the severe reduction in sample size (and consequent biases that this may present), our analysis would be essentially impossible if it were confined to the small subset of IR-detected objects. We also eschew the common alternative of using a calibrated  ``ladder'' of SFRs ranging from far-IR to optical values \citep[e.g.,][]{wuyts11a}, {because the tests in the Appendix reveal small but significant systematic differences,} and we prefer the homogeneity of having all SFR values on the same system.

\subsection{Sample Selection}\label{samp_selection}

The full GOODS-S and UDS catalogs contain 34,930 and 35,932 objects, respectively. The sample used in our analysis is constructed by applying the following selection cuts to the catalogs:

\begin{enumerate}
	\item Observed F160W magnitude $H<24.5$, as recommended by \citet{vdw14a} to ensure robust GALFIT measurements
	\item Photometry quality flag \texttt{PhotFlag} = 0 to exclude spurious sources, e.g., star spikes and hot pixels, as provided in the catalogs of \citet{santini15}
	\item SExtractor \texttt{CLASS\_STAR} $<0.9$ to reduce contamination by stars
	\item Redshifts within $0.2<z<2.5$ and stellar masses within $9.0<\log M_*/M_\odot<11.0$ to maximize the sample size while maintaining high mass completeness for the majority of our final sample \citep[e.g.,][]{tal14}
	\item Well-constrained GALFIT measurements \citep[quality flag = 0;][]{vdw12}.
\end{enumerate}

{The most important cut in choosing the sample is the $H=24.5$ magnitude limit.  In addition to ensuring GALFIT accuracy (see above), we also depend on having reliable photometric redshifts. \citet{dahlen13} saw an increase in photometric redshift errors from $\Delta z/(1+z) = 0.04$ at $H = 23.0$, 0.045 at $H=24.0$, and 0.06 at $H=25.0$.  The fraction of outliers also increased from 4\% to 5\% to 12\% at these levels. Because our goal of establishing reliable correlations at faint levels requires high-quality data, we adopt $H<24.5$ to optimize both GALFIT measurements and photometric redshifts.}

\begin{figure}

	\centerline{\includegraphics[scale=0.48]{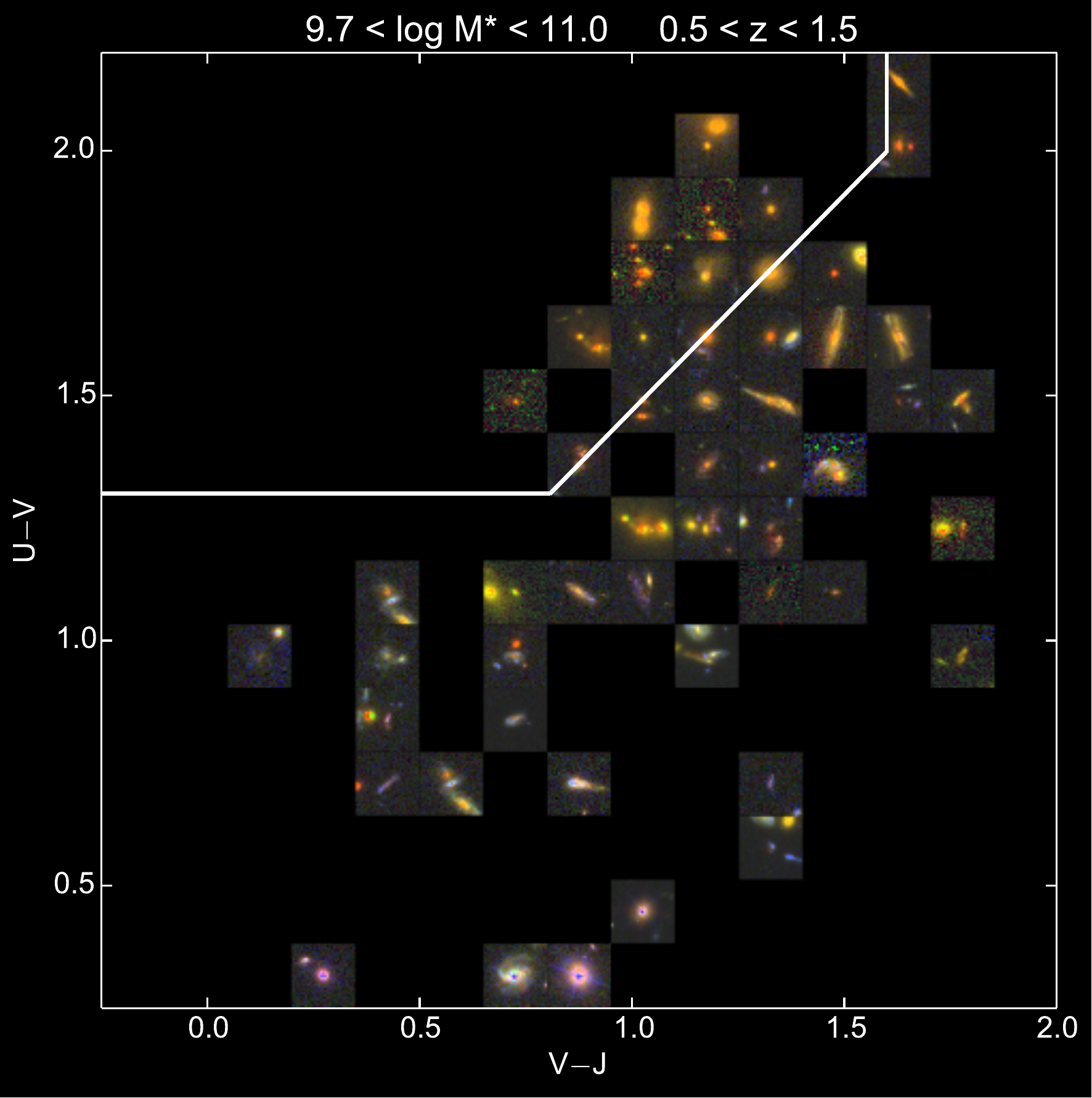}}
	
	\caption{$UVJ$ diagram showing color thumbnail images, $3''$ on a side, of a random sample of galaxies with bad GALFIT measurements \citep[i.e., GALFIT quality flag $\geq1$;][]{vdw12}. The lines delineate the quenched region as given by \citet{williams09}. Visual inspection of the images reveals that $\sim75\%$ of such galaxies are either disturbed or contaminated by neighboring objects, while the remaining  $\sim25\%$ have no obvious problems aside from being small, which may preclude reliable fits.}

	\label{badgalfit}
\end{figure}

The final sample contains 9,135 galaxies: 4,028 from GOODS-S and 5,107 from UDS, or roughly one-eighth of the original catalogs. The $H<24.5$ cut is the most restrictive, followed by the mass and redshift limits. Star-forming galaxies (based on $UVJ$) make up $88\%$ of the final sample (8,060 objects). Table \ref{uvj_sample} details the selection criteria and the resulting sample sizes after each cut. Of particular note is the GALFIT quality flag cut, which excludes $\approx13\%$ of the selected objects lying within our mass and redshift limits. Figure \ref{badgalfit} presents a $UVJ$ diagram showing color postage stamps of a sample of galaxies with bad GALFIT values that have been excluded by this cut. Visual inspection shows that $\sim75\%$ of these galaxies appear to suffer from contamination from nearby objects, some fraction of which are mergers and disturbed. The remaining $\sim25\%$ appear normal. However, many of these latter objects have small angular sizes, which may preclude reliable fits \citep{vdw12}. In this paper, we are not in general using absolute counts of objects, so the loss of these objects \emph{per se} is not an issue.  {We have also verified that the excluded GALFIT objects have almost precisely the same SFR distribution in each mass--redshift bin as the retained objects, so there is no bias created as a function of SFR. Though losing these objects has little impact on our study,} future counting studies will need to take the loss of these objects into account.

\begin{deluxetable*}{lccc} 
\tablecaption{Sample Selection Cuts \label{uvj_sample}}

\tablehead{\colhead{Cut} & \colhead{GOODS-S} &
			\colhead{UDS} & \colhead{Combined}  }

\startdata
Full catalog 					& 34,930 (100\%) & 35,932 (100\%)  & 70,862 (100\%) \\
F160W\,$<24.5$ 					& 9,904 (28.4\%) & 12,223 (34.0\%) & 22,127 (31.2\%) \\
\texttt{PhotFlag} = 0 			& 9,607	(27.5\%) & 11,392 (31.7\%) & 20,999 (29.6\%) \\
\texttt{CLASS\_STAR}\,$<0.9$ 	& 9,376	(26.8\%) & 11,090 (30.9\%) & 20,466 (28.9\%) \\
$0.2<z<2.5$ 					& 7,656 (21.9\%) & 9,534  (26.5\%) & 17,190 (24.3\%) \\
$\log\,M_*<11.0$ 				& 7,585	(21.7\%) & 9,445  (26.3\%) & 17,030 (24.0\%) \\
$\log\,M_*>9.0$ 			& 4,683	(13.4\%) & 5,810  (16.2\%) & 10,493 (14.2\%) \\
GALFIT flag = 0 				& 4,028	(11.5\%) & 5,107  (14.2\%) & 9,135 (12.9\%)\\
Star-forming					& 3,581 (10.3\%) & 4,479  (12.5\%) & 8,060 (11.4\%) \\
Quiescent						& 447 (1.3\%)    & 628 (1.7\%)     & 1,075 (1.5\%)\\
\enddata
\end{deluxetable*}

\begin{figure}

	\centerline{\includegraphics[scale=0.45]{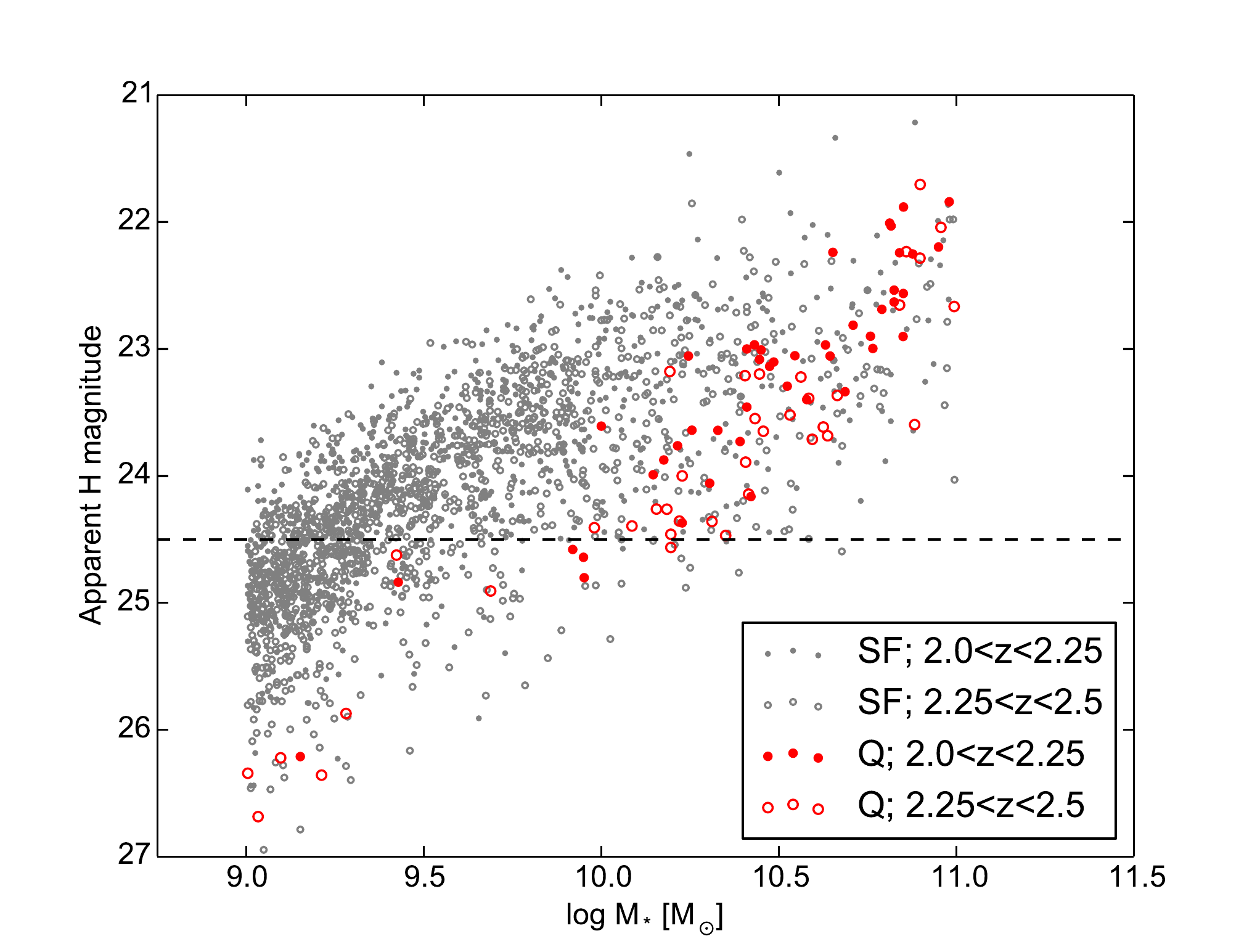}}
	
	\caption{{Apparent $H$ magnitude vs.~stellar mass $M_*$ in the highest redshift bin $z = 2.0-2.5$.  The sample magnitude limit at $H = 24.5$ is shown (dashed line).  Quiescent galaxies (which are selected using $UVJ$, see Figure \ref{uvj_whole}) are the red points.  Filled and open circles are at $z= 2.0-2.25$ and $z = 2.25-2.5$, respectively. SF galaxies are grey. Their vertical width reflects their colors due to dust reddening, with dustier galaxies trending toward fainter $H$ magnitudes.  The SF sample at this redshift is strongly biased to bluer galaxies, and dust-reddened galaxies with $V-J > 1.2$ and $M_*< 10^{10}\,M_\odot$ are largely lost. Nearly all quiescents are captured except for a handful just below the magnitude limit at $10^{9.5-10}\,M_\odot$ (the clump at $H\gtrsim26$ is dubious).}}

	\label{hmag_mass}
\end{figure}

Aside from GALFIT, our sample selected down to $H = 24.5$ is virtually 100\% complete photometrically but corresponds to different mass limits at different redshifts and colors. {This is illustrated in Figure \ref{hmag_mass}. Photometric redshift errors remain below 10\% down to $H=26$ \citep{dahlen13}, which is good enough to show which galaxies are removed by the $H<24.5$ cut.  The sample contains nearly all except the reddest galaxies above $10^{9.5}\,M_\odot$ at $z = 2.0$ but is severely limited to just the very bluest galaxies at $10^{9}\,M_\odot$ at $z = 2.5$. Quantitatively, SF galaxies with $V-J < 0.5$ are 99\% complete at  $10^{9.5}\,M_\odot$  and 51\% complete at  $10^{9}\,M_\odot$, while SF galaxies with $0.5< V-J<1.2$ are 91\% complete at  $10^{9.5}\,M_\odot$ and 29\% complete at $10^{9}\,M_\odot$. SF galaxies with $V-J>1.2$ are only 43\% (13\%) complete at $10^{9.5}$ $(10^9)\,M_\odot$. The small clump of quiescent galaxies below $H=26$ appears dubious (likely from photometric errors), but all other quiescents are captured except for a smattering near $\sim10^{9.5-10}\,M_\odot$.  In particular, the general truncation of quiescents below $10^{10}\,M_\odot$ appears real rather than a result of our magnitude cut.}

To summarize, our sample includes all SF galaxies in most bins but is $\lesssim50\%$ complete for $M_*<10^{9.5}\,M_\odot$ and $z\gtrsim2$. Red galaxies above $10^{10}\,M_\odot$  are captured everywhere, which includes nearly all of them. These estimates are consistent with the completeness limits quoted by \citet{vdw12}.

\subsection{Residuals from the SSFR--Mass and Size--Mass Relations}\label{uvj_residuals}

Some of our parameters (e.g., SSFR, SMA) show strong trends with stellar mass and/or redshift. For our analysis, we ``divide out'' these trends and use quantities that are normalized to the typical galaxy at a given mass and redshift. In particular, we calculate residuals in SSFR and SMA relative to the SSFR--mass and size--mass relations.

\begin{figure}

	\centerline{\includegraphics[scale=0.5]{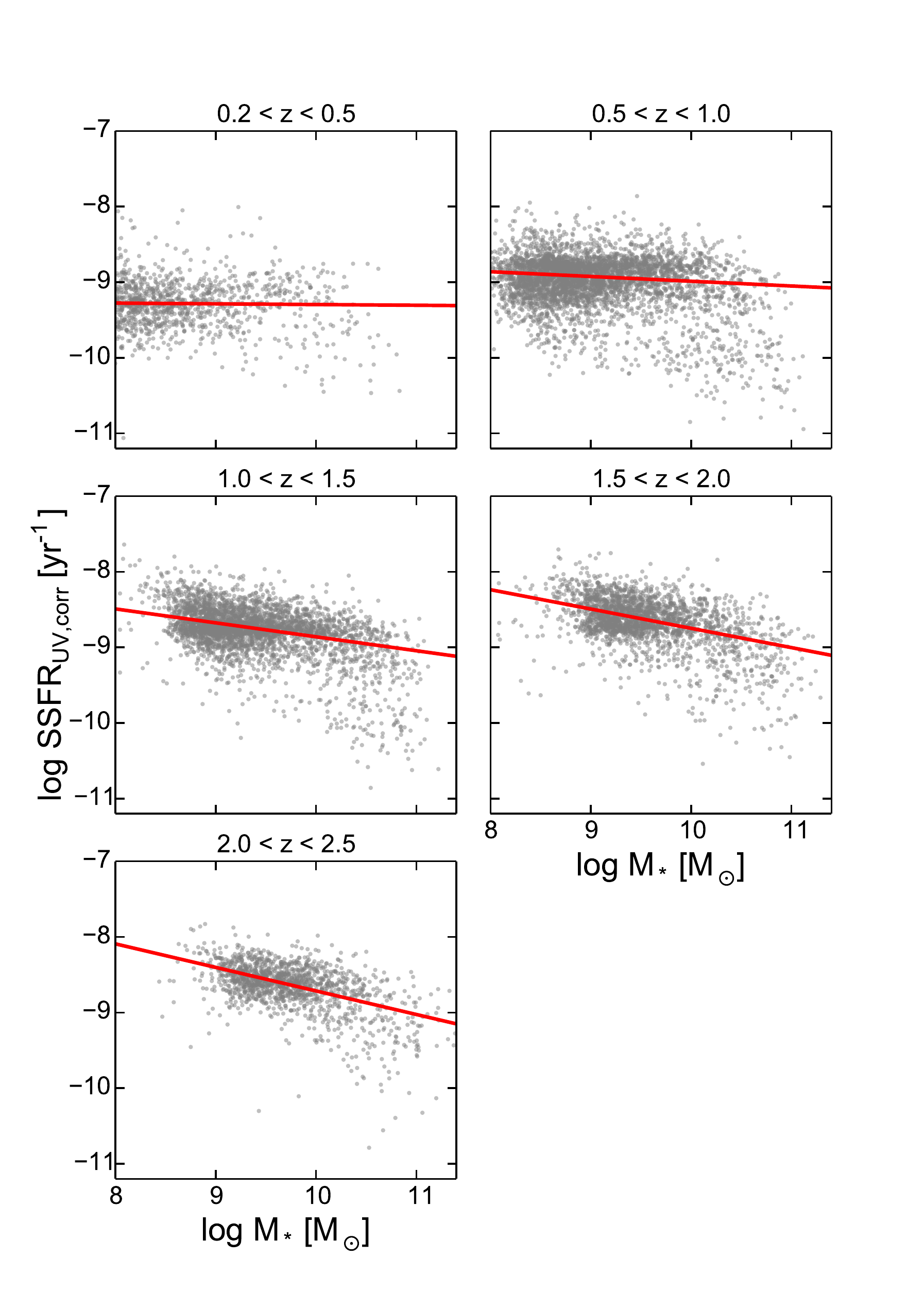}}
	
	\caption{\ssfruv\ vs.~stellar mass in the five redshift bins used in this study. Only $UVJ$-defined SF galaxies are shown (see Figure \ref{uvj_whole}). In each panel, the red line indicates the best-fit linear relation to galaxies located on the ridge line of the SFMS (see Table \ref{ssfr_fit} for fit parameters). The fits were performed by iteratively excluding outlying points (see text). Residuals from the fit, denoted \delssfr, are used later to quantify the relative star-formation activity for galaxies in a given mass and redshift bin.}

	\label{ssfr_mass}
\end{figure}

Figure \ref{ssfr_mass} plots the SSFR--mass relation for galaxies in our five adopted redshift bins, using the dust-corrected SSFRs, \ssfruv, described in Section \ref{sfr_defn}. The distributions of points in Figure \ref{ssfr_mass} clearly trace out the SFMS at these redshifts \citep[e.g.,][]{daddi07,noeske07, salim07,whitaker12}, while the ``green valley'' appears as the tail of objects below the SFMS (abbreviated because only galaxies defined as SF are used). Linear fits were made after excluding outliers, as follows. An initial fit to all SF galaxies was made, then objects greater than $1.5\sigma$ away from the fit were excluded. A second fit was made on this pruned sample, with a new estimate of $\sigma$, and galaxies greater than $1.5\sigma$ away were removed. A third fit was made using this final sample and adopted as the final fit. The parameters of the fits are provided in Table \ref{ssfr_fit}. We opt for this approach in order to obtain relations that pass reasonably close to the highest-density ridge line of the SFMS.

After the fits were in hand, vertical offsets from the relations were calculated for galaxies in each redshift bin. These residuals are denoted \delssfr\ with galaxies lying above (below) the best-fit relation defined to have positive (negative) residuals. \delssfr\ is used later to quantify the relative star-formation activity for galaxies in a given mass and redshift bin. While the fits include galaxies outside the nominal mass range of the sample, our use of relative quantities at fixed mass and redshift means that our results are generally insensitive to the exact slopes or zero points of the fits.

\begin{deluxetable}{ccc} 
\tablewidth{0in}
\tablecaption{Parameters of SSFR--Mass Fits \label{ssfr_fit}}

\tablehead{\colhead{Redshift Range} & \colhead{Slope $a$} &
			\colhead{Zeropoint $b$}  }

\startdata
$0.2<z<0.5$ & $-0.009$ & $-9.296$\\
$0.5<z<1.0$ & $-0.063$ & $-8.987$\\
$1.0<z<1.5$ & $-0.184$ & $-8.860$\\
$1.5<z<2.0$ & $-0.255$ & $-8.748$\\
$2.0<z<2.5$ & $-0.311$ & $-8.714$\\
\enddata
\tablecomments{The best-fit linear relations are of the form $\log$~SSFR = $a(\log M_*-10) + b$, with SSFR in $\mathrm{yr^{-1}}$ and $M_*$ in $M_\odot$ and were determined after excluding outliers.}
\end{deluxetable}

\begin{figure}

		\centerline{\includegraphics[scale=0.45]{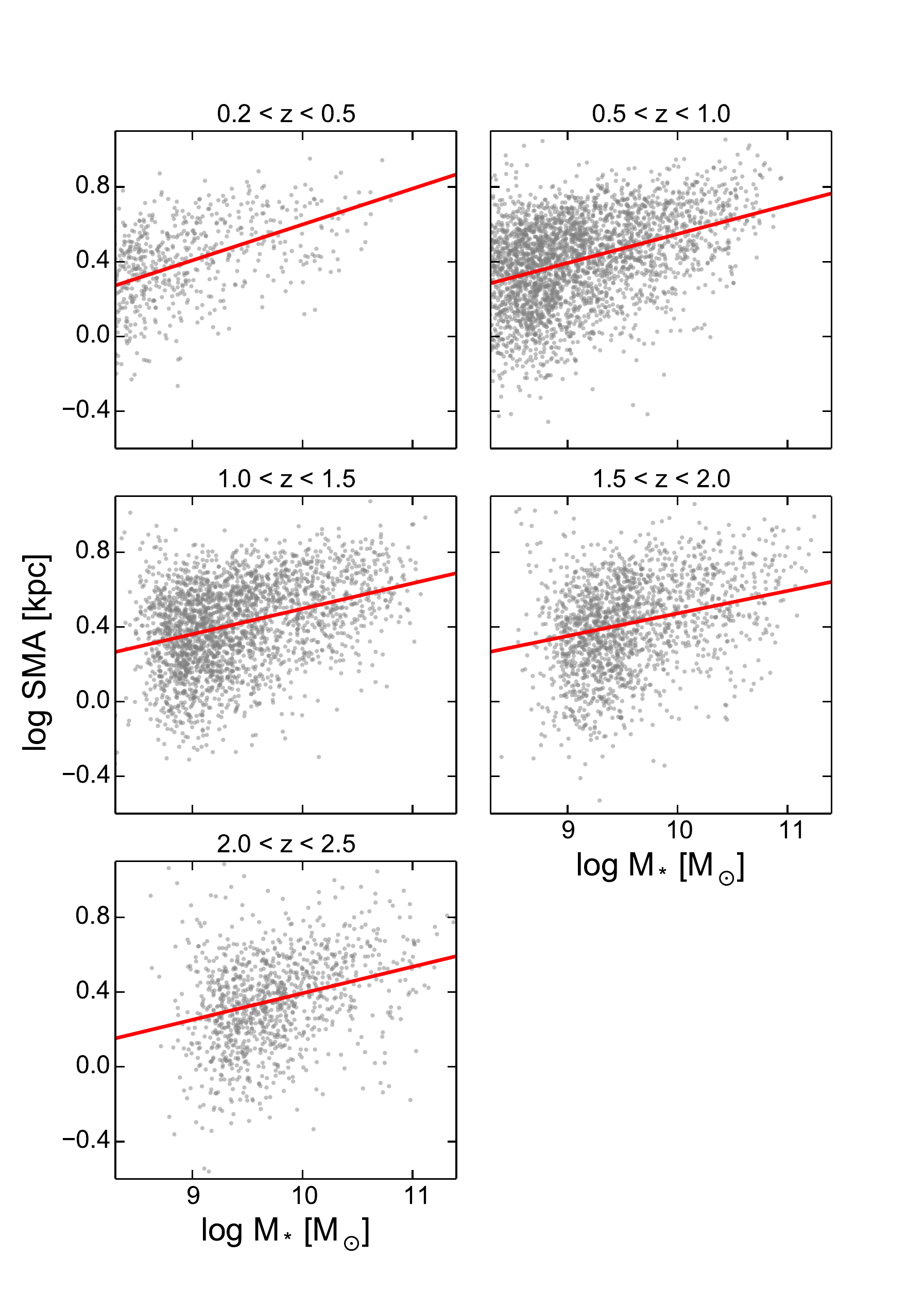}}
	
	\caption{Galaxy semi-major axis (SMA) vs.~stellar mass in the five redshift bins used in this study.  Solid red lines indicate the best-fit linear relation when outliers as well as transition galaxies (\delssfr\,$<-0.45$ dex) are excluded (see Table \ref{sma_fit} for fit parameters). The points plotted are the surviving points used to calculate the final relations.  Residuals from the fit, denoted \delsma, are used later to characterize the relative sizes of galaxies in a given mass and redshift bin.}

	\label{sma_mass}
\end{figure}

A slightly altered fitting procedure was used to calculate the size--mass relation for SF galaxies. Outliers were iteratively removed as above, but transition galaxies with \delssfr\,$<-0.45$~dex were also removed. These prove to be smaller than galaxies on the main sequence (see Figure \ref{delssfr_delsma}) and therefore need to be excluded. Figure \ref{sma_mass} shows the resulting size--mass relation for the retained sample. Fit parameters are given in Table \ref{sma_fit}. Our slopes are systematically shallower by $\sim0.05-0.1$~dex compared to the fits of \citet{vdw14a}. These discrepancies do not affect our results because we are concerned only with \emph{relative} size differences at fixed mass and redshift. Including galaxies outside the mass range of the sample would also not affect our conclusions for the same reason. 

\begin{deluxetable}{ccc} 
\tablewidth{0in}
\tablecaption{Parameters of SMA--Mass Fits \label{sma_fit}}

\tablehead{	
			\colhead{Redshift Range} & \colhead{Slope $a$} &
			\colhead{Zeropoint $b$}  }

\startdata
$0.2<z<0.5$ & $0.192$ & $0.599$\\
$0.5<z<1.0$ & $0.155$ & $0.548$\\
$1.0<z<1.5$ & $0.136$ & $0.497$\\
$1.5<z<2.0$ & $0.121$ & $0.472$\\
$2.0<z<2.5$ & $0.141$ & $0.394$\\
\enddata
\tablecomments{The best-fit linear relations are of the form $\log$~SMA = $a(\log M_*-10) + b$, with SMA in kpc and $M_*$ in $M_\odot$ and were determined after excluding outliers and transition galaxies (\delssfr\,$<-0.45$ dex).}
\end{deluxetable}


\subsection{Sources of Uncertainties in SED-Derived Quantities}\label{errors}

{Finally, various sources of uncertainties in the CANDELS SED-fitting parameters should be considered.  According to \citet{mobasher15}, the rms uncertainty in \emph{relative} stellar masses (which are of primary interest here) comes mainly from the age--dust degeneracy and is about 0.2 dex on average (their Table 9). The CANDELS $A_V$ value determines the dust correction to $L_{2800}$ and therefore affects our measured SFR.  $A_V$ depends on three assumptions: (1) that galaxy stellar populations are described by single $\tau$-models, (2) that all stars suffer the same amount of dust absorption and reddening (foreground screen), and (3) that the reddening curve is well described by a Calzetti law. With regard to (1), it turns out that $\tau$-models, so often assumed in SED fitting, are extreme in having very blue and narrow values of $V-J$, which has the effect of both increasing and narrowing the derived values of $A_V$. Other star-formation histories (such as composite models with old and young stars) would place populations to the right of $\tau$-models in $UVJ$ and yield both lower $A_V$ by a few tenths of a magnitude and lower SSFR by a few tenths of a dex \citep{wang17}. 

{The Calzetti law may also not be universal. \citet{salmon15} found a steeper reddening vector for heavily reddened galaxies. Applying the Salmon curve produces a difference in dust-corrected \csed\ (defined in Section \ref{uvj_trends}) of $-0.1$~mag for $A_V=2$~mag, resulting in SSFR that is about 0.1 dex higher than we obtain.  \citet{kriek13} found a different form of the reddening curve in certain galaxy spectral classes.  Their corrections would decrease our SSFRs by $\sim0.25$~dex on average and add additional rms scatter of $\sim0.24$ dex in quadrature to our error bars.  The effect comes principally from reducing the reddening at 2800 \AA, not from changing the stellar populations. }


In further support of our $A_V$ values, \citet{forrest16} calculated the UV slope $\beta$, infrared excess, and $A_V$ using UV--optical SED fitting for $z = 1-3$ galaxies.  They found tight relations among all three dust estimates and claimed a close relation between dust and $V-J$, in agreement with our results. Finally, Figure \ref{av} in the Appendix plots $A_V$ vs.~the ratio of \ssfruvir\ to the uncorrected UV rate, SSFR$_\mathrm{UV}$.  If $A_V$ is correctly determined, the two should agree perfectly, and it is reassuring to see a strong correlation, albeit with some systematic offsets of order 0.25~mag that merit further investigation.   

However, the real concern is not errors in $A_V$ \emph{per se} but the impacts they could have on \ssfruv.  Of prime interest in future work is the accuracy of the residual \delssfr\ about the SFMS. This is tested independently of $A_V$ in the Appendix by comparing to residual $\Delta\log$\,\ssfruvir\ from MIPS $24\,\mu$m. A total rms scatter is found of 0.24~dex, which, if assigned equally to both quantities, implies an rms error in \delssfr\ of 0.17~dex.  In addition, systematic zero point differences of order 0.2 dex appear on the main sequence that vary with redshift. However, since \emph{relative values} of SSFR on the MS ridgeline are preserved, such errors do not affect our conclusions, at least not for main sequence galaxies. Errors may be larger for transition and quenched galaxies, as mentioned in the Appendix, but these objects are not the major focus of this paper. 

\section{Systematic Trends in the $UVJ$ Diagram}\label{uvj_trends}

Figure \ref{uvj_whole} shows the rest-frame $UVJ$ diagram for all galaxies in the final sample. Most of our sample lies within the SF region. The locus of SF galaxies in Figure \ref{uvj_whole} is not a line but is rather extended in two directions, having one long axis and one short axis crosswise to it.  As is known, these two coordinates can be identified with two important parameters of galaxies, namely $A_V$ and SSFR.

To further quantify these relationships, we define rotated coordinate axes, hereafter \ssed\ and \csed, that are parallel and perpendicular to the SF sequence, respectively, as shown in Figure \ref{uvj_whole}. To determine the orientation, all the SF galaxies in Figure \ref{uvj_whole} were first fit with a linear relation ($U-V$ vs.~$V-J$), and then the slope of this line is used to define 
\begin{eqnarray}
        S_\mathrm{SED} &=& (V-J)\cos\theta+(U-V)\sin\theta\\
              &=& 0.82(V-J)+0.57(U-V)\\
	C_\mathrm{SED} &=& (U-V)\cos\theta-(V-J)\sin\theta\\
              &=& 0.82(U-V)-0.57(V-J),
\end{eqnarray}
where $\theta=34.8^{\circ}$ is the inverse tangent of the slope of the best-fit line. The naming convention of these new coordinates reflects the fact that \ssed\ and \csed\ are like principal components: \ssed\ measures the net \emph{slope} of the spectrum from $U$ to $J$ while \csed\ is approximately the \emph{curvature}, given by the slope difference above and below the 4000 \AA\ break.

\begin{figure}

		\centerline{\includegraphics[scale=0.5]{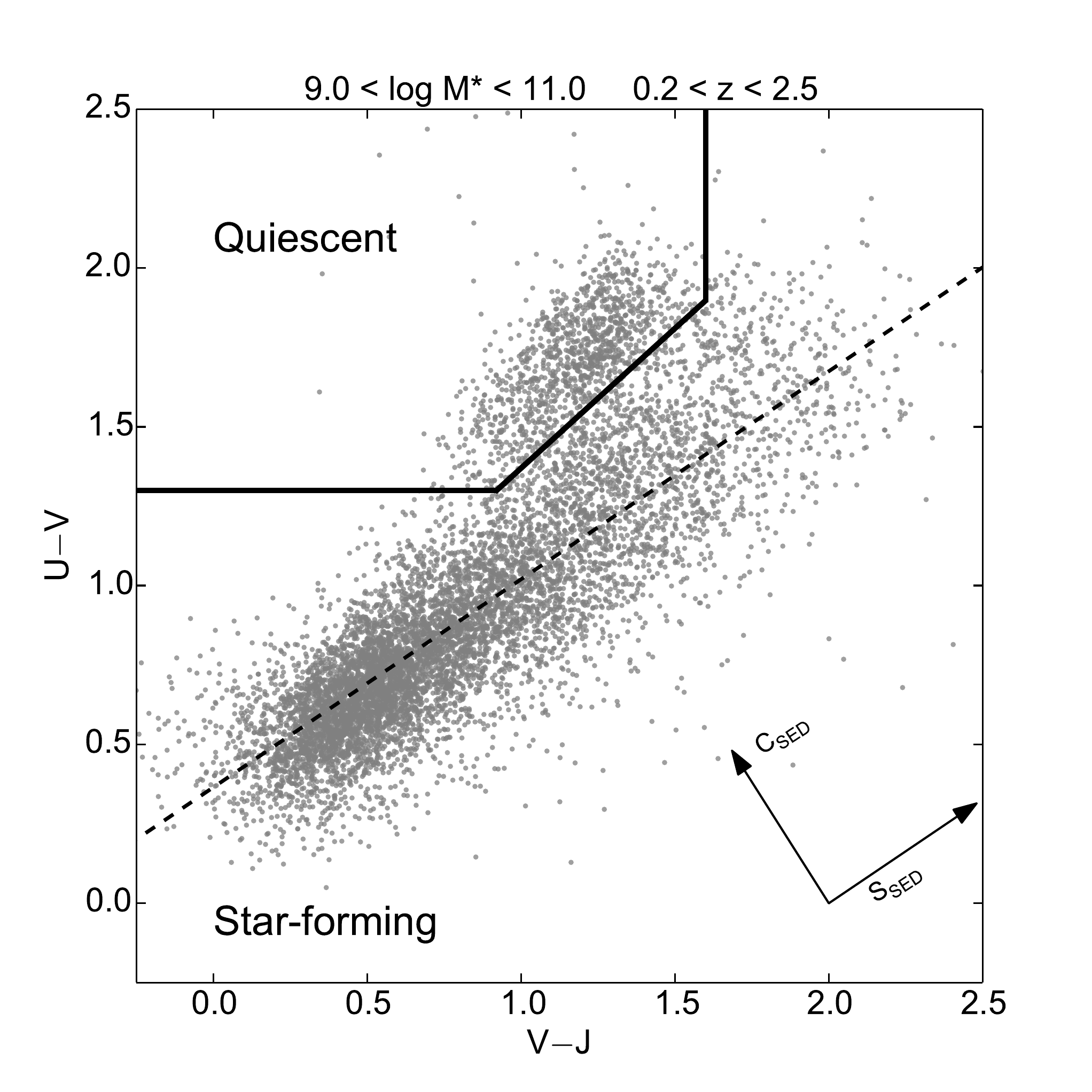}}
	
	\caption{Rest-frame $UVJ$ diagram for all 9,135 galaxies in the final sample ($9.0<\log M_*/M_\odot<11.0$ and $0.2<z<2.5$). The solid black lines separate quiescent and SF galaxies, according to the definition of \citet{williams09} for $z = 1-2$. The set of axes in the lower-right corner indicates the rotated coordinates, \vjrot\ and \uvrot, that are used to facilitate our analysis. The dashed line indicates the best-fit relation to the SF galaxies used to define \vjrot\ and \uvrot.}

	\label{uvj_whole}
\end{figure}

A feature of our analysis is dividing the sample into narrow bins of redshift and mass. By doing so, underlying systematic trends emerge more clearly that might remain hidden if all masses and redshifts are lumped together. Our basic visualization tool is a diagram showing a grid of scatter plots, each corresponding to a bin of redshift and stellar mass. The grid is divided into four, 0.5-dex-wide mass bins between $9.0<\log M_*/M_\odot<11.0$ and five redshift bins between $0.2<z<2.5$. By presenting all scatter plots in this master coordinate system, one can more easily spot evolutionary trends as a function of redshift and/or mass. 

The grid system is also a convenient way to connect galaxies in a given scatter plot with their progenitors and descendants. To illustrate this, Figure \ref{mass_tracks} shows a sample grid diagram of scatter plots overlaid with stellar mass growth tracks based on estimates of how galaxies grow in mass with time \citep{moster13,papovich15}. Scatter plots along a given trajectory then represent the evolutionary states of galaxies of the same final mass at different times.  

Variations in the steepness of the mass-growth trajectories reflect the differing growth rates of low-mass vs. high-mass galaxies after $z = 2.5$ \citep[e.g.,][]{behroozi13,moster13}. High-mass galaxies accumulate their stellar mass earlier than low-mass galaxies, a phenomenon loosely termed ``downsizing'' \citep{cowie96}. This is evident as the steeper trajectories of massive galaxies in Figure \ref{mass_tracks}, which signify little mass growth at late times. We see other manifestations of such mass-accelerated evolution throughout this work.\footnote{Because ``downsizing'' has been applied in many different contexts, some far removed from the original usage in \citet{cowie96}, we prefer the term ``mass-accelerated evolution'' to express the fact that more massive galaxies appear to evolve through their life cycles faster than smaller galaxies, and therefore that many qualities appear first in massive galaxies and later in smaller ones.} However, galaxies move upward and to the right as they grow in mass, and the general trend is that scatter plots in the lower-left corner of the grid evolve into scatter plots in the upper right corner.

\begin{figure}

		\centerline{\includegraphics[scale=0.32]{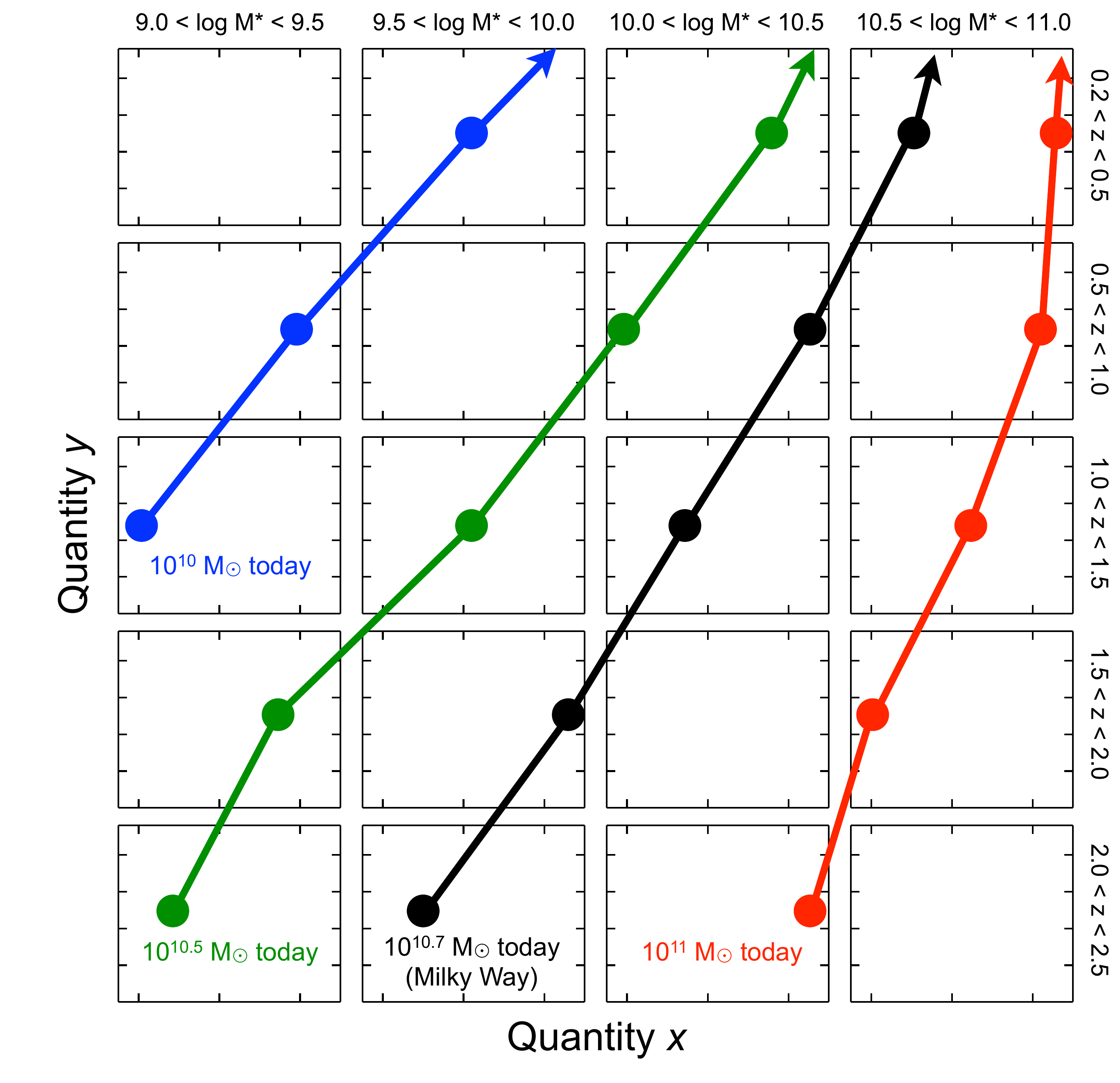}}
	
	\caption{Scatter plots of galaxies in narrow mass and redshift bins arranged in a master grid of mass vs.~redshift to illustrate the evolution of galaxy properties.   The variables in the scatter plots are arbitrary -- they can be $UVJ$ diagrams or any pair of $X,Y$ variables; here they are left empty.  Superposed on the scatter plot grid is a separate (and invisible) mass--redshift coordinate system scaled to match the mass--redshift axes of the grid. Mass growth tracks of galaxies are plotted in this separate coordinate system, illustrating how galaxies move through the mass--redshift grid as they evolve. Four representative tracks are shown, labeled by their stellar masses at $z=0$. The vertical positions of the points are the middles of the redshift intervals.  The horizontal positions assume that the vertical edges of each panel correspond to the mass limits of each bin. Scatter plots along a given growth track are progenitors and descendants of one another. Galaxies generally evolve diagonally upwards through the grid to higher masses at later times.  The Milky Way track is from \citet{papovich15}, while the others are from \citet{moster13}.  }

	\label{mass_tracks}
\end{figure}

\subsection{A (Universal) Relation between \uvrot\ and SSFR}\label{uvrot_ssfr}

\begin{figure*}

	\centerline{\includegraphics[scale=0.27]{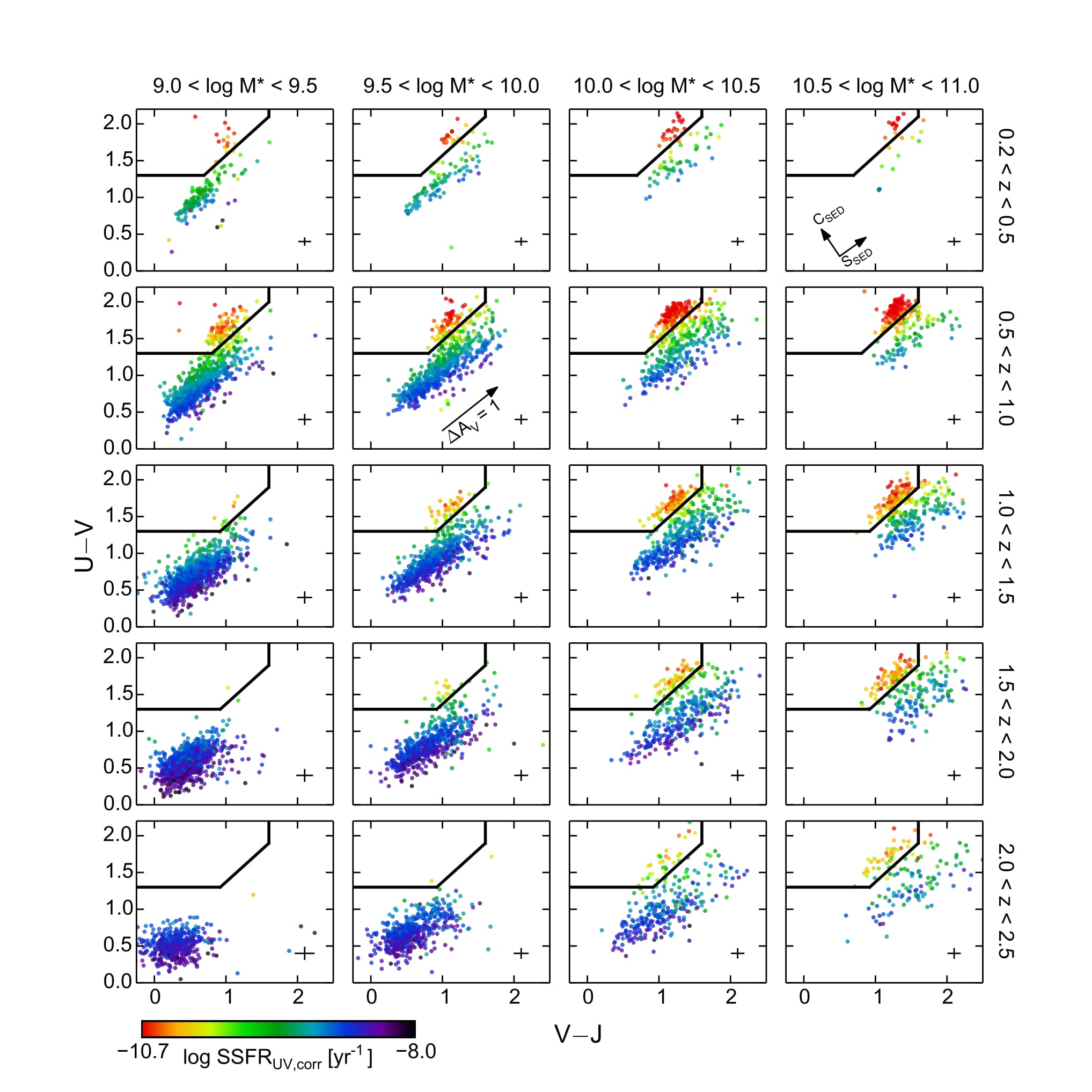}}
	
	\caption{Rest-frame $UVJ$ diagram, divided into narrow stellar mass and redshift bins. Points are color-coded by the dust-corrected, UV-based SSFR, \ssfruv. The arrow is the Calzetti reddening vector for $\Delta A_V=1$ mag. The rotated vectors at upper right show the rotated coordinates, \uvrot\ and \vjrot. Crosses indicate the median error bars in $U-V$ and $V-J$ for SF galaxies. The SF sequence shifts to redder colors as age and dust content increase. The quiescent population is seen to form first at higher mass. The shift upwards in $U-V$ (and \uvrot) is due to falling SSFRs with time, while the shift to redder $V-J$ is due to more dust. Moreover, a clear gradient in \ssfruv, running nearly parallel to \uvrot, is seen in all panels except the bottom two with $\log\,M_*/M_\odot<10.0$ and $z>2$.}

	\label{uvj_ssfr_uv}
\end{figure*}

Figure \ref{uvj_ssfr_uv} shows the $UVJ$ evolution of SF galaxies in the grid diagram from $z = 2.5$ to $z = 0.2$. Inspection of the figure reveals several trends: 

\begin{enumerate}
	\item The mean location of the SF sequence shifts toward larger \uvrot\ (toward the upper left) with increasing galaxy mass and cosmic time. These shifts are accompanied by a fall in SSFR, suggesting that \uvrot\ is closely related to SSFR.  
	\item Galaxies progressively fill in the dusty region of the SF sequence, toward larger \vjrot\ (and redder $V-J$), as they evolve. This is consistent with their having a higher dust content at late times and in more massive galaxies \citep[e.g.,][]{whitaker12b}. 
	\item The buildup of objects in the quiescent region is clearly evident at all masses; moreover, quiescent objects appear earlier at higher masses. 
\end{enumerate}

Each of the evolutionary trends enumerated above exhibits mass-accelerated evolution.  This is evident by choosing a mass--redshift bin (e.g., $\log\,M_*/M_\odot=10.0-10.5$ and $z=1.5-2.0$) and visually identifying the corresponding bin at smaller mass that best matches it. Invariably one finds the lower-mass bin at \emph{later} redshift (i.e., $\log\,M_*/M_\odot\approx9.5-10.0$ and $z\approx0.5-1.0$), confirming that higher-mass galaxies evolve more quickly. We conclude that not only the rate of quenching but also the SSFR and specific rate of dust production are both subject to mass-accelerated evolution. Dust production is naturally connected to star formation because more dust is generated as stellar nucleosynthesis produces more metals. But it is interesting that dust content, as evidenced by mean $V-J$, seems to vary more strongly with mass than with redshift compared to the other two parameters.  The dust content of galaxies is examined further in Figure \ref{uvj_av} and in future papers.

Another striking feature seen in nearly every panel of Figure \ref{uvj_ssfr_uv} is the presence of a gradient in \ssfruv\ running parallel to \uvrot. This gradient is evident as the pattern of colored stripes, each stripe representing a different value of \ssfruv. Such a gradient was previously noted by \citet{williams10, patel11, arnouts13}, {and \citet{straatman16}.} The fact that a gradient is seen in every bin suggests that \ssfruv\ is well-correlated with \uvrot\ across a large range in mass and redshift.

To more easily visualize the stripes, Figure \ref{uvjrot_ssfr_uv} plots a modified $UVJ$ diagram that uses the rotated coordinates \uvrot\ and \vjrot. The rotated coordinates do a fairly good job of capturing the tilt of the SF sequence in $UVJ$ space, i.e., lines of constant \ssfruv\ run nearly horizontally in most panels. However, there is a progressive difference between \uvrot\ and \ssfruv\ that is visible as a mild tilt in lines of constant SSFR.  The tilt increases towards higher redshift and lower mass, culminating in the two leftmost bottom panels ($M_*<10^{10} M_\odot$ and $2.0<z<2.5$), where the rotated coordinates are significantly misaligned relative to the stripes of constant \ssfruv. {Tests indicate that photometric errors are too small to cause this discrepancy, but a factor may be strong [\ion{O}{3}] emission in the $V$ filter at the lowest masses and highest redshifts.  These two panels are not used in the calibration below.}

\begin{figure}

	\centerline{\includegraphics[scale=0.14]{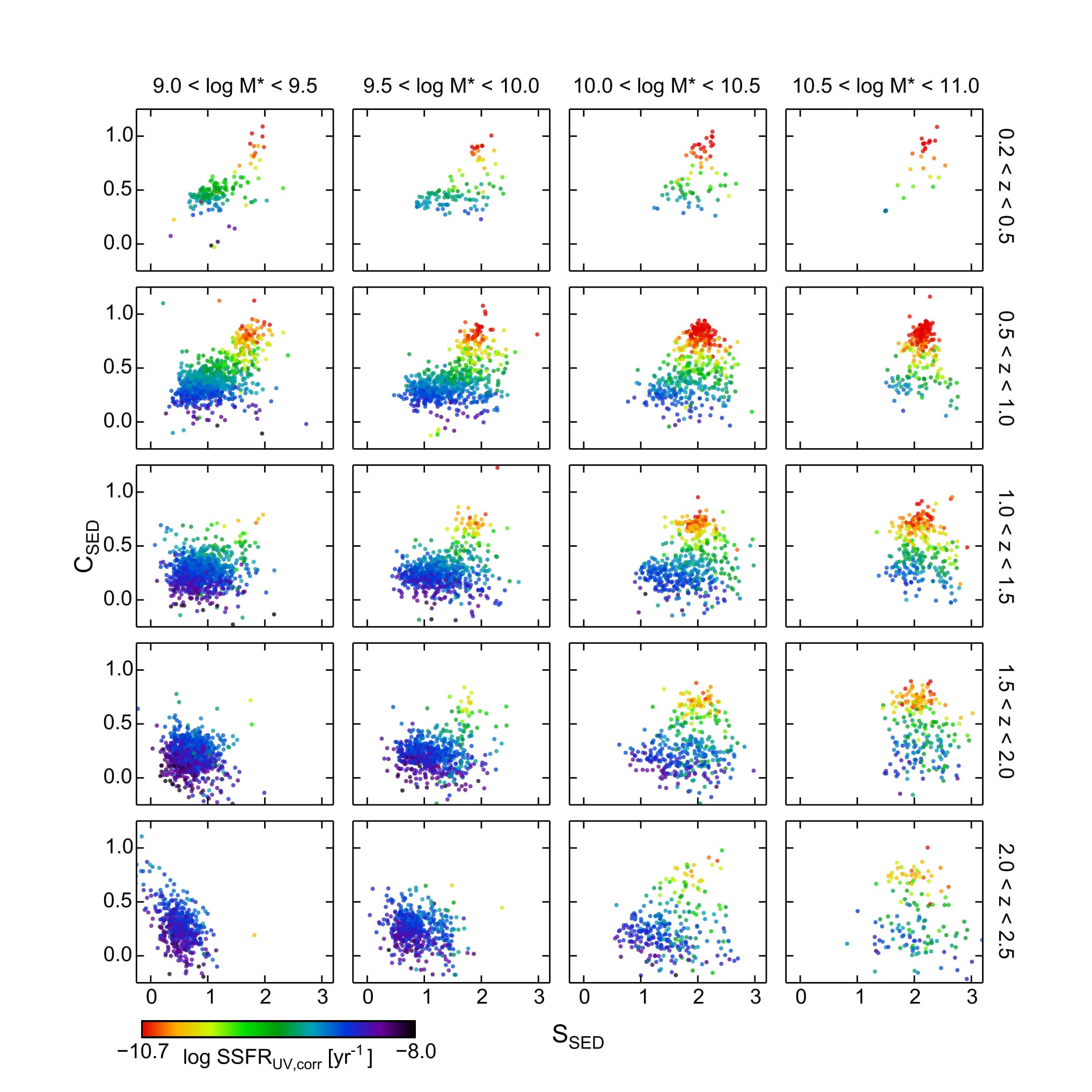}}
	
	\caption{Modified $UVJ$ grid diagram using the rotated coordinates \uvrot\ and \vjrot, divided into narrow stellar mass and redshift bins. Points are color-coded by \ssfruv. In this parameter space, lines of constant \ssfruv\ run nearly parallel to the horizontal axis, \vjrot, though the relation is not perfect, as an increasing tilt is seen towards lower mass and higher $z$. In the two leftmost bottom panels, the rotated coordinates do not accurately describe these galaxies (see Figure \ref{uvj_ssfr_uv}). }

	\label{uvjrot_ssfr_uv}
\end{figure}

Aside from the lower-left two bins, the striped \ssfruv\ pattern looks nearly fixed as a function of mass and redshift; i.e., at fixed \uvrot, the same value of \ssfruv\ is found independent of $M_*$ and $z$. To investigate this, Figure \ref{uvjrot_all_stddev} presents two versions of the ``rotated'' $UVJ$ diagram. The gradient in \ssfruv\ is clearly seen, even when galaxies over a wide range in mass and redshift are included. This combined with the small dispersion supports the hypothesis that the gradient is basically a fixed pattern ``embedded'' in $UVJ$ space through which galaxies move as they evolve.

\begin{figure}

	\centerline{\includegraphics[scale=0.35]{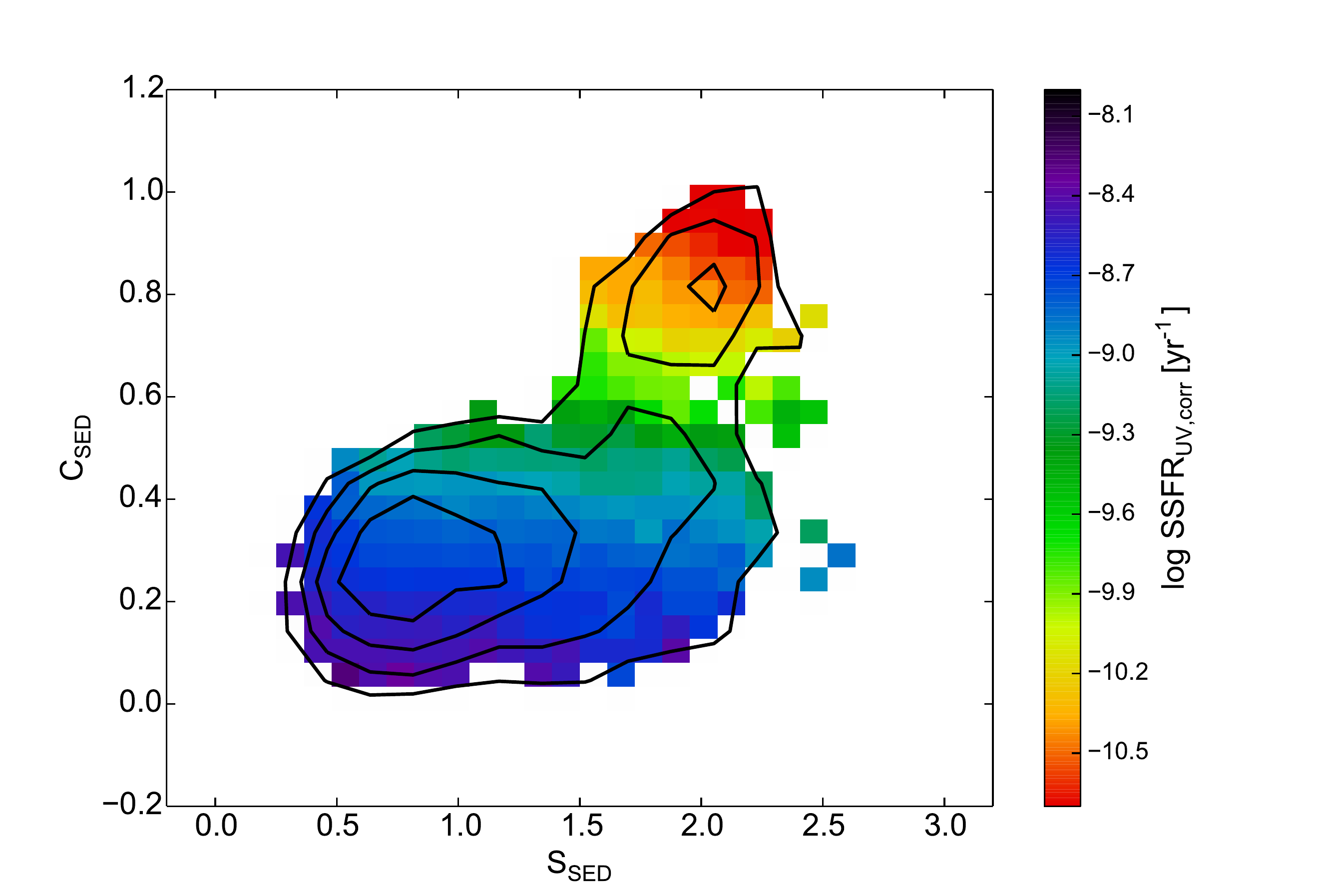}}
	\centerline{\includegraphics[scale=0.35]{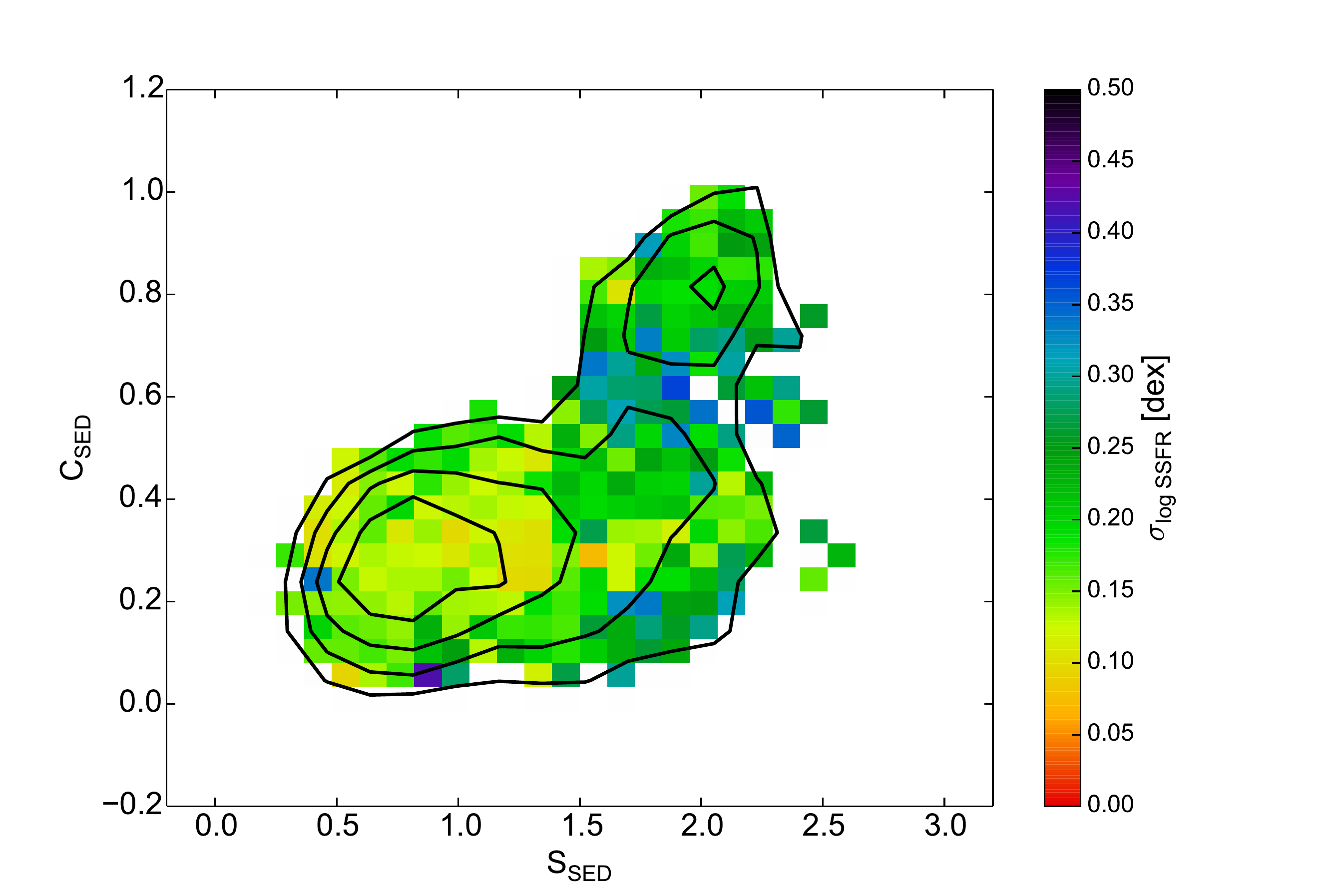}}

	\caption{Stacked diagrams using the rotated coordinates \uvrot\ and \vjrot. The entire sample is plotted except galaxies with $M_*<10^{10} M_\odot$ and $2.0<z<2.5$ (see Figure \ref{uvjrot_ssfr_uv}). Black contours are logarithmically spaced and indicate the density of points. The colored 2D pixels show the median value (top panel) and the $1\sigma$ dispersion (bottom panel) in log~\ssfruv\ in each pixel. Only pixels containing $\ge10$ objects are plotted. The top panel shows that \ssfruv\ is correlated with \csed, while the bottom panel indicates that the typical dispersion in log~\ssfruv\ in a pixel is generally $\lesssim0.25$~dex. This small value suggests that the SSFR gradient is a fixed pattern ``embedded'' in the $UVJ$ diagram, through which galaxies move as they evolve. }

	\label{uvjrot_all_stddev}
\end{figure}

Finally, Figure \ref{ssfr_uvrot} illustrates this even more directly. Very little systematic dependence of the residuals on either redshift or stellar mass is seen, indicating that the relation, to first order, is independent of mass and redshift. The dashed line roughly divides the sample into SF and quiescent objects. A quadratic fit to the data (obtained after excluding $>2\sigma$ outliers) is:  
\begin{equation}
	\label{uvrot_ssfr_fit}
	\log \mathrm{SSFR_{UV,corr}}= -1.95C_\mathrm{SED}^2-0.82C_\mathrm{SED}-8.35,
\end{equation}
with \ssfruv\ in $\mathrm{yr}^{-1}$. The $1\sigma$ vertical scatter about the fit is 0.20~dex (after excluding $>2\sigma$ outliers). {Including the two problematic bins would not change the fit significantly.}

Figure \ref{ssfr_uvrot} indicates that the gradient in \ssfruv\ seen in the $UVJ$ diagram is a (nearly) fixed relation: for SF galaxies, there is little systematic offset from the backbone of the relation between galaxies of different masses or redshifts.  This tightness means that a galaxy's SSFR can be estimated to first order just by knowing its location in the $UVJ$ diagram. In more detail, some systematic residuals are seen with mass and redshift, which can be demonstrated quantitatively by including mass and redshift as additional parameters in the fit:  
\begin{eqnarray}\label{uvrot_all_fit}
	\log \mathrm{SSFR_{UV,corr}}=& -1.81C_\mathrm{SED}^2-0.91C_\mathrm{SED}\\ 
								 &-0.18\log\,M_*+0.094z-6.83.\nonumber
\end{eqnarray}
The mass term contributes more than the redshift term, but both are subdominant in the fit compared to the color terms. Moreover, including $M_*$ and $z$ only decreases the scatter in $\log$~\ssfruv\ by an additional 0.02~dex (to 0.18~dex) relative to Equation \ref{uvrot_ssfr_fit}. (The final scatter of 0.18~dex corresponds to a factor of 1.5.) Use of either Equation \ref{uvrot_ssfr_fit} or \ref{uvrot_all_fit} is thus a convenient way to estimate SSFR when only $UVJ$ colors are available. 

Finally, we call attention to the diffuse cloud of aberrant objects lying $>2\sigma$ below the mean relation at blue values of \csed\ in Figure \ref{ssfr_uvrot}.  Nearly all turn out to have bluer-than-normal FUV continua (see Figure \ref{SEDs_aberrant}) and return systematically low values of $A_V$, which in turn cause low \ssfruv.  We have examined their photometric redshift uncertainties (68\% confidence intervals) and find that, while some of the most extreme outliers have larger uncertainties, the rest have uncertainties comparable to the main sample. These objects are discussed further in Sections \ref{uvj_sed} and \ref{uvj_dc}.

\begin{figure}

	\centerline{\includegraphics[scale=0.35]{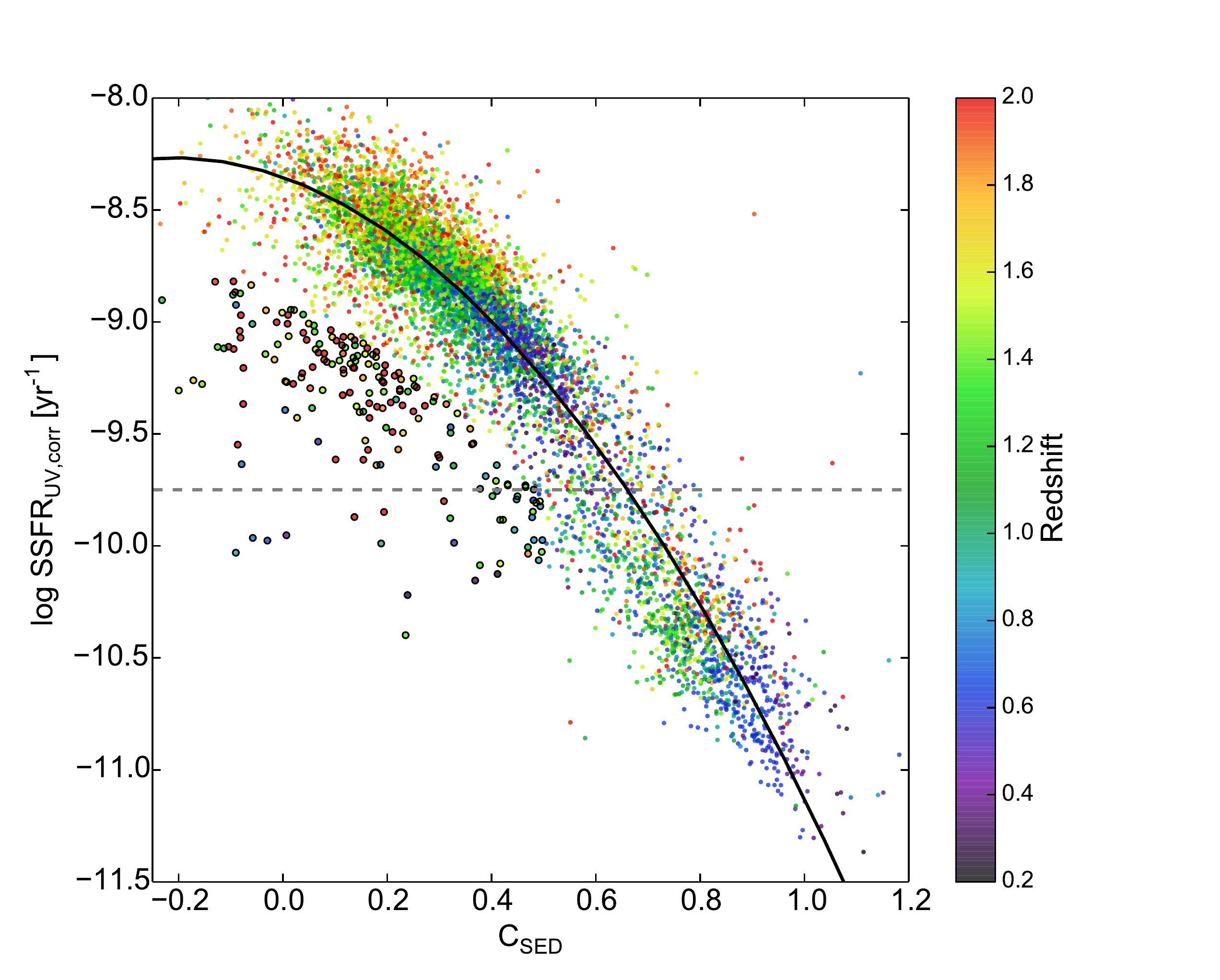}}
	\centerline{\includegraphics[scale=0.35]{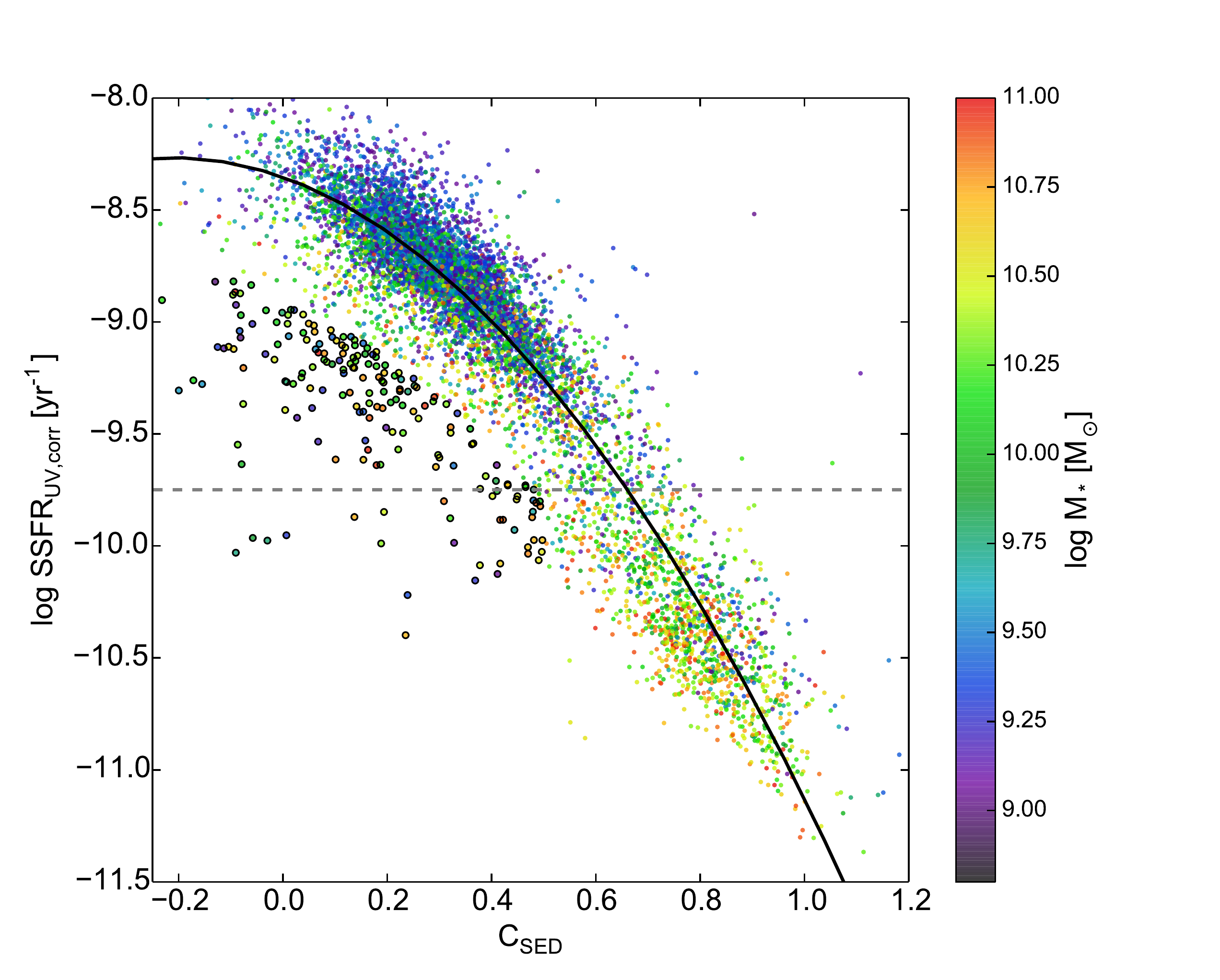}}

	\caption{\ssfruv\ vs.~\uvrot\ for our sample (excluding objects in the lowest-mass, highest-redshift bins, for which \uvrot\ is not reliable; Figure \ref{uvjrot_ssfr_uv}). Galaxies are color-coded by redshift (top panel) and stellar mass (bottom panel). The gray dashed line roughly divides the sample into SF and quiescent objects. The black curve indicates the quadratic fit to the data (Equation \ref{uvrot_ssfr_fit}). The rms scatter in \ssfruv\ about the fit is 0.20~dex (after excluding $2\sigma$ outliers). Including $M_*$ and $z$ in the fit reduces the scatter to 0.18~dex. {Points outlined in black are objects $>2\sigma$ below the relation with \uvrot\,$<0.5$. Most of these aberrant objects turn out to have brighter-than-normal FUV spectra and} are discussed further in Sections \ref{uvj_sed} and \ref{uvj_dc}.}

	\label{ssfr_uvrot}
\end{figure}

\subsection{Previous Work on SSFR in the $UVJ$ Diagram}\label{stripes_previous}

The first mention of SSFR stripes in $UVJ$ was by \citet{williams10}, who also used SFRs based on UV--optical SED fitting.\footnote{By ``striped'', we mean that contours of constant SSFR are parallel to \ssed\ and that SSFR falls with increasing \csed.}  They lumped all masses ($\log\, M_*/M_\odot = 9.5-11.5$) and all redshifts ($z = 1-2$) together and did not test for stability of the pattern with time and mass or calibrate it as a quantitative measure of SSFR. \citet{patel11} analyzed a mixture of field and cluster galaxies in the redshift range 0.6--0.9 using similar techniques, and their conclusions regarding $UVJ$ and SFR were similar. {\citet{patel12} further noted that the highly reddened end of the SF locus tends to be dominated by galaxies with high inclinations, providing evidence for dust.} \citet{whitaker12} analyzed the UV--optical SEDs of $z = 0-2.5$ galaxies from the NEWFIRM survey \citep{whitaker11} and obtained similar stripes, albeit in stellar age, not SSFR.

The above studies all determined SFR from SED fitting, which is highly influenced by the $UVJ$ colors \emph{per se}.  A desirable check is whether independent, IR-based SFR values give the same pattern. \citet{arnouts13} analyzed the infrared excess, $L_\mathrm{IR}/L_\mathrm{UV}$, in NUV$-r$ vs.~$r-K$ and found that the vector {\boldmath$NRK$} (analogous to our \vjrot) can recover IR luminosity (inferred from $24\,\micron$) with a scatter of 0.22--0.27~dex. They also identified a vector running cross-wise to the long axis of the SF distribution in $NUVrK$ that is analogous to \uvrot, but they did not present a quantitative calibration of it versus SSFR.

{The most recent work \citep{straatman16} used ground-based photometry from the ZFOURGE survey augmented with \emph{Spitzer}/MIPS $24\,\mu$m data.  This sample goes deeply to high redshift, and it is reassuring to see that prominent stripes are still seen in $UVJ$ using these IR-based SSFRs.  Interestingly, their stripes also tend to fall apart beyond $z\sim2$, analogous to the two low-$M_*$/high-$z$ panels in Figure \ref{uvj_ssfr_uv}.  This is further evidence that the different distributions in these two panels are real and are not caused by photometric errors. \citet{straatman16} also provided a partial calibration of the diagram using a sideways coordinate similar to \csed\ vs.~SSFR, and the results agree with ours in Equations \ref{uvrot_ssfr_fit} and \ref{uvrot_all_fit} to within 0.1 dex at $z$ = 1.25.  However, their work treats only heavily reddened galaxies ($V-J > 1.0$) whereas our calibration is valid for all reddenings.  Their redshift term is also roughly three times larger than ours, which may reflect systematic differences between \ssfruv\ and \ssfruvir\ versus redshift. Such differences are discussed further in the Appendix.}

The net result of these several works is that the diagonal locus of SF galaxies in $UVJ$ (or $NUVrK$) always shows \emph{finite width} in SSFR, which must in turn reflect the width of the SFMS \citep[0.3~dex,][]{whitaker12b}. The locus is therefore a map of SFMS residuals spread out by different amounts by dust reddening, a point first made by \citet{patel11}.  The scatter in \uvrot\ is a clue to how galaxies evolve through the $UVJ$ diagram, which we return to in Section \ref{uvj_dc}.

{\subsection{Dust in the $UVJ$ Plane}\label{sect_uvj_av}}

Figure \ref{uvj_av} replots the $UVJ$ grid diagram in Figure \ref{uvj_ssfr_uv} but this time color-coding points by $A_V$. The typical dust offset increases along the galaxy evolutionary tracks in Figure \ref{mass_tracks}. This signals a growth in dust in individual galaxies, likely due to a growth in interstellar medium (ISM) metallicity with time \citep[e.g.,][]{reddy10}.  {Figure \ref{uvj_av} may be the first time that a dust indicator, in this case $A_V$, is followed as a function of mass and time long enough to see the growth of dust along actual evolutionary tracks.} Closer inspection confirms the conclusion in Section \ref{uvrot_ssfr} that dust forms first in massive galaxies at high redshift and forms in smaller galaxies at later times, in agreement with previous work \citep[e.g.,][]{martis16}. This suggests that heavy element synthesis is yet another example of mass-accelerated evolution in galaxies, in agreement with studies of the mass--metallicity relation \citep[e.g.,][]{zahid11,henry13a,henry13b,sanders15}, which always shows that massive galaxies are more metal-rich at every redshift.

\begin{figure*}

	\centerline{\includegraphics[scale=0.27]{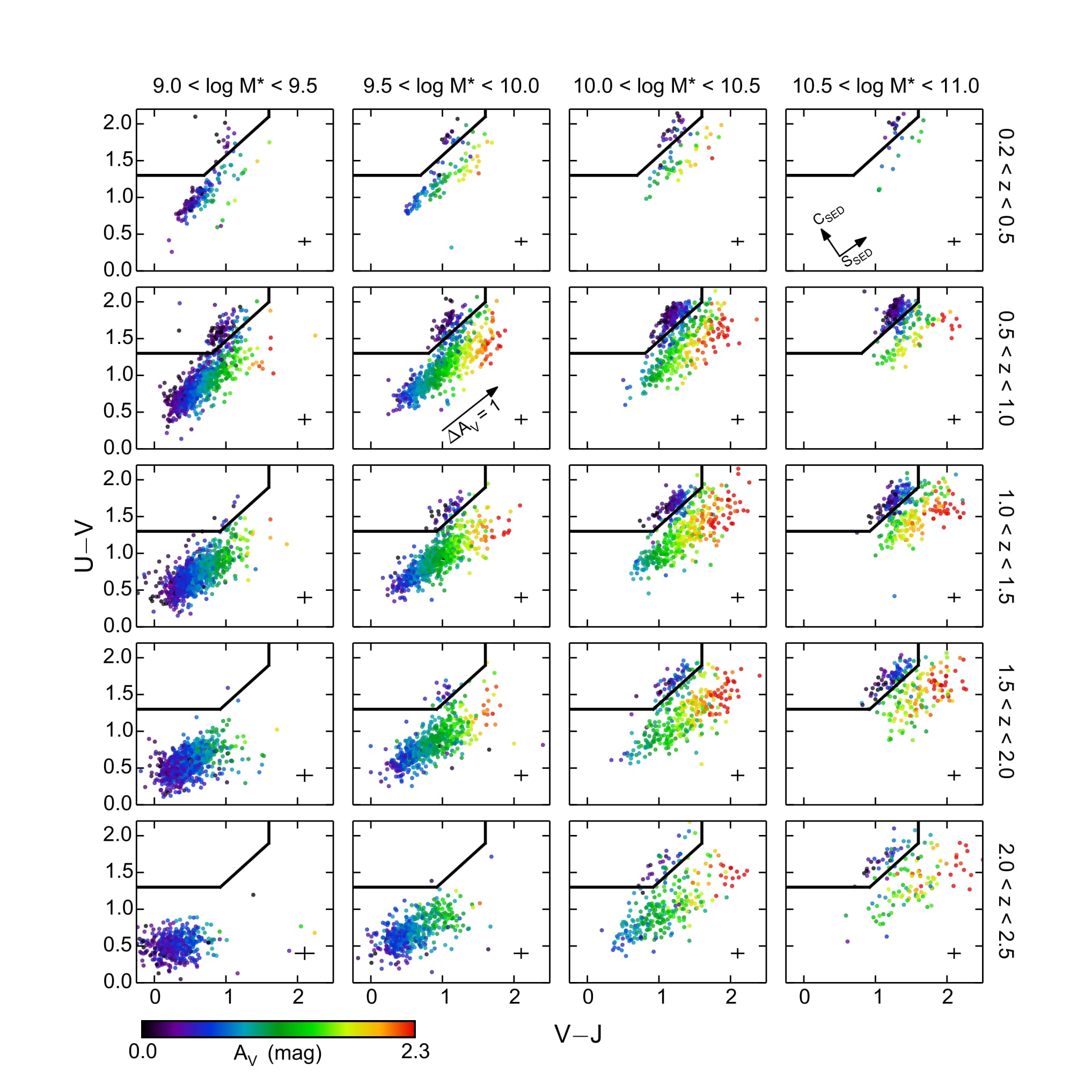}}
	
	\caption{Rest-frame $UVJ$ diagram, divided into narrow stellar mass and redshift bins. Points are color-coded by visual attenuation $A_V$. The arrow is the Calzetti reddening vector for $\Delta A_V=1$ mag. Crosses indicate the median error bars in $U-V$ and $V-J$ for SF galaxies. The overall attenuation for SF galaxies increases with mass and time, and dust is seen to form first in massive galaxies (an example of mass-accelerated evolution). Contours of constant $A_V$ run nearly vertically, i.e., parallel to $V-J$, except for a population of low-$A_V$ objects running along the top of the SF distribution, which are identified as transition galaxies in Section \ref{fading_gal}. Quiescent galaxies have uniformly low attenuation. }

	\label{uvj_av}
\end{figure*}

In each panel of Figure \ref{uvj_av}, the contours of equal reddening are seen to run approximately vertically for strongly SF galaxies. We show below in Figure \ref{uvj_bins_models} that the vertical nature (and narrow width) of the iso-$A_V$ contours is a consequence of our estimating dust by fitting to $\tau$-models (see Section \ref{sect2dot3}).  However, the point here is that the SED-fitting procedure used by CANDELS yields contours of iso-SSFR and iso-$A_V$ that are not orthogonal: the former follow the reddening vector, while the latter follow approximately constant $V-J$. {A similar trend between $V-J$ and $A_V$ was observed by \citet{price14}, \citet{forrest16}, and \citet{martis16}.}

{Also visible in Figure \ref{uvj_av} is a collection of low-$A_V$ but SF objects running along the top of the SF distribution between the SFMS and the quiescent region. These low-$A_V$ objects are identified in Section \ref{fading_gal} with a population of ``transition galaxies'' below the main sequence. As discussed there, their small radii, low SSFRs, and low $A_V$ are consistent with their losing ISM as star formation ends.}

\subsection{Rest-frame SEDs Across the UVJ Diagram}\label{uvj_sed}

\begin{figure*}

	\centerline{\includegraphics[angle=0,scale=0.5]{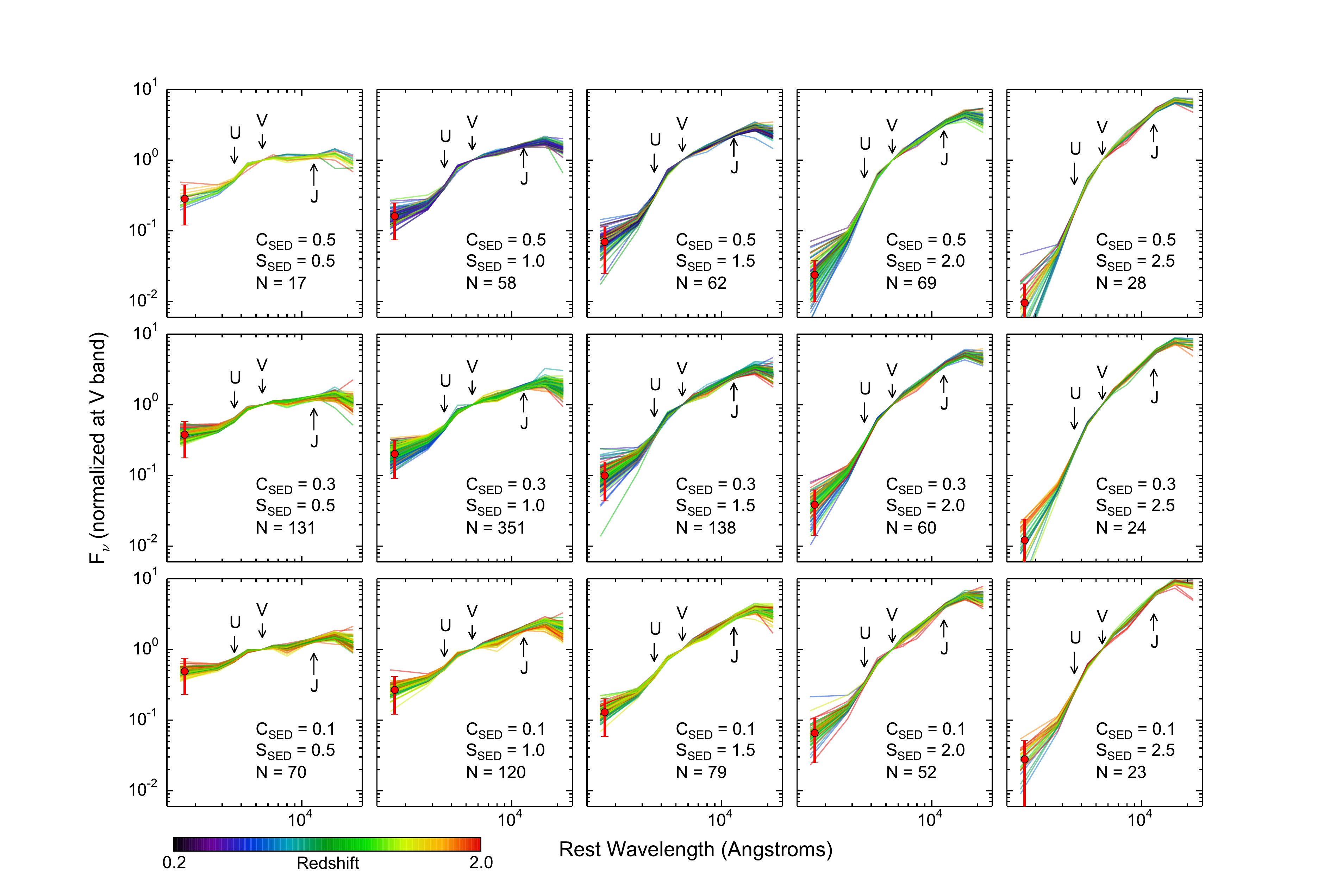}}

	\caption{Rest-frame SEDs (before dust correction) of SF galaxies from the same regions of the $UVJ$ diagram, color-coded by redshift. Galaxies are divided into bins 0.1 mag wide in \uvrot\ and 0.2~mag wide in \vjrot, centered around the values indicated in each panel; the number of objects in each bin is given. Galaxies in the same bin are predicted to have similar values of SSFR and $A_V$ based on having similar \ssed\ and \csed, and the observed uniformity of SEDs is consistent with this.  Only a subset of all bins are included.  SEDs are in $F_\nu$, normalized at $V$. Points with error bars indicate the median uncertainties in the rest-frame FUV fluxes in each bin; the observed FUV scatter is consistent with the error bars. $U$, $V$, and $J$ passbands are marked with arrows.  \ssed\ and \csed\ reflect the overall tilt and curvature (``convexity'') across these three filters.  Dust reddening causes a steepening of the SED from left to right as \vjrot\ increases. Increasing SSFR causes a decreasing convexity from top to bottom as \csed\ decreases. Among the reddest SEDs (rightmost column), UV slopes steepen toward lower redshift, consistent with aging stars and increasing dust with time. }

	\label{sed_redshift}
\end{figure*}

The small dispersion in \ssfruv\ across the $UVJ$ diagram (Figure \ref{uvjrot_all_stddev}) implies that galaxies with the same $UVJ$ colors have similar SSFRs that are (nearly) independent of mass and redshift. Does this similarity extend to their entire SEDs? To explore this question, we study the SEDs, spanning the rest-frame far-UV to the near-IR, of SF galaxies in small bins in the $UVJ$ plane \citep[see also][]{reddy15}. The rest-frame photometry is derived from EAZY (Section \ref{uvj_data}), and we use magnitudes in the FUV, NUV, $U$, $B$, $V$, $R$, $I$, $J$, $H$, and $K$ bandpasses to construct the SEDs. 

Figures \ref{sed_redshift} and \ref{sed_mass} present a montage of SEDs for a sampling of $UVJ$ bins for SF galaxies. The SEDs get steeper with increasing \vjrot, which mainly reflects the increased reddening due to dust (Figure \ref{uvj_av}). Second, galaxies have lower redshift toward increasing \uvrot\ (Figure \ref{sed_redshift}), reflecting the aging of the overall stellar populations with time. Third, the average mass increases with \vjrot\ (Figure \ref{sed_mass}), consistent with increasing dust content in more massive galaxies \citep[e.g.,][]{reddy10,whitaker12}. Moreover, the scatter in SED shape (Figure \ref{sed_redshift}) correlates mildly with $z$, particularly at redder \vjrot: the UV slope becomes steeper toward lower redshift. This may be consistent with the increase in dust content (and metallicity) in galaxies with time. However, no residual trends are seen with $M_*$ within the bins (Figure \ref{sed_mass}).

\begin{figure*}

	\centerline{\includegraphics[angle=0,scale=0.5]{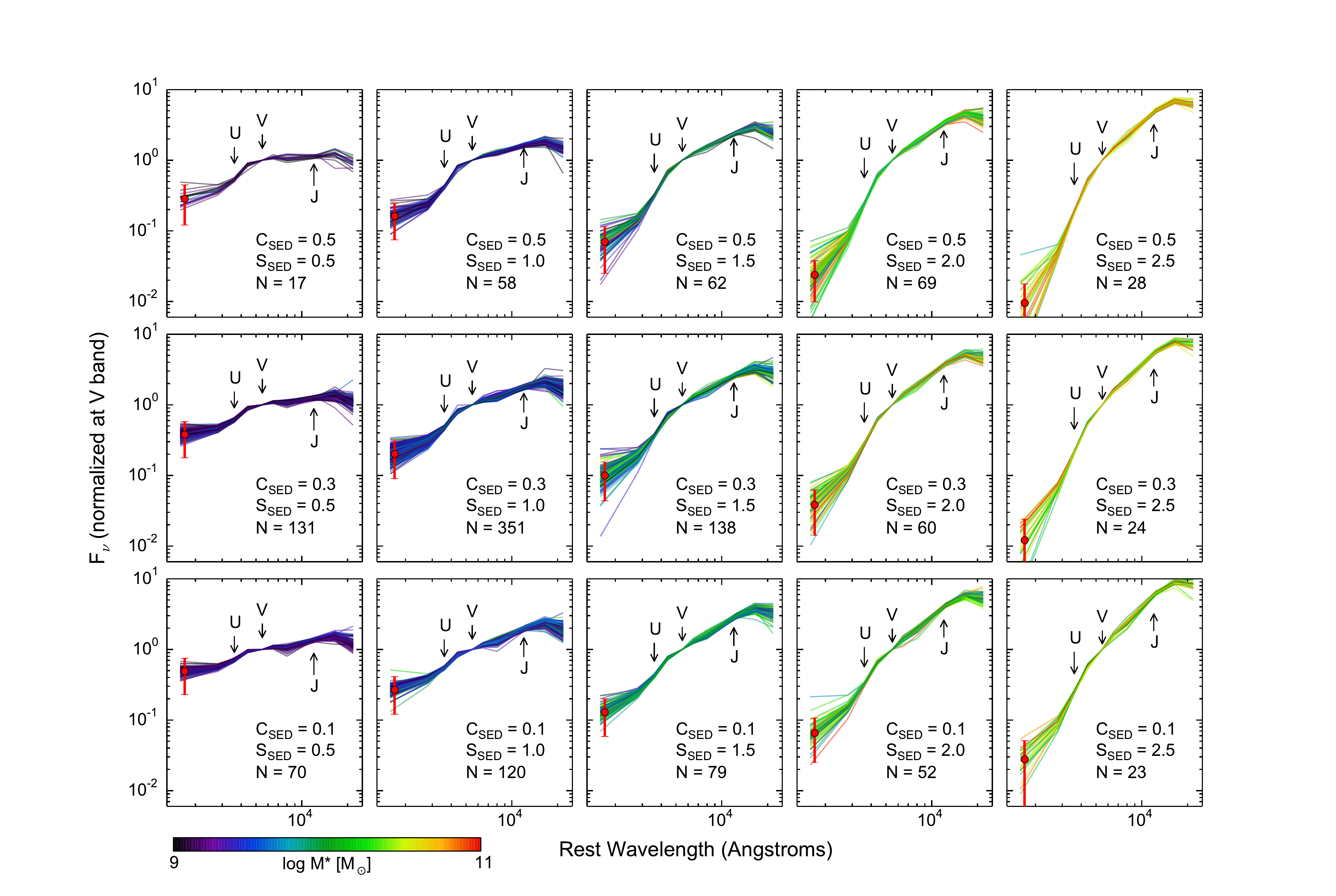}}

	\caption{Identical to the SEDs in Figure \ref{sed_redshift} but now color-coded by stellar mass.  \vjrot, and thus dust content, is strongly correlated with stellar mass. No residual trend with mass is seen within a bin. }

	\label{sed_mass}
\end{figure*}

The broad conclusion from these figures is that the overall dispersion in the SEDs at a given location in $UVJ$ is generally small, and hence that the $UVJ$ colors are a good predictor of the rest of the spectrum for most galaxies.  {However, there is at least one aberrant population, namely the low-lying points below the main relation in Figure \ref{ssfr_uvrot}, which were mentioned in Section \ref{uvrot_ssfr}. Most of them turn out to have brighter FUV continua than average for their location in $UVJ$. This is shown explicitly in Figure \ref{SEDs_aberrant}, where their SEDs are plotted in comparison to the average SED at that location in $UVJ$. The SED fitting process has returned low values of $A_V$ for them (see Section \ref{uvj_dc}), and thus low values of \ssfruv.  Perhaps these objects have composite stellar populations due to a recent small burst of star formation that is not well-matched by $\tau$-models. Aside from these objects, the observed FUV scatter seems comparable to the error bars, though there is room to hide more aberrant cases like the galaxies just discussed.  Future studies should look for further correlations that may be hidden in the FUV residuals.}

\begin{figure}

	\centerline{\includegraphics[angle=0,scale=0.4]{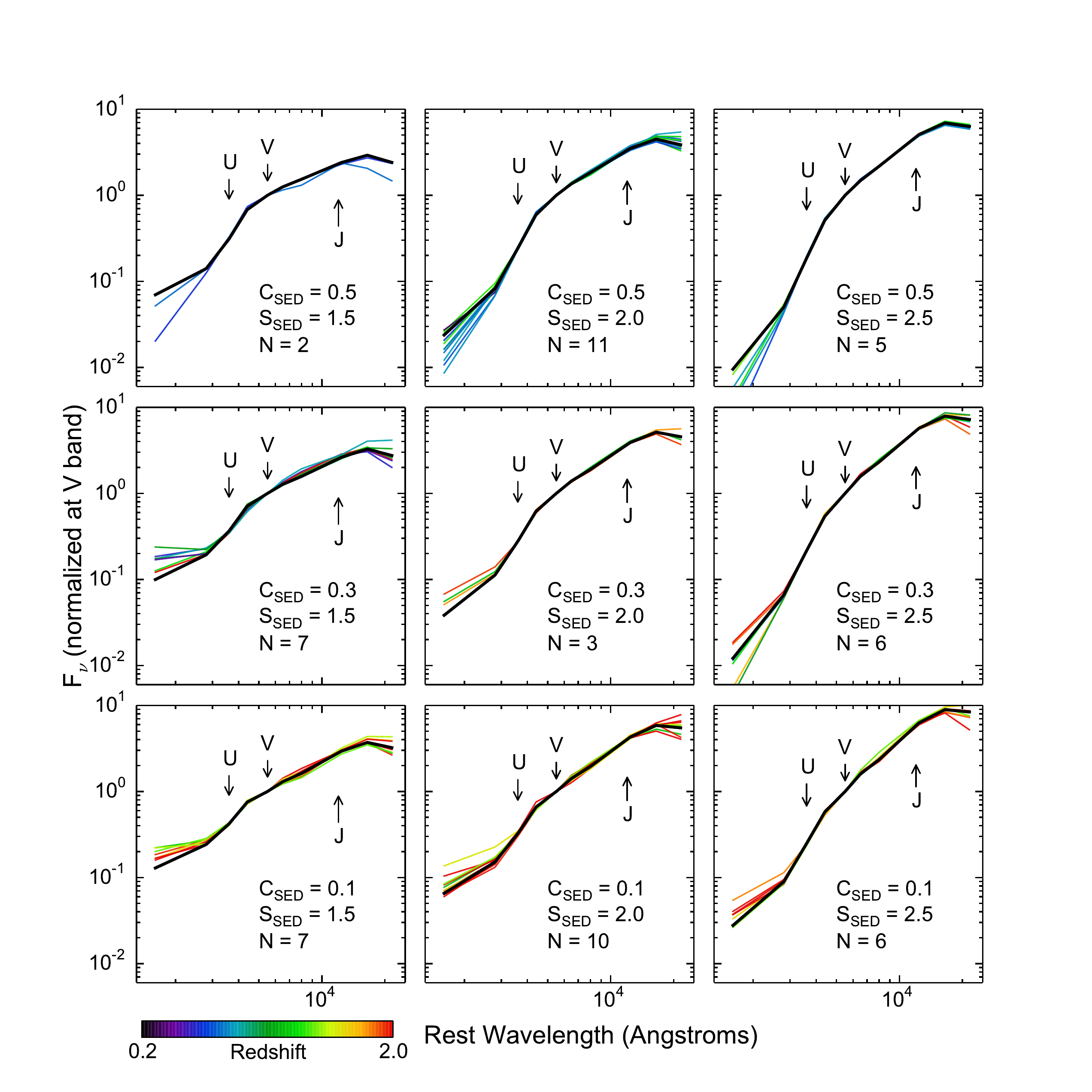}}

	\caption{Rest-frame SEDs (before dust correction) of the aberrant galaxies identified in Figure \ref{ssfr_uvrot}, in bins of \vjrot\ and \uvrot and color-coded by redshift. The black curve in each panel is the mean SED for SF galaxies at the indicated location in $UVJ$ space. In general, the aberrant galaxies have elevated FUV fluxes relative to the overall SF population. Objects in the top row have lower-than-average FUV fluxes, consistent with their being transition (green valley) objects, rather than SF galaxies. }

	\label{SEDs_aberrant}
\end{figure}

\section{Dust-corrected $UVJ$ Diagram}\label{uvj_dc}

The dispersion of SF galaxies in $V-J$ is mainly due to varying amounts of dust reddening \citep[e.g.,][]{wuyts07,patel11,patel12}. Indeed, the existence of stripes in Figure \ref{uvj_ssfr_uv} shows that galaxies with the same SSFR can have different amounts of $A_V$. To what extent, then, are the \emph{intrinsic}, dust-free colors of SF galaxies similar? Because we have estimates of dust attenuation ($A_V$), we can correct the observed colors and examine the resulting distributions in the $UVJ$ diagram.  Undoing the effect of dust will also help us to understand how the methodology of using $\tau$-models has shaped the derived values of SSFR and $A_V$.

To remove dust reddening, we take the measured $A_V$ and apply the \citet{calzetti00} attenuation law to determine the appropriate attenuation in $U$ and $J$, i.e., $A_U=1.5A_V$, $A_J=0.35A_V$. The resulting dust-corrected $UVJ$ diagrams are shown in Figure \ref{uvj_ssfr_corr}. SF galaxies populate a fairly narrow locus that extends diagonally upward, with \ssfruv\ decreasing along the sequence. Broadly speaking, the locus of dust-corrected points conforms to the theoretical dust-free model at all masses and redshifts. However, the scatter about the model track is not zero, which could reflect intrinsic variations in galaxy properties that are not adequately captured by the SED fitting methods used thus far. {In particular, we highlight the aberrant galaxies from Figure \ref{ssfr_uvrot} with black circles. As discussed earlier and highlighted in Figure \ref{SEDs_aberrant}, these objects have brighter FUV continua than other galaxies of the same $UVJ$ colors. They are discussed more below.}

\begin{figure*}

	\centerline{\includegraphics[scale=0.27]{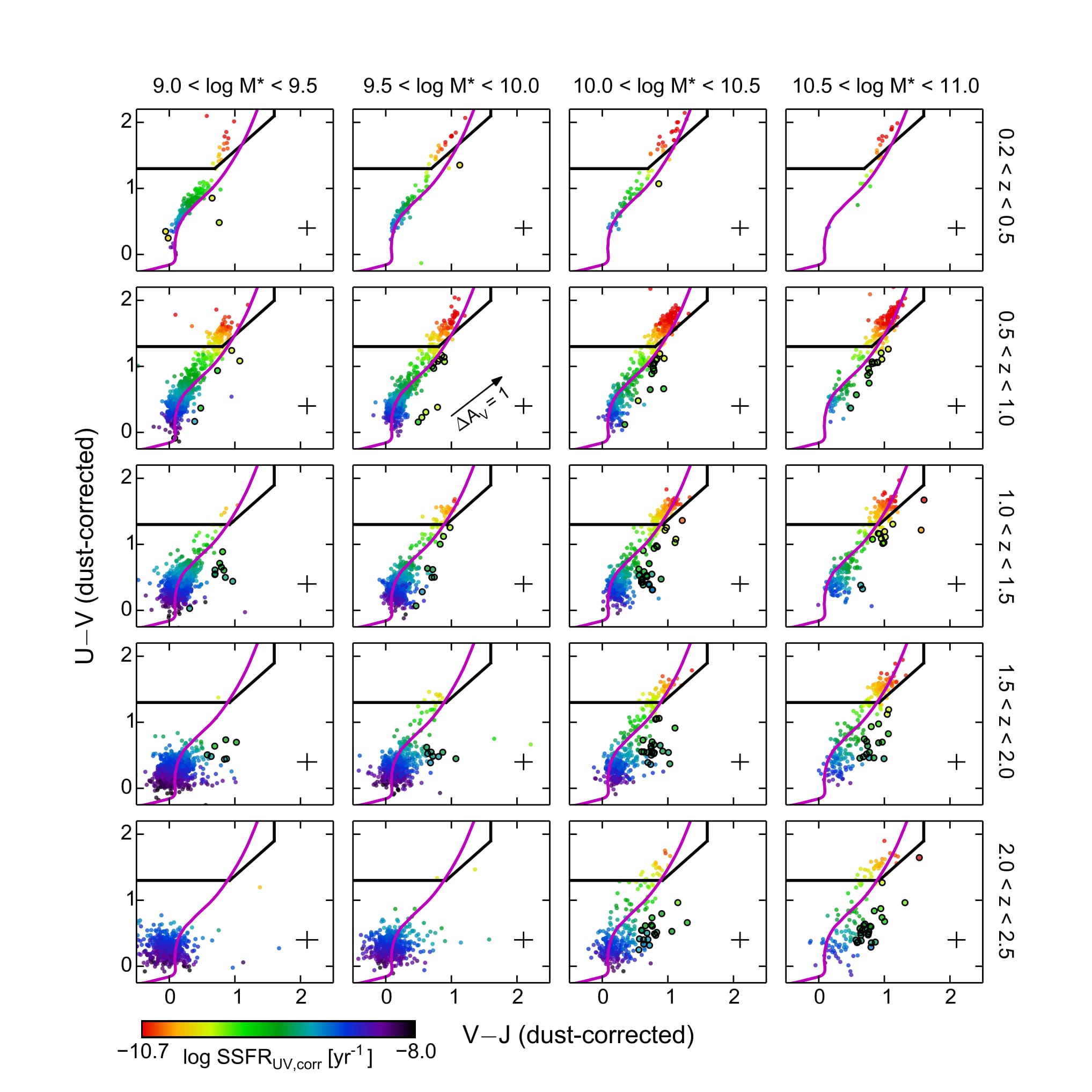}}
	
	\caption{Dust-corrected $UVJ$ diagram, divided into narrow stellar mass and redshift bins. Individual galaxies have been corrected for $A_V$.  Points are color-coded by \ssfruv. Error bars represent median uncertainties in the photometry and the derived $A_V$ values. The magenta curve shows the evolutionary track for a dust-free, $\tau=3$~Gyr, solar-metallicity stellar population model from \citet{bc03}. Applying the dust correction shifts points blueward in both colors, resulting in a narrow locus of points that roughly coincides with the stellar population model. {Points outlined in black are the bright-FUV aberrant galaxies that lie low in Figure \ref{ssfr_uvrot}. They account for nearly all the points that scatter to the right of the dust-free track, due to their having low returned values of $A_V$.}}

	\label{uvj_ssfr_corr}
\end{figure*}

Initially, the model track runs nearly vertically in the $UVJ$ diagram because at early ages, the amount of $U$-band light (from young stars) falls more rapidly than the redder light (from older stars) as the SFR decreases with time. This agrees with the dust-corrected data: the gradient in \ssfruv\ is essentially vertical. After a while, both $U-V$ and $V-J$ increase together, and the track bends toward the upper right as it passes into the quiescent region. Given that there is scatter in the data, we consider how the model track changes if its parameters are adjusted. Figure \ref{uvj_bins_models} plots the dust-corrected $UVJ$ diagram for galaxies with $10.0<\log\,M_*/M_\odot<10.5$ in two redshift bins. Varying $\tau$ results in only slight differences in the shape of the model trajectories; what changes most is the rate at which galaxies move along the track. 

Now suppose we adopt a delayed-$\tau$ model for the star-formation history. If the rise time is short, a galaxy's colors would remain blue, and it would hover near the $t=0$ point in the trajectory. Then as the SFR declines, the galaxy would trace the same path as a standard $\tau$-model. In other words, it is not easy to distinguish between delayed-$\tau$ and normal $\tau$-models from $UVJ$ colors alone. More generally, $UVJ$ alone may not be able to say much about a galaxy's previous star-formation history or its duration ($\tau$ value). In addition, it is apparent from Figure \ref{uvj_bins_models} that variations in $\tau$ alone cannot reproduce the scatter in the observed data. 

We sound a final cautionary note about the assumed star-formation histories.  $\tau$-models, though commonly used, turn out to be extreme in yielding the bluest possible $V-J$ values and thus the largest reddenings.  Alternative star-formation histories tend to lie to the right of the  $\tau$ tracks.  For example, constant star-formation models evolve \emph{along} the reddening vector \citep{patel11}, while \emph{mixtures} of very old and very young stars also lie to the right \citep{wang17}.  With such models as starting points, the derived $A_V$ can be much smaller.  It is wise to keep in mind that the generic $\tau$-model assumption, so often made, is in fact extreme in the amounts of reddening it yields.   

\begin{figure}

	\centerline{\includegraphics[scale=0.45]{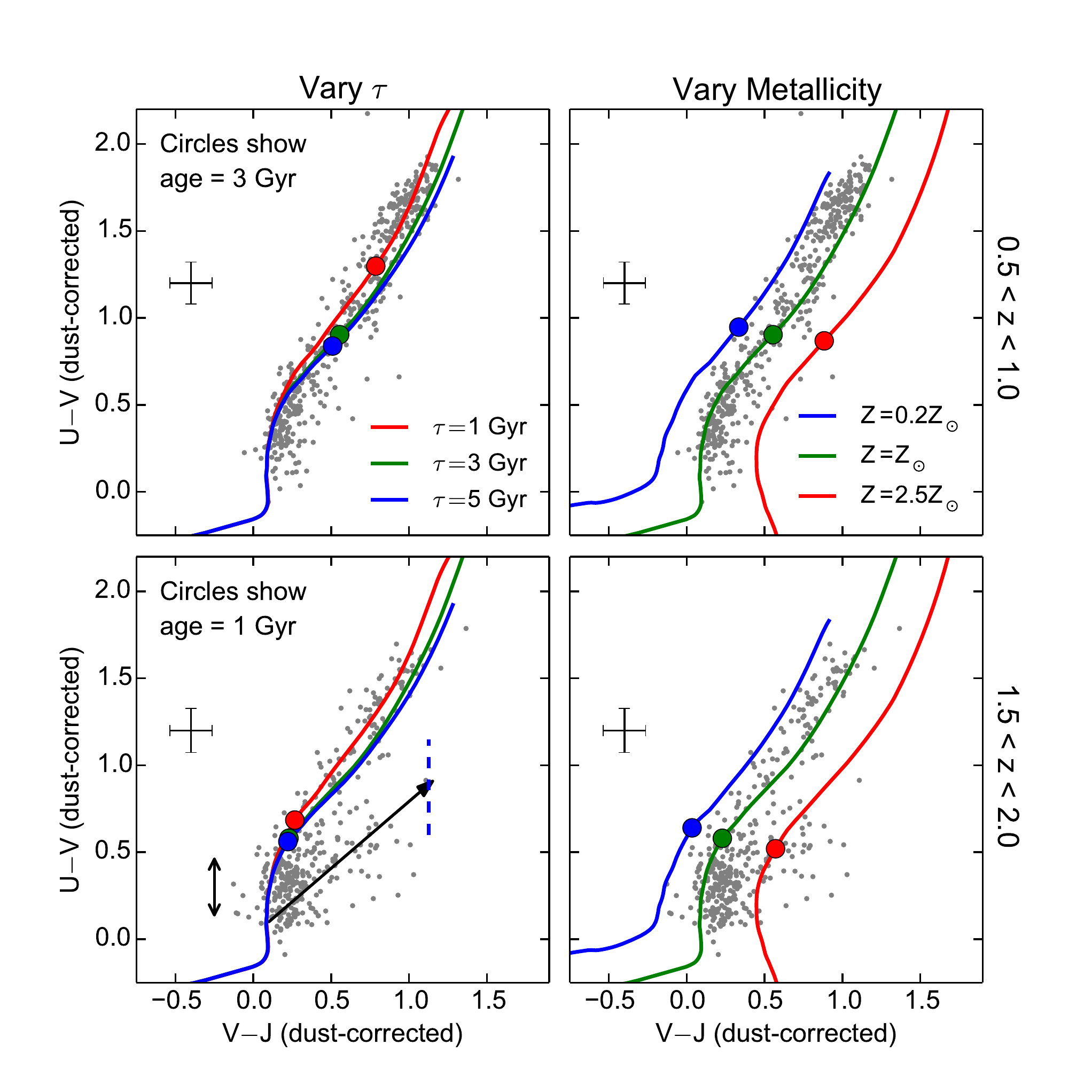}}
	
	\caption{Dust-corrected $UVJ$ diagram for galaxies with $10.0<\log\,M_*/M_\odot<10.5$ in two redshift bins (gray points). Error bars indicate median uncertainties in the colors for SF galaxies in each bin. Various \citet{bc03} stellar population models are plotted in each panel. Colored curves in the left column are dust-free, solar metallicity tracks with different values of $\tau$. In the right column, the tracks are dust-free, $\tau=3$~Gyr models of different metallicities. Colored circles indicate an age of 3~Gyr (top row) and 1~Gyr (bottom row). The locations of the tracks are rather insensitive to the star-formation timescale $\tau$, but vary significantly with metallicity. The arrow in the lower-left panel illustrates 1.5~mag of $A_V$ reddening applied to strongly SF galaxies on the $\tau$ track. The small double arrow schematically represents $\tau$-models undergoing SFR fluctuations.}

	\label{uvj_bins_models}
\end{figure}

Whereas the tracks were relatively insensitive to $\tau$, metallicity has a much stronger effect on the location of the tracks (Figure \ref{uvj_bins_models}). In particular, model $V-J$ colors are reddened with increasing metallicity (line blanketing). This means that $V-J$ could, in principle, serve as a metallicity indicator. However, dust reddening is degenerate with metallicity, making the two effects hard to disentangle without independent information.

The degeneracy between dust and metallicity leads to two related issues when dust-correcting $UVJ$ colors. First, the values of $A_V$ derived from SED fitting depend on the chosen metallicity of the models. The common assumption of solar metallicity means that $A_V$ is overestimated for galaxies more metal-rich than this and underestimated for more metal-poor galaxies. This is important, for example, when interpreting $A_V$ values for low-mass galaxies at high redshift, which presumably have sub-solar metallicities (e.g., galaxies in the lower-left panels of Figure \ref{uvj_ssfr_corr}). 

The second issue relating to the dust--metallicity degeneracy arises when averaging the outputs of several of the CANDELS SED-fitting codes, as we have done to obtain $A_V$ (actually the median value, see Section \ref{sect2dot3}). While most of the codes fix $Z$ to solar, a few allow $Z$ to vary. This means that the ``best-fit'' templates for a given galaxy may have different metallicities, depending on the code, and, consequently, different values of $A_V$. Hence the median $A_V$ will be skewed higher or lower than if all the codes used only solar-metallicity models. This is important because the resulting dust-corrected colors would show dispersion like that seen in the lower panels of Figure \ref{uvj_bins_models}, specifically, the plume of points extending toward redder dust-corrected $V-J$. (These points are among those outlined in black in Figure \ref{uvj_ssfr_corr}.) We have separately verified that the metallicities of these objects, as inferred from the SED-fitting codes that allow $Z$ to vary, are generally super-solar. Consequently, the $A_V$ values from these same codes are \emph{smaller} than those that assume solar metallicity, driving the derived median $A_V$ for these galaxies down. {A possible explanation for this is the fact that these aberrant galaxies have bluer-than-normal FUV continua than other galaxies with the same $UVJ$ colors (Sections \ref{uvrot_ssfr} and \ref{uvj_sed}).  This would cause the fitting codes to return lower $A_V$ values and possibly higher metallicities for those codes that vary $Z$.}

Finally, the morphology of the unreddened $\tau$-model tracks allows us to understand why the iso-$A_V$ contours in Figure \ref{uvj_av} are vertical.  This is explained by the arrow in the lower-left panel of Figure \ref{uvj_bins_models}, which applies 1.5~mag of $A_V$ reddening along the Calzetti vector to a set of galaxies on the strongly SF portion of the solar-$Z$ $\tau$-model tracks.  Because this portion is nearly vertical, the act of parallel translation creates a new set of reddened galaxies that is also vertical. The SED-fitting undoes the dust translation and brings a galaxy back to an assumed unreddened stellar population model.  It is thus no wonder that the dereddened galaxies closely follow solar-$Z$ $\tau$-models in most panels---it would be very surprising had it been otherwise. 

A final important point is the distribution of galaxies \emph{along} the $\tau$-model tracks, specifically the $U-V$ locations of the bluest ones.  As noted in Section \ref{stripes_previous}, the scatter in \uvrot\ at a given mass and redshift arises from the scatter of galaxy residuals about the SFMS, which is observed to be approximately $\pm0.3$~dex rms \citep{whitaker12}.  {Because of this scatter, most mass/redshift bins contain blue galaxies with $U-V < 0.50$.  A galaxy's age must be $<0.8$~Gyr for it to remain this blue, regardless of $\tau$ \citep[e.g., Figure \ref{uvj_bins_models} and][]{wang17}.  But the age of the Universe increases by 6~Gyr from $z = 2.5$ to $z = 0.5$. Therefore, if all galaxies started out as blue $\tau$-models at $z = 2.5$, \emph{even the slowest-evolving} objects should have aged away into the redder regions of the $UVJ$ diagram by $z = 0.5$, yet this is not seen.}  Nevertheless, there \emph{is} a strong net flow from blue to red as some galaxies peel off the SFMS to enter the quiescent region.

These two features---weak \emph{average} color evolution of the blue cloud itself combined with a strong \emph{net} flow from the blue cloud to the red sequence---{may perhaps} be reconciled by imagining continual modest fluctuations in the SFR as long as galaxies remain on the main sequence, followed by eventual quenching of star formation. Occasional upward fluctuations in SFR would continuously repopulate the blue end of the $\tau$ track. These excursions are represented schematically by the vertical double arrow in the lower-left panel of Figure \ref{uvj_bins_models}. The long SF period would then be followed by some (separate) event that plucks galaxies out of the blue cloud and directs them toward the quiescent region. A similar picture of galaxies bobbing up and down randomly about a slowly falling SFMS ridge line has been advanced by \citet{tacchella15b}.  This picture has significant implications for the structural properties of galaxies on the SFMS and will be explored further in future papers.

\section{Transition Galaxies}\label{fading_gal}

A population of galaxies is located in the SF region of the $UVJ$ diagram but whose SSFRs lie well below the main sequence (Figure \ref{ssfr_mass}). These transition objects represent a bridge between SF and quiescent galaxies. For any particular galaxy, it is not known whether it is moving from the main sequence to the quiescent region because its star formation is going out, or whether it is moving backwards towards the main sequence due to some ``rejuvenation'' process \citep[e.g.,][]{martin07,fang12,salim12}.  However, because the net flow of galaxies is from SF to quiescent, the majority of these objects must be fading.  At low redshift, such objects are known to have properties distinct from actively SF objects. In particular, their disks appear to be fading as their bulges build up, and their visible-light radii are consequently shrinking \citep{fang13}. {At higher redshifts, galaxies are observed to undergo similar transformations, though at a more rapid pace. Such quenching is triggered by a compaction event that transforms the galaxy into a compact ``blue nugget'' that then turns into a quenched ``red nugget'' \citep{barro14,barro17}. In either case, quenching is associated with a shrinkage in size \citep[e.g.,][]{pandya16}. }

Figure \ref{delssfr_delsma} shows the relation between SSFR and size for SF galaxies. Most objects lie on or close to the SFMS, but a tail of low-SSFR objects is seen in most panels. Moreover, these objects have smaller sizes and lower dust attenuation on average than those on the SFMS itself. Indeed, the average attenuation \emph{continues to fall} as SSFR declines, which suggests that the decline in SFR is due to the loss of ISM, consistent with fading galaxies. {Our results are consistent with \citet{patel11} and \citet{cava15}, who found a decline in ISM tracers such as MIPS $24\,\mu$m and [\ion{O}{2}] in and near the quenching boundary for galaxies at $z\sim1$. \citet{forrest16} also showed that objects near the quenching boundary exhibit lower $A_V$ and SSFR, as we find. A similar trend between $A_V$ and SFR is also seen in nearby massive galaxies \citep{zahid13}.} The relative number of transition galaxies increases with stellar mass and with decreasing redshift, consistent with the overall growth of the quiescent population with cosmic time \citep[e.g.,][]{bell04,faber07,pgp08b,muzzin13}.

\begin{figure}

	\centerline{\includegraphics[scale=0.14]{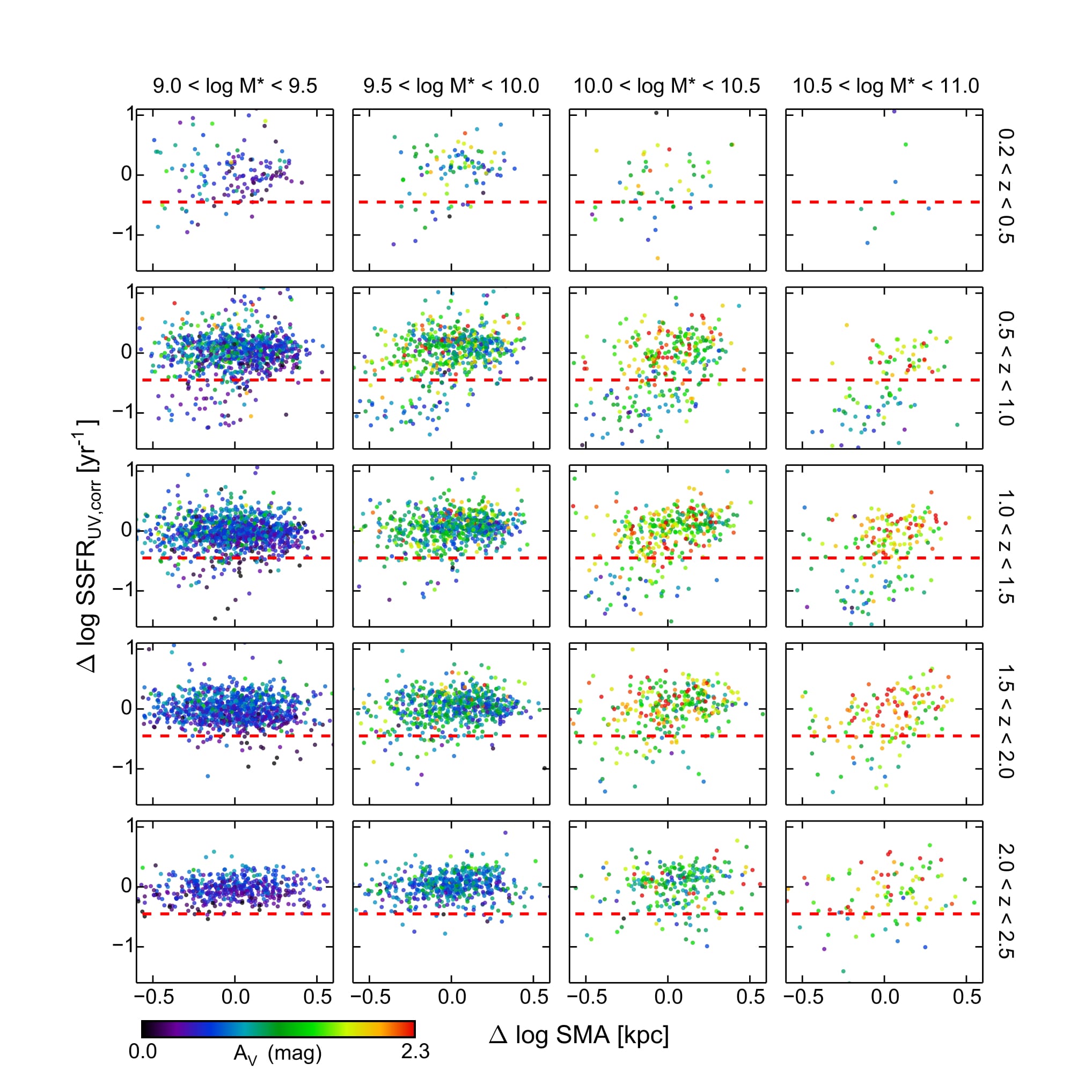}}
	
	\caption{\delssfr\ vs.~\delsma\ (semi-major axis) for galaxies that lie within the SF region of the $UVJ$ diagram, divided into narrow stellar mass and redshift bins. Points are color-coded by the SED-derived visual attenuation, $A_V$. Red dashed lines show our criterion to select transition galaxies, \delssfr\,$<-0.45$~dex. This value was chosen by eye to best separate ridge-line main-sequence galaxies from transition objects, which turn out to have low $A_V$ and small radii.}

	\label{delssfr_delsma}
\end{figure}

Having identified transition galaxies, we show in Figure \ref{uvj_fading} where they lie in the $UVJ$ diagram. By construction, these galaxies are chosen to have low SSFRs, and therefore we expect to find them inside the SF region but close to the quiescent boundary.  Most are indeed found there.  The scatter of transition galaxies further from the boundary at higher redshift could be due to larger photometric errors rather than to real evolution among the transition galaxy population. {In addition, some of the scatter is due to the aberrant objects previously identified as having bluer-than-normal FUV continua and lower \ssfruv\ than other galaxies with similar $UVJ$ colors. As a result, they would be misidentified as transition galaxies based on their (abnormally low) \delssfr.} \citet{pandya16} have identified transition galaxies in CANDELS data using similar criteria to ours. They likewise found smaller radii, higher mass densities, as well as higher S\'ersic indices, also consistent with disk fading. They note, as we do, that such galaxies' $UVJ$ colors seem to represent a bridge between SF and quiescent galaxies.\footnote{On the other hand, within a given mass--redshift bin, the transition galaxies are in the process of fading whereas the quiescent galaxies faded at earlier times and the SF galaxies will fade at later times. This should be taken into account when comparing transition galaxies to quiescent and SF galaxies at the same epoch.} 

In summary, if most transition galaxies are fading, their locations represent the quenching paths taken by galaxies between the SF and quiescent regions, and Figure \ref{uvj_fading} suggests that these paths are mass-dependent. The most massive galaxies appear to move into the quiescent region from right to left horizontally, consistent with the simultaneous shutting down of star formation and loss of interstellar dust \citep{barro14}. The least massive galaxies move upward and to the right, roughly parallel to the unreddened 3-Gyr $\tau$-model track, which is consistent with the lower dust content of low-mass galaxies.  Intermediate-mass galaxies have transition paths that lie between these extremes.

\begin{figure}

	\centerline{\includegraphics[scale=0.35]{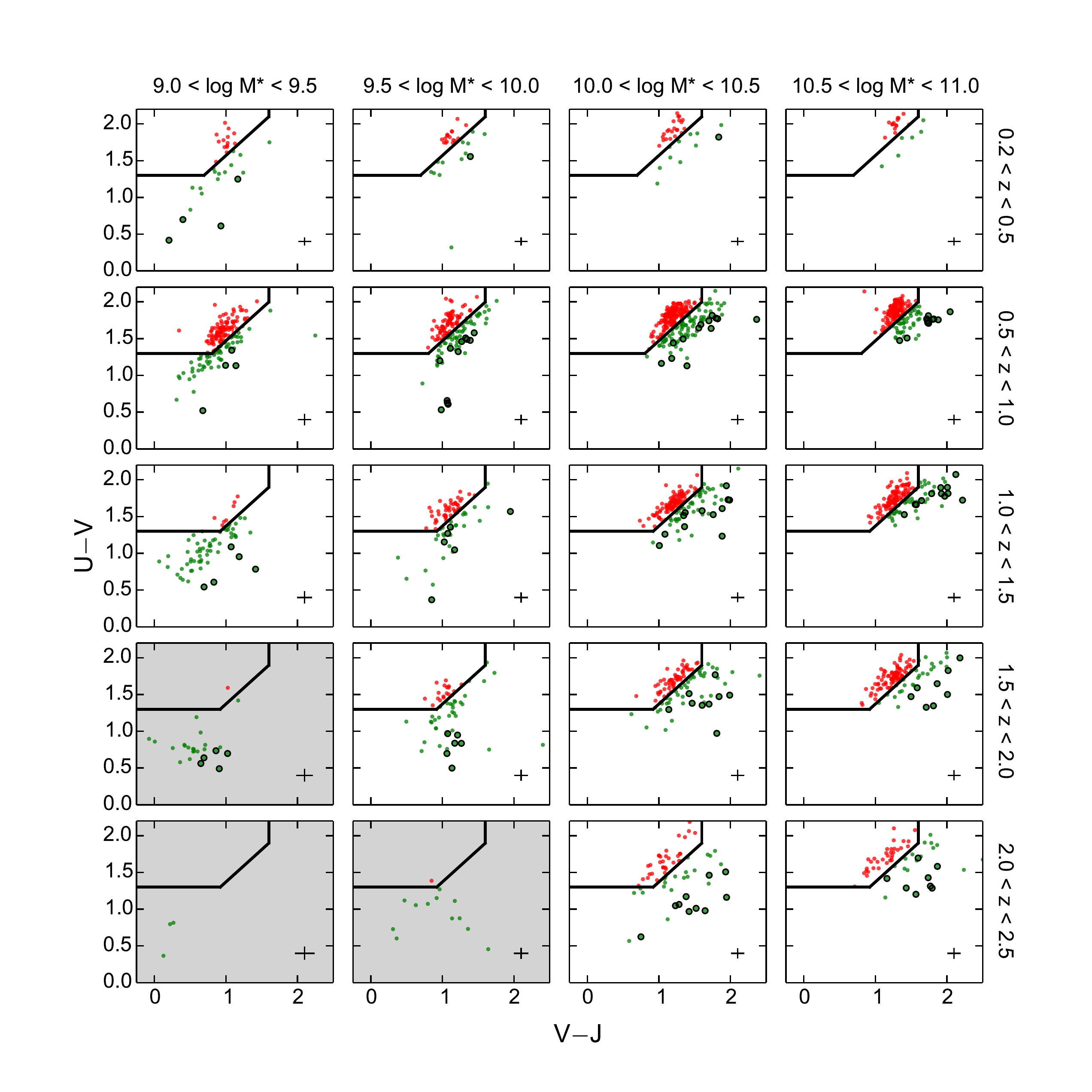}}
	
	\caption{Rest-frame $UVJ$ diagram, divided into narrow stellar mass and redshift bins. Only transition (green points) and quiescent (red points) galaxies are shown. Shaded panels indicate bins where the transition population is probably strongly contaminated by SFMS galaxies that are scattered into the transition region by photometric errors or by strong emission-line contamination of the broad-band photometry. Error bars indicate median uncertainties in the rest-frame colors for transition galaxies. The locations of transition galaxies vary as a function of mass in a way that suggests that the typical quenching path in $UVJ$ is modulated by the disappearance of dust. Many of the most deviant transition galaxies turn out to be members of the aberrant FUV-bright population highlighted in Figure \ref{ssfr_uvrot}, which return low values of $A_V$ and \ssfruv\ under SED fitting.  They may not be normal transition galaxies. }
	

	\label{uvj_fading}
\end{figure}

\section{Blue Cloud, Green Valley, and Red Sequence Fractions versus Mass and Redshift}\label{fractions}

The $UVJ$ diagram provides valuable information on the changing frequencies of galaxies in various stages of SF activity due to evolutionary effects. We therefore close this overview of the $UVJ$ diagram by examining how the populations of blue, SF galaxies and red, quiescent objects evolve as a function of time and mass.  

The ``blue cloud'' was originally defined as a relative overdensity in color--magnitude diagrams consisting of strongly SF galaxies on the SFMS. Quiescent, quenched galaxies populate another overdensity called the ``red sequence''. And the transition galaxies identified in Section \ref{fading_gal} are often called ``green valley'' galaxies, so named because they exhibit intermediate color corresponding to a dip in the galaxy number density \citep[e.g.,][]{balogh04,martin07}. The blue cloud and red sequence were originally defined using single colors, such as $U-V$ \citep{bell04} and $U-B$ \citep{faber07}. However, dust can also redden galaxies, sending strongly SF galaxies into the green valley and even onto the red sequence.  The extent of the contamination when only one color such as $U-V$ is used was quantified by \citet{brammer09}.  Subsequent works \citep[e.g.,][]{patel11,arnouts13} stressed the importance of using two-color diagrams such as $UVJ$ and NUV$rK$ to properly isolate red sequence and green valley galaxies; Figure \ref{uvj_ssfr_uv} in this paper validates this technique.

\begin{figure*}

	\centerline{\includegraphics[scale=0.7]{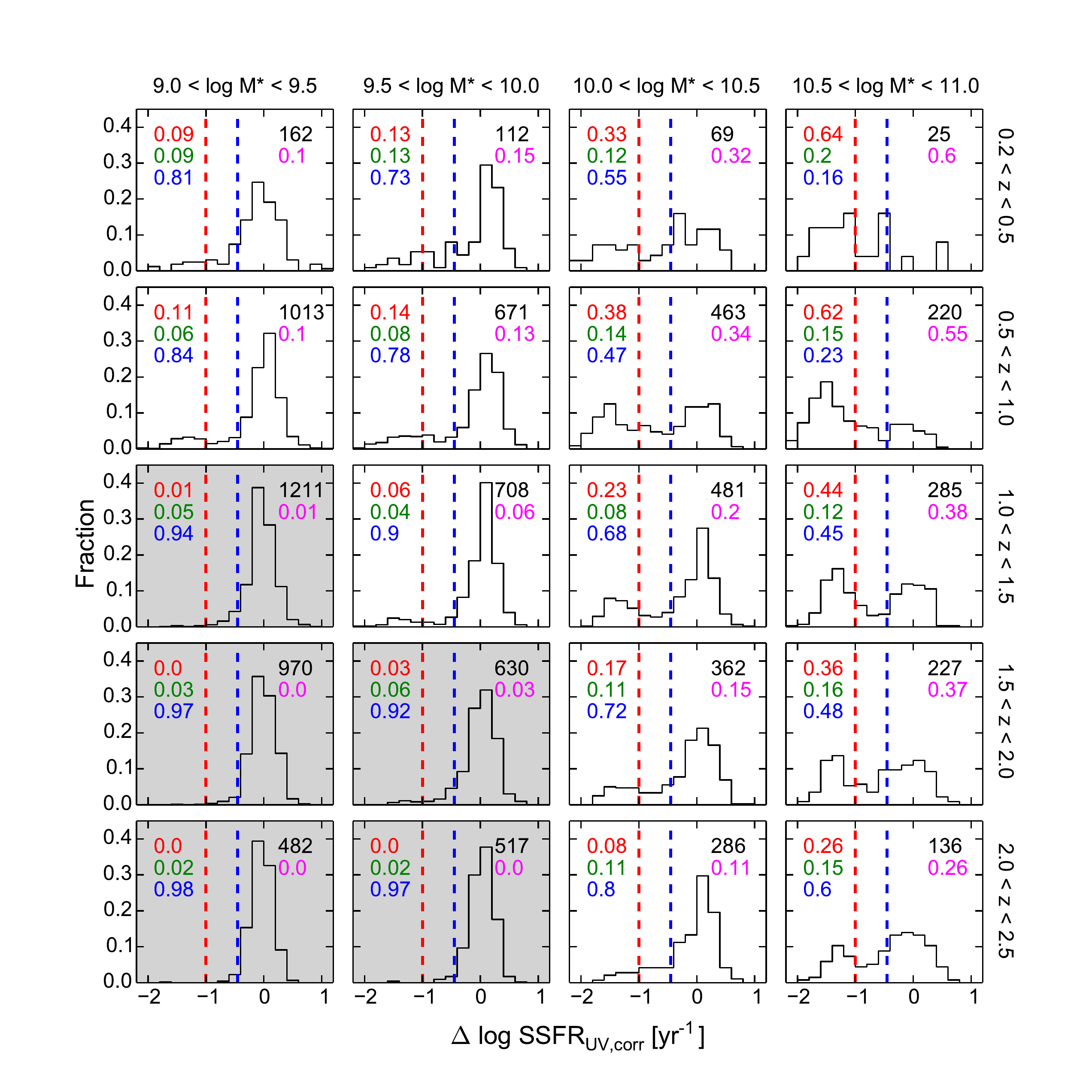}}
	
	\caption{\delssfr\ histograms of galaxies divided into mass and redshift bins. Red sequence galaxies are defined to lie to the left of the red lines (\delssfr$\,<-1.0$~dex), green valley objects in between the blue and red lines, and blue cloud galaxies to the right of the blue line (\delssfr$\,>-0.45$~dex). The total number of galaxies in each bin is given in black. Color-coded numbers in the upper-left indicate the fraction of red sequence, green valley, and blue cloud galaxies in each panel. For comparison, the magenta numbers at upper-right are the fractions of $UVJ$-defined quiescent galaxies from  Figure \ref{q_frac}(a). {Grey shaded panels indicate bins that are $<90\%$ complete in SF and/or quiescent galaxies.} The relative number of red galaxies increases at the expense of the blue galaxies as a function of time and stellar mass. A clear bimodality is particularly evident at higher masses and lower redshifts.}

	\label{uvj_histograms}
\end{figure*}

Our sample is highly complete to $H=24.5$~mag, enabling a census of galaxies in various evolutionary stages.  However, Section \ref{samp_selection} also drew attention to the roughly 15\% of the galaxies that were excluded due to bad GALFIT flags. Such objects tend to be disturbed, have nearby neighbors, and/or are very small in radius. However, separate tests show that the excluded fraction has the same \ssfruv\ distribution as the retained galaxies at all masses and redshifts, and so the use of the sample to derive \emph{relative} numbers of blue, red, and green valley galaxies should be reasonably reliable.

Figure \ref{uvj_histograms} shows histograms of the number of galaxies versus \delssfr. The various fractions vary smoothly with mass and time with little sign of measurement noise. A clear bimodality is also evident in several panels, particularly at higher masses and lower redshifts. Incompleteness notwithstanding, the fraction of galaxies on the red sequence increases with mass and time from small values to a maximum of $\sim60\%$ at high mass and late times. Second, the fraction of galaxies in the blue cloud decreases with mass and time from a maximum of nearly 100\% to a minimum of $\sim20\%$ at high mass and late times. The typical fraction of galaxies in the green valley increases with mass and time from a minimum of a few percent to a maximum of $\sim15\%$ at high mass and late times. Mass-accelerated evolution is also strongly evident in this figure, the red sequence being first established at high redshift in the most massive galaxies.  

\begin{figure*}

	\centerline{\includegraphics[scale=0.45]{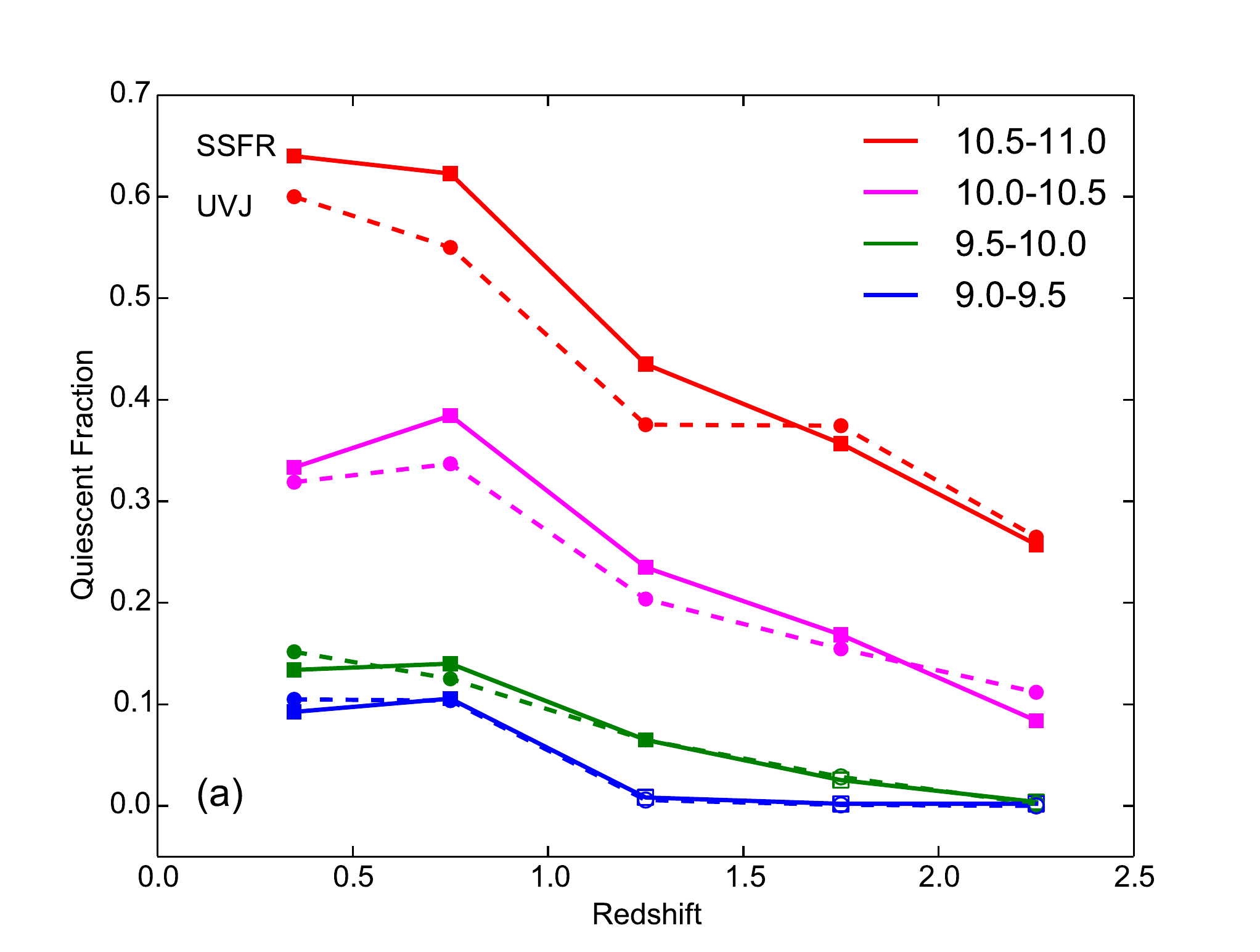} \includegraphics[scale=0.45]{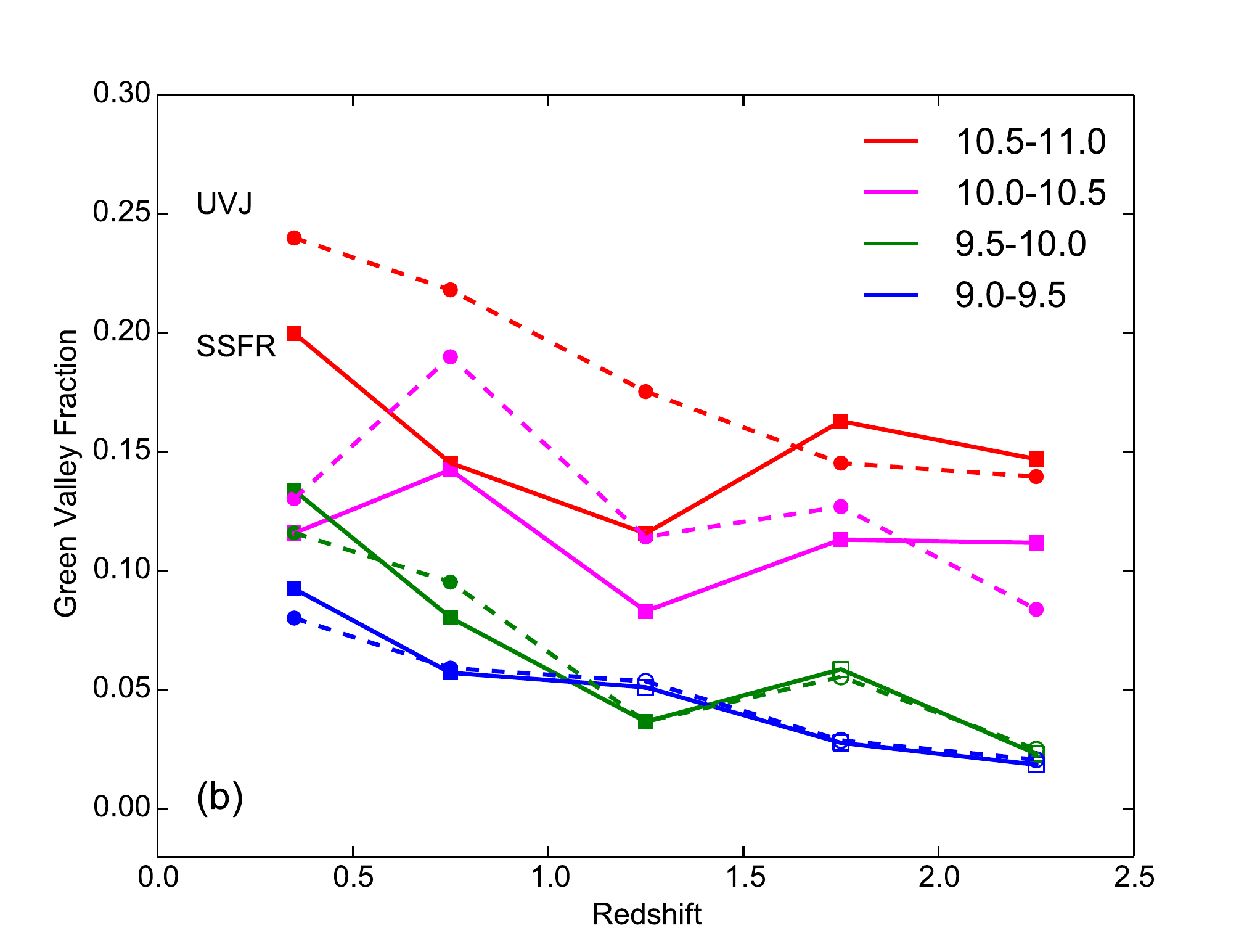} }
	\centerline{\includegraphics[scale=0.45]{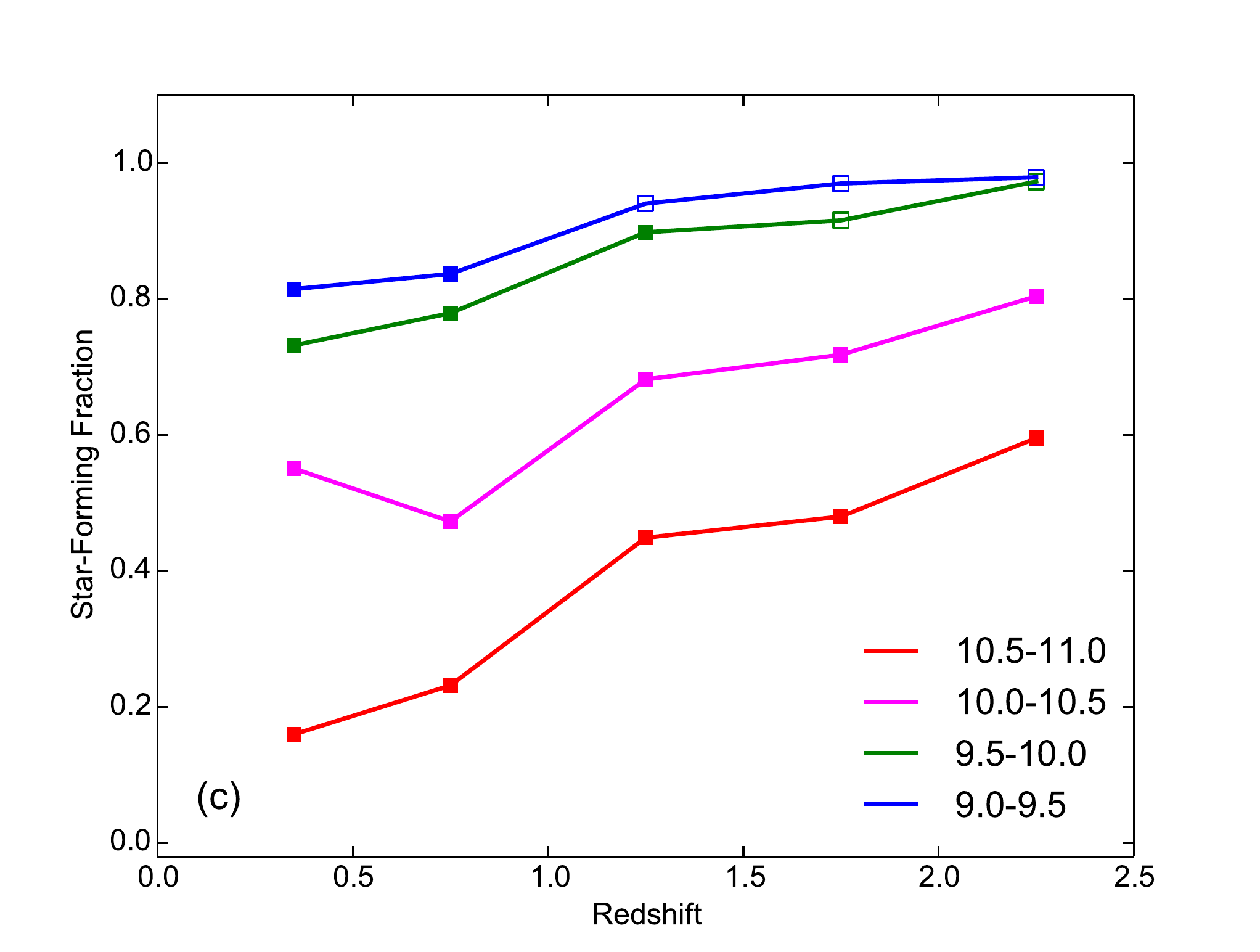}}
	
	\caption{{Panel (a): Quiescent fraction as a function of redshift in four stellar mass bins (colored lines). Quiescent galaxies are identifed in two ways: \delssfr$<-1.0$~dex (squares/solid lines) or objects lying within the quiescent region in the $UVJ$ diagram (circles/dashed lines). Panel (b): The green valley fraction, defined as the objects that remain after subtracting off the quiescent and SF galaxies. Panel (c): The SF fraction, defined as the fraction of galaxies with \delssfr$>-0.45$~dex. In all panels, open symbols denote bins that suffer from incompleteness (the shaded panels in Figure \ref{uvj_histograms}). }}

	\label{q_frac}
\end{figure*}

A summary of these numbers is shown in Figure \ref{q_frac}, which plots the fraction of quiescent, {green valley, and SF galaxies} as a function of mass and redshift. We identify quiescent galaxies in two ways: (1) the definition used in Figure \ref{uvj_histograms}, \delssfr$<-1.0$~dex and (2) galaxies lying in the quiescent region in the $UVJ$ diagram in Figure \ref{uvj_ssfr_uv}. The quiescent fraction increases with time and is greater in more massive galaxies (mass-accelerated evolution). In general, our two definitions of quiescence give similar results. {That they differ somewhat is to be expected, as the relation between color and SSFR is not perfect. This can be seen from the scatter present in Figure \ref{ssfr_uvrot}, which plots \uvrot\ against \ssfruv. Galaxies identified as quiescent according to SSFR may not be included in a $UVJ$-selected sample, and vice versa.} Agreement could be improved by adjusting the quiescent boundaries in $UVJ$ to better capture low-SSFR galaxies. One modification would be to eliminate the vertical cut at $V-J=1.6$, which may be excluding dustier quiescent galaxies. Also, our \uvrot--\ssfruv\ calibration (Equation \ref{uvrot_all_fit}) can be used to optimize the diagonal quiescent boundary in $UVJ$ as a function of mass and redshift.   
{Our findings here are qualitatively consistent with \citet{ownsworth16}, who also found mass-accelerated evolution in the quiescent fraction when selecting galaxies at a constant number density and tracking their evolution with time.} {Similar trends in the quiescent fraction were also observed by \citet{martis16}, who used the $UVJ$ criteria to select quiescent galaxies in the larger UltraVista DR1 sample.} 

{Moving on to the green valley galaxies, Figure \ref{q_frac}(b) plots their fraction as a function of mass and redshift. Green valley objects are defined to be galaxies that are not SF or quiescent, i.e., $F_\mathrm{GV}=1-F_\mathrm{Q}-F_\mathrm{SF}$. $F_\mathrm{Q}$ is defined in two ways, as stated above. The difference between the two is a measure of the uncertainty in $F_\mathrm{Q}$. Our adopted $F_\mathrm{SF}$ is based on a strict cut in \delssfr\ and is well-defined by the fall-off in numbers below the main sequence (Figure \ref{uvj_histograms}). Using these fractions, we obtain the values shown in Figure \ref{q_frac}(b) for $F_\mathrm{GV}$. The agreement between the $UVJ$- and \delssfr-based fractions is generally good, with a typical discrepancy of only a few percent. This difference can be viewed as a measure of the uncertainty in $F_\mathrm{GV}$. Finally, for completeness, we show in Figure \ref{q_frac}(c) the fraction of SF galaxies, defined to have \delssfr$>-0.45$~dex.}


{The results in this section depend on the reliability of our SSFR values. Readers are referred to the Appendix for further discussion of differences between our preferred \ssfruv\ and UV+IR-based rates and the possible impact on the fractions presented here. } 

\section{Summary and Conclusions}\label{conclusions}

This paper, the first in a series, utilizes a rich database of $\sim\,$9,100 galaxies with $0.2<z<2.5$ and $9.0<\log M_*/M_\odot<11.0$, taken from the GOODS-S and UDS regions of the CANDELS program, to study the overall demographics of SF galaxies, focusing on the $UVJ$ diagram. By dividing the sample into narrow mass and redshift slices, we have uncovered some new regularities in galaxy evolution and clarified and strengthened previously known ones. Our major findings are as follows:

\begin{enumerate}
	
	\item SF galaxies in the $UVJ$ diagram trace out a slanting two-dimensional distribution.  As modeled here using $\tau$-models reddened by a Calzetti foreground screen, these two dimensions are interpreted as variations in SSFR and dust reddening. Loci of constant SSFR trace out ``stripes'' that run along the long axis of the distribution.  The value of SSFR in each stripe is closely related to the coordinate \uvrot, which runs perpendicular to the long axis. We find a nearly universal trend between SSFR and \uvrot, indicating that a galaxy's SSFR can be estimated just from $U-V$ and $V-J$ (Equation \ref{uvrot_ssfr_fit}) with a scatter of only $\sim0.2$~dex. 
	
	\item The diagonal extent of the SF locus in the $UVJ$ diagram is mainly a dust sequence: galaxies with redder $V-J$ suffer higher visual attenuation ($A_V$). Moreover, dust attenuation (and, presumably, gas-phase metallicity) increase steadily with mass and time. The observed increase in $A_V$ is an example of ``mass-accelerated evolution'', i.e., more massive galaxies reach higher $A_V$ earlier. 

	\item The full UV--near-IR rest-frame SEDs of galaxies with the same $UVJ$ colors are strikingly similar, being (nearly) independent of mass and redshift. {A small population of galaxies is identified with brighter-than-average FUV continua.  SED fitting for these objects returns low values of $A_V$ and \ssfruv, moving them below the calibration in Equation \ref{uvrot_ssfr_fit}. Perhaps their stellar populations are not well fit by single $\tau$-models because they are composite (mixtures of old and young stars).}
	
	\item Galaxies in the dust-corrected $UVJ$ diagram generally lie close to solar-metallicity $\tau$-model tracks, but this is required by the SED-fitting procedure.  Over long times, galaxies flow from blue to red along the tracks, resulting in both the global decline of average SSFR on the SFMS and the gradual buildup of the quiescent population. Both trends occur faster in massive galaxies and are thus additional examples of mass-accelerated evolution. However, the persistent presence of SF galaxies with blue $U-V$ colors and young \emph{nominal} ages ($\lesssim0.8$~Gyr) at all masses down to $z = 0.5$ suggests that SFRs {may} fluctuate while galaxies are on the main sequence, broadening the SFMS and maintaining a population that looks blue and young until late times.
	
	\item A population of transition galaxies is identified with SSFRs more than a factor of three below the SFMS ridge line, between the SF and quiescent regions in the $UVJ$ diagram. Given the net flow of galaxies to the quiescent region, the majority of these objects must be fading.  They have systematically smaller radii and lower dust attenuation than main-sequence galaxies of the same mass and redshift, which suggests that falling SFR is associated with a loss of ISM. Transition galaxies enter the quiescent region from different directions depending on their dust contents: dusty galaxies enter from larger to smaller $V-J$ at nearly constant $U-V$ while dust-free galaxies become redder in both colors.  Galaxies with intermediate dust content move on tracks between these two extremes.
	
	\item The fractions of SF, quiescent, and transition galaxies are computed as a function of time and mass. The fraction of red galaxies increases smoothly with time with similar mass-accelerated evolution seen in the other parameters studied here. The basic aspects of galaxy evolution, at least after $z\sim2.5$, are fairly well-described as a function of time and mass. 
	
	\item An Appendix investigates agreement between our adopted measure of SFR, which uses the dust-corrected $L_{2800}$ luminosity, and SFR determined from UV+IR luminosities. {In addition to absolute rates, we also compare residuals about the SFMS, which is a more stringent test than used in previous studies.  The total random scatter within a given mass--redshift bin is 0.24~dex for the residual--residual comparison (0.17~dex for each quantity separately), but systematic zero point offsets of order 0.2~dex, varying with redshift, are also seen.  The far-IR luminosities of transition galaxies exceed their low SFRs, perhaps because of dust heating by older stellar populations.}
		
\end{enumerate}

In conclusion, it is worth cautioning yet again that certain key findings, notably $A_V$ and hence \ssfruv, depend on the CANDELS SED-fitting process, which assumes declining-$\tau$ stellar population histories and Calzetti reddening by a foreground screen.  In reality, the dust is not in a foreground screen, and actual stellar population histories are not $\tau$-models.  It will be interesting to see how conclusions based on $A_V$ change as refinements to both of these assumptions are made.

\acknowledgements

We thank the referee for extensive comments that led to major improvements and clarifications in the paper.

The main CANDELS \emph{HST} observations were supported under program number HST-GO-12060, provided by NASA through a grant from the Space Telescope Science Institute, which is operated by the Association of Universities for Research in Astronomy, Incorporated, under NASA contract NAS5-26555. The present work is also based in part on observations with the Spitzer Space Telescope, operated by the Jet Propulsion Laboratory, California Institute of Technology, under NASA contract 1407. \emph{Herschel} is an ESA space observatory with science instruments provided by European-led Principal Investigator consortia and with important participation from NASA. PACS on \emph{Herschel} has been developed by a consortium of institutes led by MPE (Germany) and including UVIE (Austria); KU Leuven, CSL, IMEC (Belgium); CEA, LAM (France); MPIA (Germany); INAF-IFSI/OAA/OAP/OAT, LENS, SISSA (Italy); IAC (Spain). This development has been supported by funding agencies BMVIT (Austria); ESA-PRODEX (Belgium); CEA/CNES (France); DLR (Germany); ASI/INAF (Italy); and CICYT/MCYT (Spain). SPIRE \emph{Herschel} has been developed by a consortium of institutes led by Cardiff University (UK) and including University of Lethbridge (Canada); NAOC (China); CEA, LAM (France); IFSI, University of Padua (Italy); IAC (Spain); Stockholm Observatory (Sweden); Imperial College London, RAL, UCL-MSSL, UKATC, University of Sussex (UK), Caltech, JPL, NHSC, University of Colorado (USA). This development has been supported by national funding agencies: CSA (Canada); NAOC (China); CEA, CNES, CNRS (France); ASI (Italy); MCINN (Spain); SNSB (Sweden); STFC, UKSA (UK); and NASA (USA).  JJF and members of the CANDELS team at UCSC acknowledge support from NASA HST grant GO-12060.10-A and NSF grant AST-0808133.

The \emph{Rainbow} database is operated by the Universidad Complutense de Madrid (UCM), partnered with the University of California Observatories at Santa Cruz (UCO/Lick,UCSC). The Rainbow database is partially funded by: the Spanish Programa Nacional de Astronom\'ia y Astrof\'isica under grants AYA2015-63650-P and AYA2015-70815-ERC and US NSF grant AST-08-08133.

Finally, we acknowledge the support of hundreds of individuals involved in the planning, execution, and reduction of the CANDELS observations, and to the development and installation of the new instruments on \emph{HST} that made them possible. We further acknowledge the work of several dozen CANDELS collaborators who produced the photometric, structural, and other value-added data catalogs that form the foundation of this work. Finally, we highlight the contributions of ground-based observatories, too numerous to mention, that provided critical photometric and spectroscopic ground-based data. The CANDELS catalogs are the culmination of a decades-long, successful partnership between space- and ground-based facilities to map the Universe.

{\it Facilities: HST, Spitzer, Herschel}

\appendix                              

{\section{Comparison of Dust-Corrected UV-Based SSFRs to IR-Based Values}}\label{uvj_appendix}
 
{Our use of \ssfruv\ values throughout this paper is mandated by the fact that traditional IR-based SFRs do not go deep enough to reach our target population of $M_* = 10^9 M_\odot$ at $z=2-2.5$.  The sources of the IR data were described in Section \ref{irdata}.  We are using the deepest observations that exist in these fields, and the reductions have been done to the faintest reliable levels.  Nevertheless, the far-IR data (even PACS $100\,\mu$m) are not deep enough to provide an unbiased sample. \emph{Spitzer}/MIPS $24\,\mu$m values are available for more objects but give a less reliable SFR. We demonstrate the limits of the IR data first and then compare our \ssfruv\ values to values based on MIPS $24\,\mu$m.}

{Figure \ref{histograms} shows the number of available galaxies in our four redshift ranges. The figure illustrates the scarcity of IR data at our target mass limit. For example, in the redshift bin $z = 0.5-1.0$, MIPS coverage in the smallest mass bin $\mathrm{log}\,M_* =9.0-9.5$ is only 11\%, and PACS coverage is practically zero.  At $z = 2.0-2.5$, both are essentially zero. The mass of the Milky Way at $z = 2.5$ was  $\sim10^{9.5} M_\odot$ \citep{papovich15}, and our goal of studying the star-formation histories of Milky Way progenitors at this redshift is impossible with existing IR data. }

\begin{figure}

	\centerline{\includegraphics[scale=0.5]{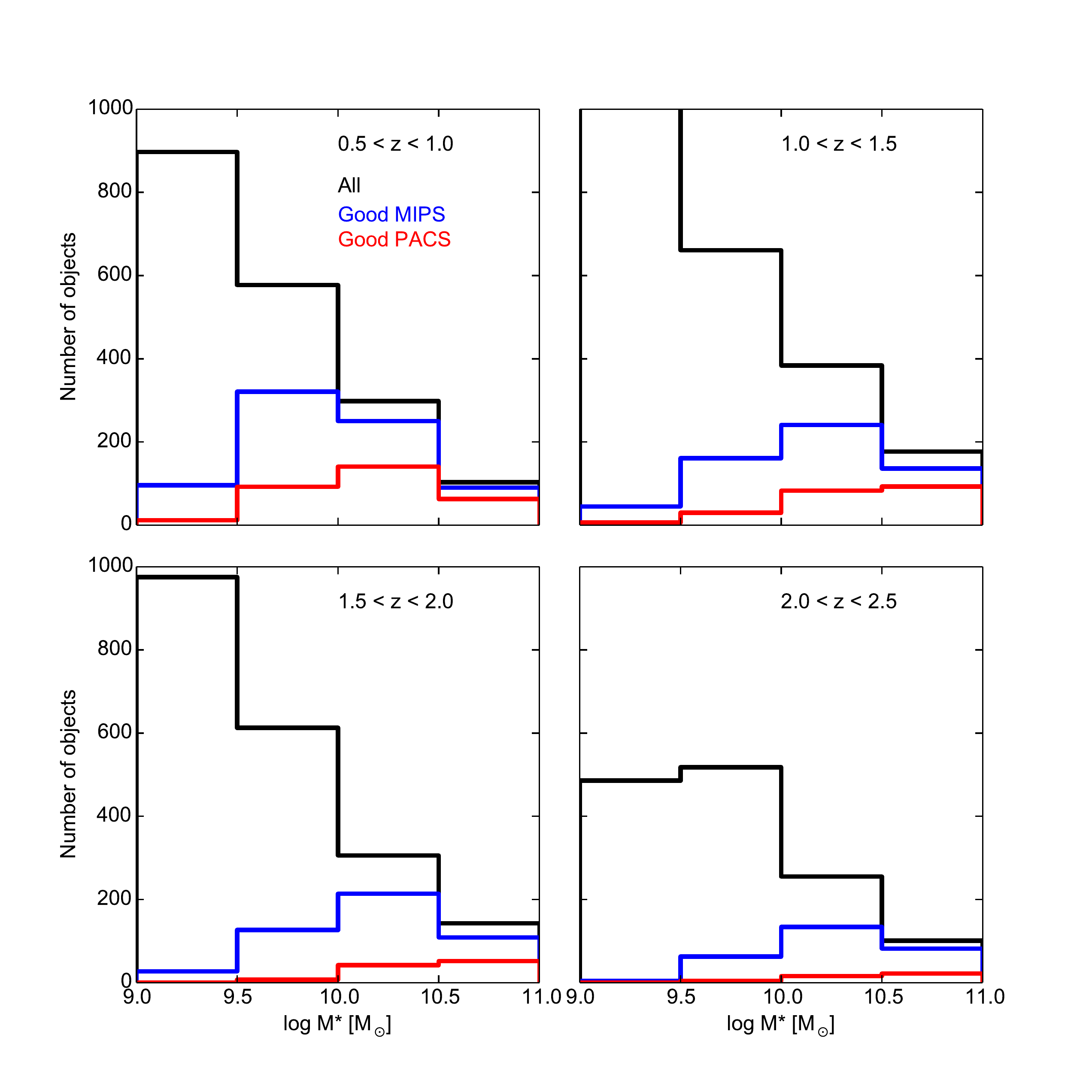}}
	
	\caption{{Histograms showing numbers of galaxies in four redshift bins versus stellar mass. The black histogram (``All'') shows the whole sample of 8,060 SF galaxies from Table \ref{uvj_sample} (this includes transition galaxies as well as SFMS galaxies). The overplotted blue histogram shows the galaxies in the All sample that have good \emph{Spitzer}/MIPS $24\,\mu$m data, and the overplotted red histogram shows the number of galaxies with good \emph{Herschel}/PACS data at $100\,\mu$m.  } }

	\label{histograms}
\end{figure}

{It is still of interest to ask how well values of \ssfruv\ agree with IR data when the latter are available. For this comparison, we use MIPS $24\,\mu$m values because that sample is larger. Our adopted \lir\ is determined from the $24\,\mu$m-based conversion of \citet{rujopakarn13}, denoted \lir\ (R13).  The choice of \lir\ (R13) as a standard is motivated by their claim that it provides a more robust estimate of \lir\ when only $24\,\mu$m is available. The R13 method corrects \lir\ downward as compared to values from local templates, especially above $z = 1.5$, where observed $24\,\mu$m probes PAH emission \citep[rest-frame 8$-$12\,$\mu$m;][]{tielens08}. The justification for the correction given by \citet{rujopakarn13} is that PAH emission strength increases at high redshift due to the fact that distant SF regions are physically more extended than local ones, which increases the available surface area of the photodissociation regions from which PAH emission originates. To account for this effect, the R13 method rescales the local IR templates to correct for the redshift evolution in the SED shape. This conversion produces \lir\ values in reasonable agreement (0.13~dex scatter) with those based on direct far-IR measurements from \emph{Herschel} (Figures 3 and 4 of R13).}

The \lir\ (R13) values were converted to \ssfruvir\ using the formula of \citet{wuyts11a}:
\begin{equation}
	\mathrm{SFR_{UV+IR}}\ [M_\odot\,\mathrm{yr}^{-1}]=1.09\times10^{-10}\left(L_\mathrm{IR}+3.3L_\mathrm{NUV}\right)\,[L_\odot],
	\label{sfr_uvir}
\end{equation}
where \lir\ is the integrated 8$-$1000\,$\mu$m luminosity, and $L_\mathrm{NUV}\equiv\nu L_\nu(2800\,\mathrm{\AA})$ is the rest-frame near-UV luminosity measured at 2800\,\AA. The effective coefficient in front of the UV term in Equation \ref{sfr_uvir} is $3.60\times10^{-10}$, whereas our adopted conversion factor (Equation \ref{sfr_uv}) to compute \ssfruv\ is $2.59\times10^{-10}$, which is $\approx25\%$ smaller. This translates to a  $\sim$0.1~dex offset between the two rates, which is small compared to the total scatter in \ssfruvir. Because we use the R13 method to derive \lir, we label these rates as \ssfruvir\ (R13).


{The Calzetti attenuation curve implies
\begin{equation}
   A_{2800} = 1.8A_V = 2.5 \log\left(\frac{\mathrm{SSFR_{UV+IR}}}{\mathrm{SSFR_{UV}}}\right),
	\label{a2800}
\end{equation}
where $\mathrm{SSFR_{UV}}$ is the raw UV SSFR uncorrected for absorption. That means that $A_{2800}$ versus the ratio on the right-hand side should follow the 1-to-1 line, or alternatively, that $A_V$ versus the ratio should follow a line of slope $2.5/1.8=1.4$. This prediction is tested in Figure \ref{av}. Agreement for the deeper GOODS-S sample is quite good: the correlations are strong in most panels with an rms scatter in $A_V$ of about 0.35 mag (after rejecting $3\sigma$ outliers and without making any correction for errors in the SSFR ratio). The shallower UDS points  also follow the relations but with larger scatter. The points scatter more at high redshift, and varying systematic offsets for GOODS-S of about $\pm0.3$~mag are evident, but it appears overall that $A_V$ from fitting the UV--optical spectrum is capable of \emph{rank ordering} galaxies by $A_V$ in a given mass--redshift bin to better than 0.35~mag. This agrees with the findings of \citet{arnouts13} and \citet{forrest16}, who likewise compared $A_V$ to other reddening measures.}

\begin{figure}

	\centerline{\includegraphics[scale=0.5]{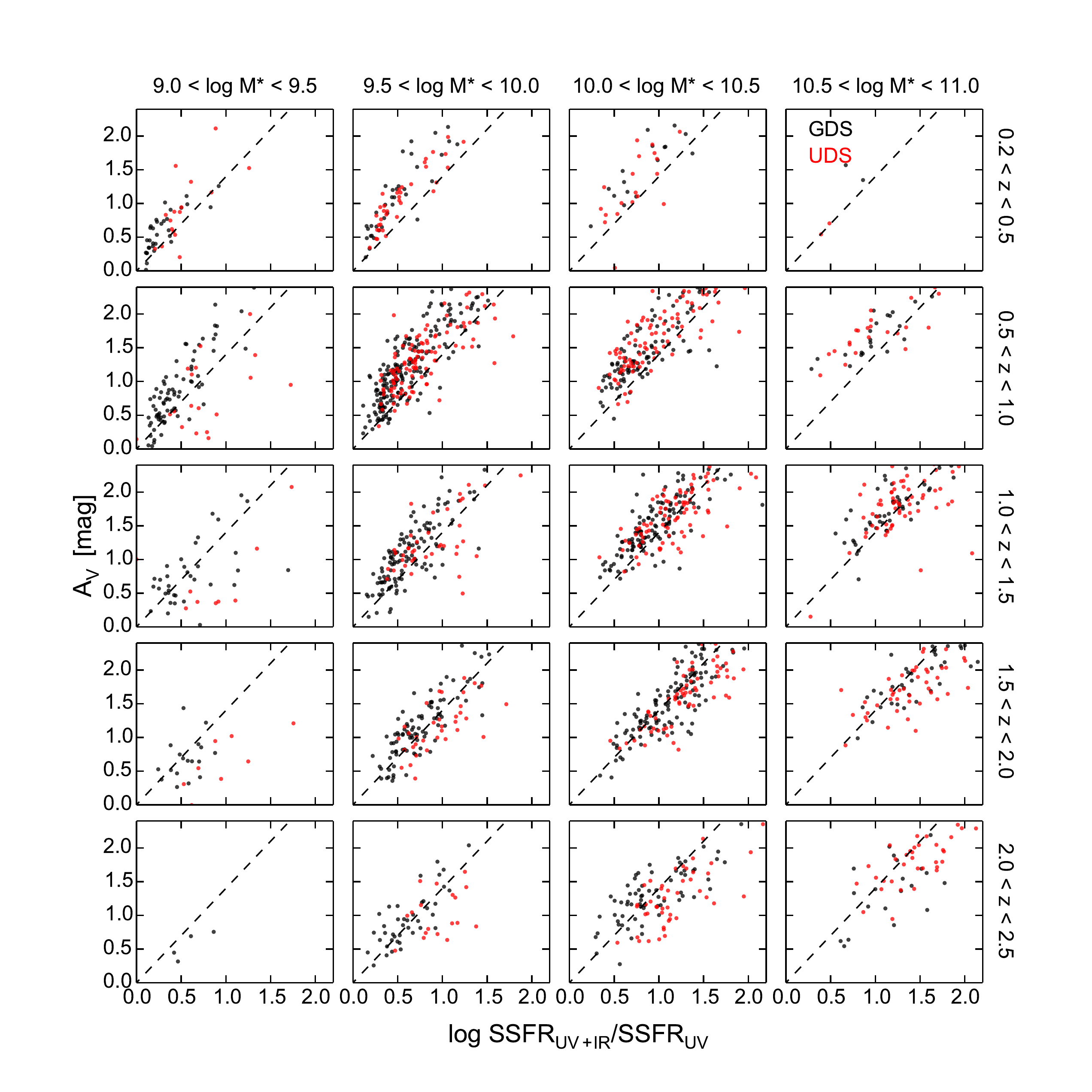}}
	
	\caption{{$A_V$ values from SED fitting to SF galaxies vs.~the ratio of specific star-formation rate \ssfruvir\ to the raw, uncorrected UV rate, $\mathrm{SSFR_{UV}}$. The sample used is galaxies on the ridge line of the SFMS (\delssfr\,$> -0.45$~dex in Figure \ref{delssfr_delsma}) with good MIPS 24\,$\mu$m \lir\ values.  GOODS-S galaxies are in black; the shallower UDS sample is in red. Dashed lines show the predictions of Equation \ref{a2800}. Correlations are good in most panels, with offsets of about $\pm0.3$~mag, depending on redshift. The rms scatter about the lines for the GOODS-S sample is typically $\sim0.35$~mag.  This is the maximum scatter in $A_V$ per galaxy, not allowing for any error in the SSFR ratio.} }
	\label{av}
\end{figure}

{Figure \ref{sfruv_sfruvir} now compares \sfruvir\ versus \sfruv. The dashed line is the one-to-one line.  Agreement is again good with an offset of $-0.03$~dex and a total rms scatter of 0.23~dex for GOODS-S. (UDS scatters slightly more, 0.3~dex.) Assigning error bars equally to both quantities would yield 0.16~dex for \sfruv\ alone. This scatter is consistent with what we would predict based on the scatter in $A_V$ in Figure \ref{av}. That is, adopting an average 0.35-mag scatter in $A_V$ ($0.63$ mag scatter in $A_{2800}$) results in a scatter of 0.25~dex in SFR$_\mathrm{UV,corr}$. This is generally where most comparisons leave off, using total SFRs. But this is not adequate for studying galaxy properties above and below the SFMS, which is the goal of future papers. For this, accurate \emph{residuals} are needed, which is more challenging. Absolute values can produce good-looking correlations because they cover several dex, and yet they may fail to properly rank galaxies by their residuals, which are much smaller. }

\begin{figure}

	\centerline{\includegraphics[scale=0.6]{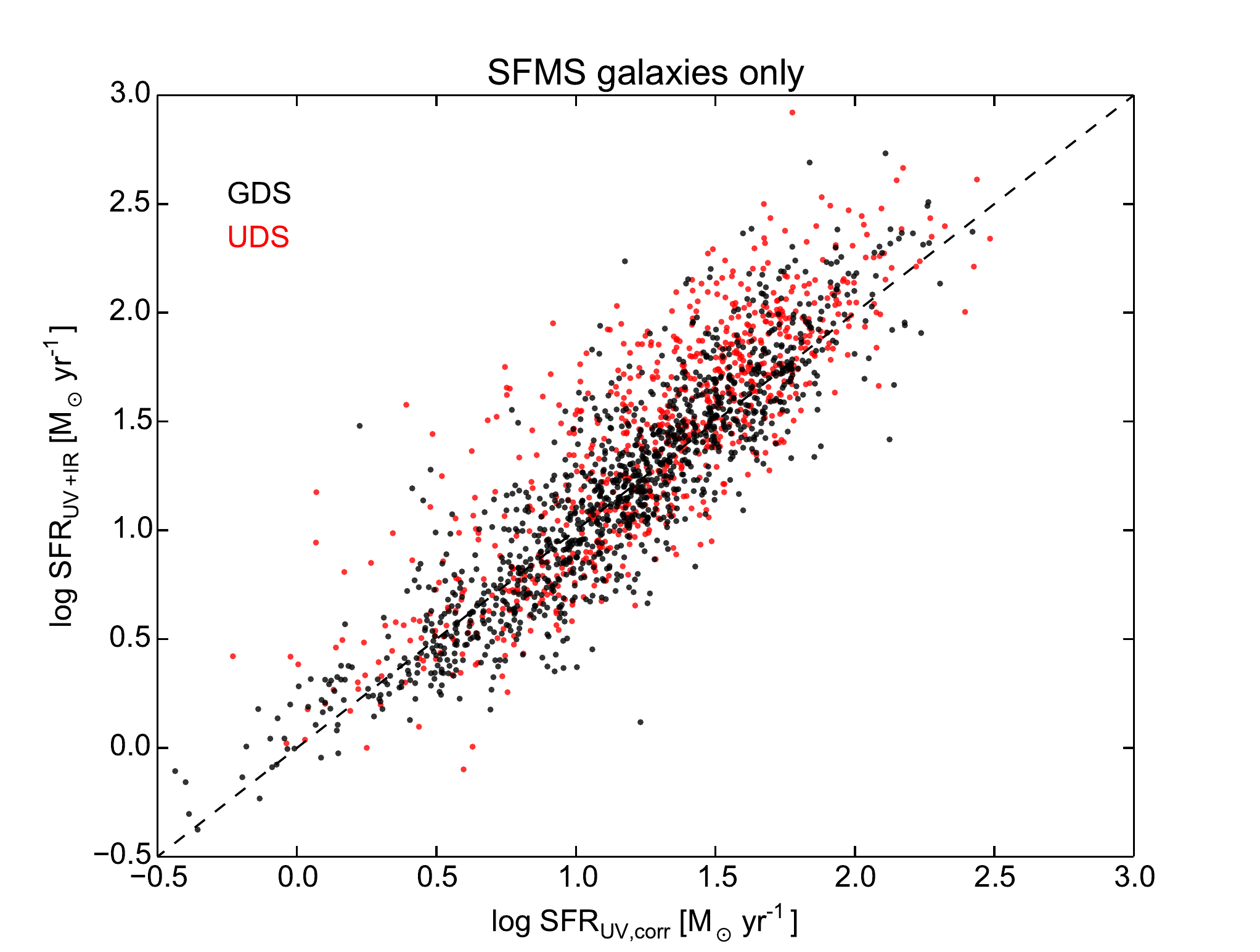}}
	
	\caption{{\sfruvir\ based on MIPS 24$\mu$m vs.~\sfruv\ from this paper. The samples are the same main-sequence ridge line galaxies as in Figure \ref{av}. The black points are from GOODS-S; the red points are from UDS.  The dashed line shows equality between the two measures. The total rms scatter of the black points is 0.23~dex; the shallower UDS sample scatters slightly more (0.3~dex).} }
	 
	\label{sfruv_sfruvir}
\end{figure}

{The more stringent test is shown in Figure \ref{ssfruv_ssfruvir}, which compares residuals in both measures of SSFR with respect to the main-sequence ridgeline. To our knowledge, this test has never been shown before. Three conclusions are evident. First, agreement using GOODS-S data is reasonable in most panels; UDS as usual scatters more. The rms residual scatter per panel is 0.24 dex for GOODS-S (for SFMS galaxies, after rejecting $3\sigma$ outliers).  Assigning this scatter equally yields an rms internal scatter of 0.17~dex for \ssfruv\ alone.  This is the uncertainty that is relevant to ranking galaxies by their residual in a given mass--redshift bin.  The level of agreement is actually remarkable, given the high dust content of many of these objects, for which the $L_{2800}$ corrections approach 2~dex (see Figure \ref{av}). That said, \ssfruv\ does tend to overestimate the SFR at low redshift but underestimate it at high redshift.  The offsets reach up to $\sim$0.2 dex. Similar trends are seen in Figure \ref{av}.  Finally, the scatter is perceptibly larger for massive galaxies at high redshift.  These tend to be very dusty, and it is possible that much of the star formation is simply not revealed in UV--optical light and that $A_V$ is too low.  This merits further follow-up.  All in all, this test confirms acceptable agreement between \ssfruv\ and \ssfruvir\ (R13) \emph{for main-sequence ridge line galaxies} and establishes the utility of \ssfruv\ to study properties above and below the SFMS. }

{Although both measures of SSFR agree reasonably well, the systematic trends seen in Figures \ref{av} and \ref{ssfruv_ssfruvir} could suggest the need for a redshift-dependent correction to \ssfruv. Figure \ref{av} indicates that at the highest redshifts, $A_V$ is underestimated by $0.3$~mag, while at the lowest redshifts it is overestimated by $0.3$~mag. Making this correction boosts \ssfruv\ by $\sim0.2$~dex at the highest redshifts and reduces it by the same amount at the lowest redshifts, consistent with the offsets seen in Figure \ref{ssfruv_ssfruvir}. Galaxies in between these redshifts suffer smaller corrections. The net result is to broaden the dynamic range of \ssfruv\ across redshift. It is possible that corrections should be applied to \ssfruv\ to account for this effect. In particular, the offsets at low redshift seen in Figure \ref{av} may be due to the presence of composite (young+old) stellar populations, which the SED fitting methods used here do not include in their models. Indeed, \citet{wang17} find that fitting a composite model with a $\tau$-model results in an overestimate of both $A_V$ (by up to $\sim1$ mag) and SSFR (by up to a factor of $\sim3$). We note in passing that this effect is stronger for transition galaxies (SSFR is overestimated by $\sim5\times$), meaning that the width of the green valley in \delssfr\ may be compressed relative to the SFMS. On the other hand, at high redshift, the offsets seen in Figure \ref{ssfruv_ssfruvir} go the other way. This may be due to uncertainties in modeling the PAH region of the IR SEDs, making it challenging to accurately recover the IR flux based on e.g., 24$\mu$m measurements alone. It is also worth noting that there is a long-standing discrepancy between integrating the instantaneous cosmic SFR density compared to the cosmic stellar mass density, in the sense that integrating the SFR density overproduces stars by $\sim0.2$~dex by $z\sim1$ \citep{madau14}. Using our (lower) values of SSFR at $z\sim2$ would bring these two into better agreement. }  

{Given the uncertainties discussed above, it is not obvious which system, \ssfruvir\ or \ssfruv, is the ``truth''. Hence, we opt not to apply any corrections to \ssfruv\ in this paper, pending further investigations into the systematics of SED fitting at high and low redshifts. For now, it is good to know that there appear to be SSFR ``systems'' (analogous to the photometric systems of old), that corrections among them are of order 0.2~dex, but that these corrections are not large enough to disturb the relative rankings of SSFR from one object to another, especially if these are done in restricted bins of mass and redshift. } 

\begin{figure}

	\centerline{\includegraphics[scale=0.5]{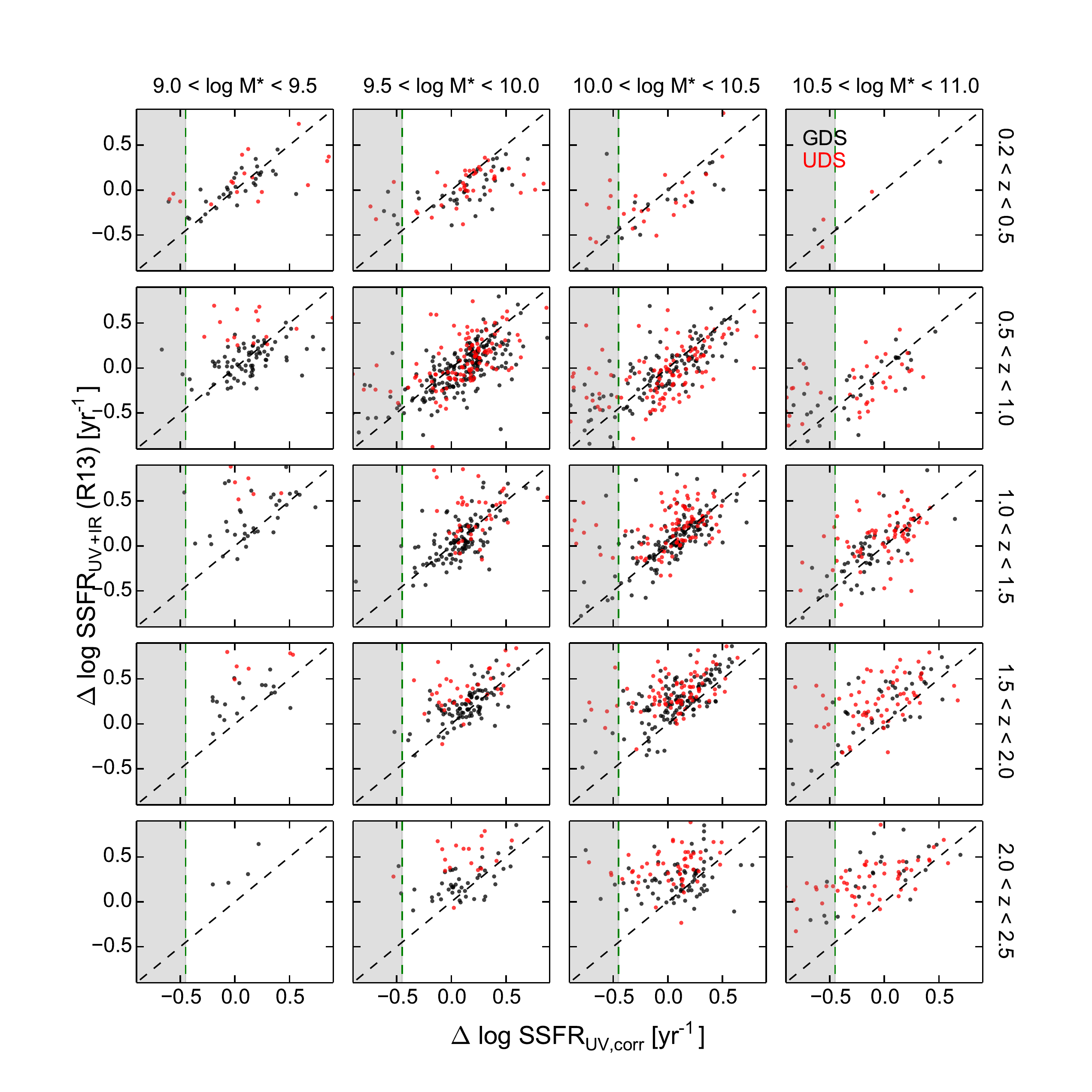}}
	
	\caption{{Residuals in \ssfruvir\ based on MIPS 24$\mu$m vs.~values of \ssfruv\ from this paper.  Residuals are calculated by subtracting the linear fits of SSFR vs.~$M_*$ in Table \ref{ssfr_fit} from both quantities, and so any systematic zeropoint offset in each panel is preserved. The samples are the same main-sequence ridge line galaxies as in Figures \ref{av} and \ref{sfruv_sfruvir} except that transition galaxies more than $-0.45$~dex below the fits in Figure \ref{ssfr_mass} are also added (grey regions). Black points are from GOODS-S; red points are from UDS. For SFMS galaxies, systematic offsets of up to $\sim0.2$~dex are apparent that vary with redshift. The rms scatter is 0.24~dex, or 0.17~dex if assigned equally to each quantity. In contrast to the main sequence, transition galaxies have systematically high values of \ssfruvir\ compared to \ssfruv\ by up to 1~dex.} }
	\label{ssfruv_ssfruvir}

\end{figure}

{We turn attention now to the transition galaxies.  For them, \ssfruvir\ (R13) overestimates SSFR by 0.16 dex on average and by more than 1~dex in some cases.  In general, the traditional $24\,\mu$m \lir\ method seems to overestimate SFRs for SF galaxies well below the SFMS.} A similar increasing offset between \ssfruvir\ and \ssfruv\ at low SFR has been seen at least four times in previous investigations. \citet{patel11} employed three different SFR estimators to measure the decline in SFR as galaxies fall into dense environments.  Good agreement was obtained between the SED-fitting method and [\ion{O}{2}] strength (both Calzetti-corrected), but a much smaller decline was seen using \lir-based SFRs.  They hypothesized that \lir\ overestimates SFR at low star-formation levels.  A similar conclusion was reached by \citet{salim09}, who compared SFRs from SED fitting to \emph{Spitzer}/MIPS $24\,\micron$ fluxes.  The MIPS values seemed consistently too high, especially for low-SFR galaxies.  Next, \citet{arnouts13} developed a process to estimate \lir\ from rest-frame NUV$-r$ and $r-K$ colors. Their method effectively calibrated \lir\ as a function of these two colors based on active SF galaxies, but it gave implausibly high \lir\ values when applied to low-SFR galaxies. The failed objects lie adjacent to the quiescent region, which is identified in the present paper with galaxies in transition (Figure \ref{uvj_fading}). The fourth finding is by \citet{utomo14}, who stacked optical--MIPS SEDs for NEWFIRM galaxies in various SFR bins.  MIPS IR luminosities at $24\,\micron$ overestimated SFRs by up to one dex compared to NUV--near-IR SED stellar population models with the discrepancy increasing smoothly toward lower SFRs.  Rates for the highest-SF galaxies agreed.  This is consistent with what we see in Figure \ref{ssfruv_ssfruvir}.

Some of the above authors have hypothesized that the $24\,\mu$m flux in low-SFR objects comes, at least in part, from sources other than dust heated by star formation. For example, dust may be heated by old stars \citep[e.g.,][]{helou86,sauvage94,calzetti95,kennicutt98,draine01,salim09} or it may be hotter (and thus radiate more efficiently at $24\,\mu$m) when SSFR is low \citep{skibba11}. At $z\sim2$, where observed $24\,\mu$m is dominated by PAH features, the effect is compounded by the fact that PAH molecules can also be excited by cooler stars in the diffuse ISM \citep{li02,calzetti07}. Finally, at \emph{very} low SSFR, $24\,\mu$m flux can come directly from old stars themselves \citep[Figure 1 of][]{skibba11}.

{In conclusion, we find a population of transition galaxies for which \ssfruvir\ is higher than \ssfruv. These objects have log SSFR values between $-9.5$ to $-10.0$ in Figure \ref{ssfr_uvrot}, with \csed\ values that match.  Increasing their SSFR values by $\sim$0.2~dex (to bring them in line with \ssfruvir) would place them off the relation in that figure established by both bluer and redder galaxies, raising the question of where this deviation comes from. As discussed above, several works have provided abundant evidence from different directions that \ssfruvir\ is likely overestimated for transition galaxies.}


{The structural properties of the transition galaxies also are consistent with their having low SSFR. Figure \ref{delssfr_delsma} shows that transition galaxies have lower $A_V$, as would be expected if their ISM is disappearing as star formation ends. Their radii are also smaller, consistent with star formation fading in more extended disks \citep{fang13}. \citet{pandya16} detected higher S\'ersic indices in transition galaxies, which likewise would occur naturally when fading disks are outshone by central bulges.  }

{\section{The SFMS Derived from \ssfruv\ vs.~Other Values}}\label{mainsequence_appendix}

{An alternative way to assess the accuracy of \ssfruv\ values is to compare to SF main sequences found by others.  For this, we use the extensive database on mean main-sequence measurements in the literature, as tabulated and described by \citet{rodriguezpuebla17}. Figure \ref{ssfr_mass_aldo} plots SSFR versus mass in redshift bins, while Figure \ref{ssfr_redshift_aldo} plots SSFR versus redshift in mass bins. Overall, agreement between \ssfruv\ and other values is good with the zero points agreeing well on average.  However, there are systematic zero point offsets that vary with redshift.  In Figure \ref{ssfr_mass_aldo}, \ssfruv\ is approximately 0.2~dex higher at low redshift and 0.3~dex lower than average at high redshift. This means we tend to underestimate the increase in average SFR from low to high redshift, as shown more directly in Figure \ref{ssfr_redshift_aldo}. Our slopes also tend to be too flat at low $z$ and too steep at high $z$, in contrast to slopes found by others, which turn over strongly at high mass and low redshift \citep[e.g.,][]{whitaker12, lee15, barro17}. Part of the latter effect is because we define the main-sequence ridge line narrowly, taking only galaxies $>-0.45$~dex while others have retained all SF galaxies according to $UVJ$.   This typically includes some transition galaxies at high mass, which pulls the ridgeline down. These systematic residuals, though small, are at a level where they could easily confuse attempts to order galaxies by their residuals about the main sequence, {but not, as we noted above, if comparisons are restricted to narrow mass and redshift ranges.} Previous studies have often filled in missing IR values with SED-based values \citep[the so-called ``ladder" approach,][]{wuyts11a}. Such a mixture clearly has the potential to introduces significant systematic errors, and it is precisely for this reason that we have preferred to use a single SSFR method, \ssfruv, throughout this paper. }

\begin{figure}

	\centerline{\includegraphics[scale=0.75]{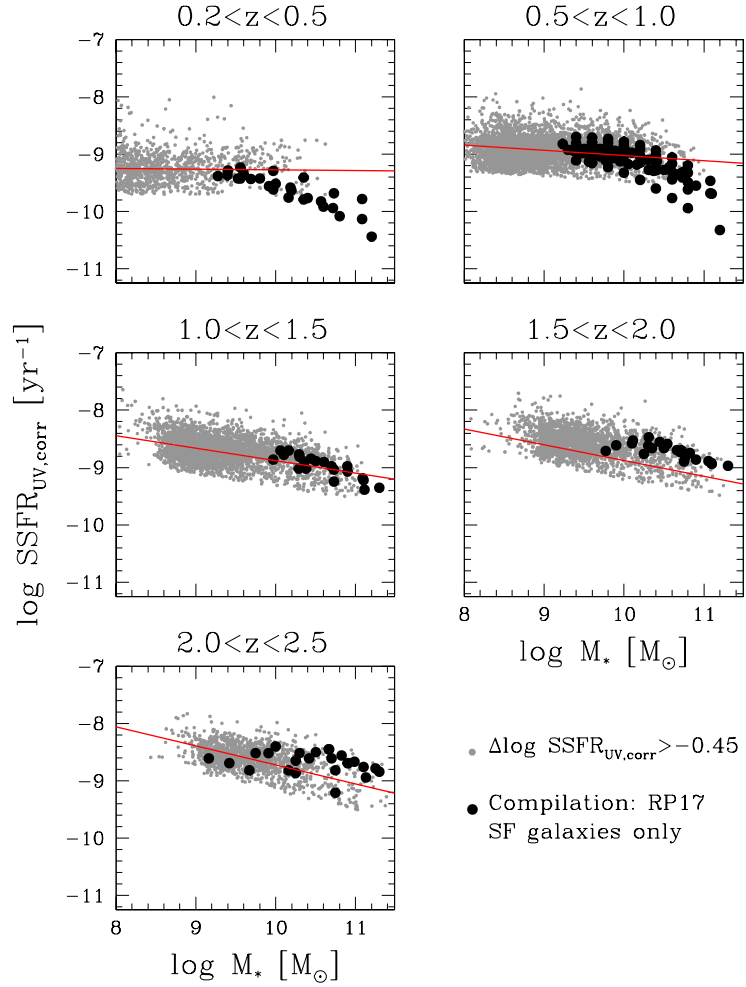}}
	
	\caption{{\ssfruv\ for galaxies within 0.45~dex of the ridge line in Figure \ref{ssfr_mass}  plotted vs.~stellar mass in redshift bins.  Main-sequence ridge line galaxies with \delssfr$>-0.45$~dex are shown in grey. Mean main-sequence data points from the literature \citep{rodriguezpuebla17} are shown as black filled circles. Overall zero point agreement is good, but systematic offsets of order $0.2-0.3$ dex vary with mass and redshift.}}
	\label{ssfr_mass_aldo}
\end{figure}


\begin{figure}


	\centerline{\includegraphics[scale=0.75]{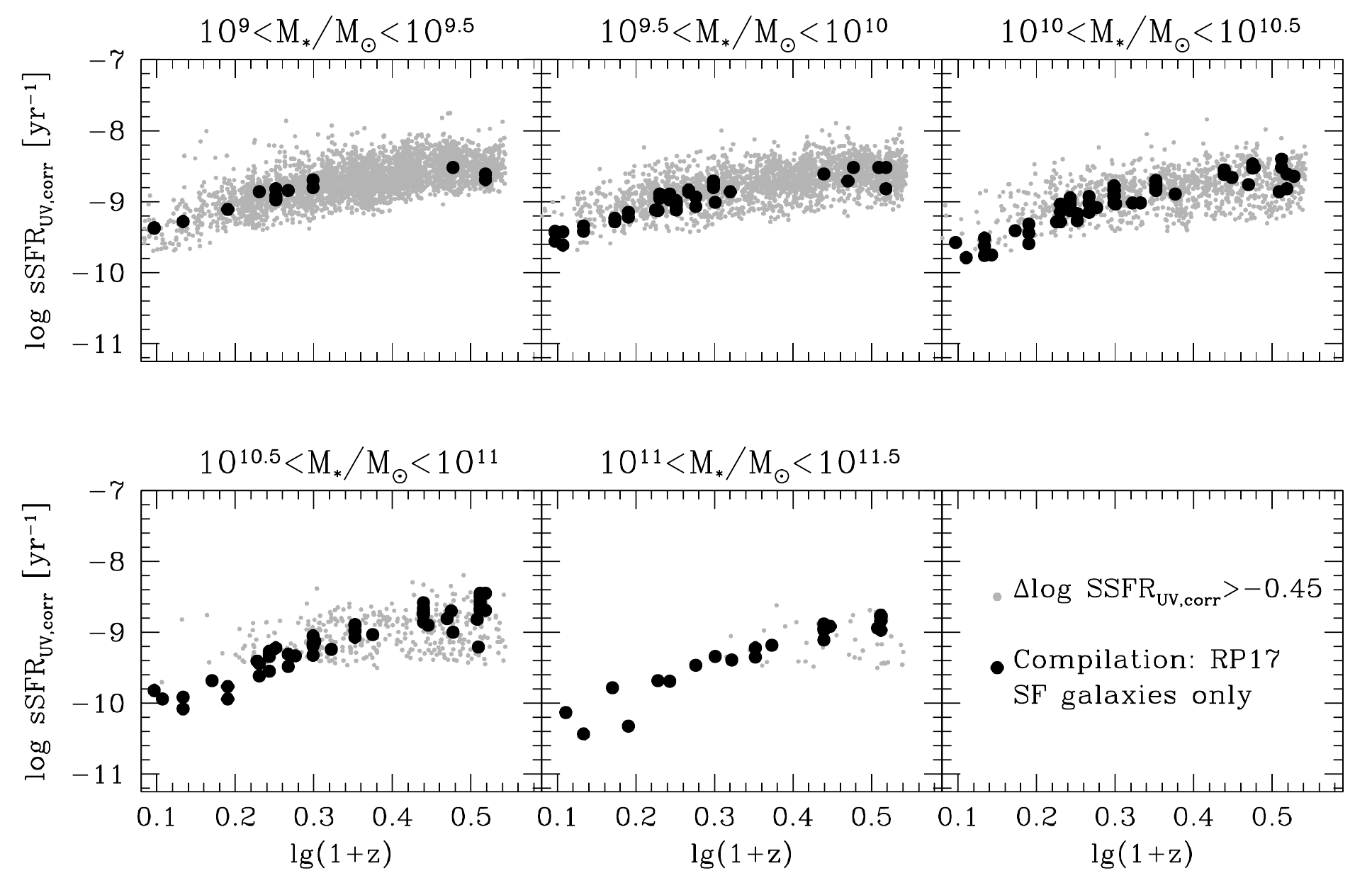}}
	
	\caption{{Same points as in Figure \ref{ssfr_mass_aldo} but now binned by mass and plotted vs.~redshift.  Overall agreement is again good, but the tendency of \ssfruv\ to underestimate the increase in SFR back in time is evident, especially at higher masses.}}
	
	\label{ssfr_redshift_aldo}
\end{figure}



\end{document}